\documentclass{tac}

\def\Uppercase#1{\def\i{I}\def\j{J}\def\ss{SS}\uppercase{#1}} 

\newtheorem{observation}{Observation}

\usepackage{stmaryrd}

\usepackage{amssymb}
\usepackage{amsmath}
\usepackage{latexsym}
\usepackage{paralist}
\usepackage{calrsfs}
\usepackage[curve,matrix,arrow]{xy}

\usepackage{pstricks-add}

\usepackage{hyperref}

\newdimen\dercldim                                
\newdimen\derccdim                                
\newdimen\dercrdim                                
\newdimen\derldim                                 
\newdimen\dercdim                                 
\newdimen\derrdim                                 
\newdimen\derdim                                  
\newdimen\derdldim                                
\newdimen\derdrdim                                
\newbox\derboxone                                 
\newbox\derboxtwo                                 
\newbox\derboxthree                               
\newbox\derboxfour                                
\newdimen\derquad\derquad=\fontdimen6\textfont2
\newdimen\deropen\deropen=\fontdimen5\textfont2\divide\deropen by3
\def\leaf #1{\global\setbox\derboxone=\hbox{\strut$#1$}%
   \global\derldim=0pt                            
   \global\dercdim=\wd\derboxone                  
   \global\derrdim=0pt                            
   }%
\def\rootaux #1#2#3{\setbox\derboxtwo=\hbox{\unhbox\derboxone}%
   \setbox\derboxthree=\hbox 
      {$\smash{\lower\fontdimen22\textfont2\hbox{$#1$}}$}%
   \setbox\derboxfour=\hbox 
      {$\smash{\lower\fontdimen22\textfont2\hbox{$#2$}}$}%
   \leaf{#3}
   \derdim=\dercdim\advance\derdim by-\derccdim\divide\derdim by2 
   \global\derldim=\dercldim\global\advance\derldim by-\derdim
   \global\derrdim=\dercrdim\global\advance\derrdim by-\derdim
   \deropen=\fontdimen5\textfont2\divide\deropen by3
   \setbox\derboxone=\hbox{\vbox{\offinterlineskip
         \hbox{\ifdim\derldim<0pt\kern-\derldim\fi
               \box\derboxtwo
               \ifdim\derrdim<0pt\kern-\derrdim\fi}%
         \kern\deropen
         \hbox{\ifdim\dercldim>\derldim
                  \ifdim\derldim>0pt\kern\derldim\fi
                  \else\kern\dercldim\fi
               \hbox to0pt{\hss\copy\derboxthree}%
               \vbox{\ifdim\derccdim>\dercdim\hsize=\derccdim
                                        \else\hsize=\dercdim \fi
                    \hrule height.2pt depth.2pt width\hsize}%
               \hbox to0pt{\copy\derboxfour\hss}%
               \ifdim\dercrdim>\derrdim
                  \ifdim\derrdim>0pt\kern\derrdim\fi
                  \else\kern\dercrdim\fi}%
         \kern\deropen
         \hbox{\ifdim\derldim>0pt\kern\derldim\fi
               \box\derboxone
               \ifdim\derrdim>0pt\kern\derrdim\fi}}}%
   \ifdim\derldim<0pt\global\derldim=0pt\fi       
   \ifdim\derrdim<0pt\global\derrdim=0pt\fi       
   \derdldim=\wd\derboxthree\advance\derdldim by-\dercldim
   \derdrdim=\wd\derboxfour \advance\derdrdim by-\dercrdim
   \ifdim\derdim<0pt
      \ifdim\derdldim<0pt
         \derdldim=0pt                            
      \fi
      \ifdim\derdrdim<0pt
         \derdrdim=0pt                            
      \fi
   \else
      \ifdim\derldim>0pt
         \ifdim\derdldim>-\derdim
            \advance\derdldim by\derdim           
         \else			                          
            \derdldim=0pt                         
         \fi                                      
      \else
         \advance\derdldim by\dercldim            
      \fi
      \ifdim\derrdim>0pt
         \ifdim\derdrdim>-\derdim
            \advance\derdrdim by\derdim           
         \else			                          
            \derdrdim=0pt                         
         \fi                                      
      \else
         \advance\derdrdim by\dercrdim            
      \fi
   \fi
   \global\setbox\derboxone=\hbox
      {\kern\derdldim\unhbox\derboxone\kern\derdrdim}%
   \global\advance\derldim by\derdldim            
   \global\advance\derrdim by\derdrdim            
   }%
\def\rootr #1#2#3#4{{#4}%
   \dercldim=\derldim
   \derccdim=\dercdim
   \dercrdim=\derrdim
   \rootaux{#1}{#2}{#3}}%
\def\rrootr #1#2#3#4#5{\derquad=\fontdimen6\textfont2
   {#4}%
           \dercldim  =\derldim
   \setbox\derboxtwo=\hbox{\unhbox\derboxone\kern\derquad}%
           \derccdim  =\dercdim
   \advance\derccdim by\derrdim
   \advance\derccdim by\derquad
   {#5}%
   \setbox\derboxone=\hbox{\unhbox\derboxtwo\unhbox\derboxone}%
   \advance\derccdim by\derldim
   \advance\derccdim by\dercdim
           \dercrdim  =\derrdim
   \rootaux{#1}{#2}{#3}}%
\def\rrrootr #1#2#3#4#5#6{\derquad=\fontdimen6\textfont2
   {#4}%
           \dercldim  =\derldim
   \setbox\derboxtwo=\hbox{\unhbox\derboxone\kern\derquad}%
           \derccdim  =\dercdim
   \advance\derccdim by\derrdim
   \advance\derccdim by\derquad
   {#5}%
   \setbox\derboxtwo=\hbox{\unhbox\derboxtwo\unhbox\derboxone\kern\derquad}%
   \advance\derccdim by\derldim
   \advance\derccdim by\dercdim
   \advance\derccdim by\derrdim
   \advance\derccdim by\derquad
   {#6}%
   \setbox\derboxone=\hbox{\unhbox\derboxtwo\unhbox\derboxone}%
   \advance\derccdim by\derldim
   \advance\derccdim by\dercdim
           \dercrdim  =\derrdim
   \rootaux{#1}{#2}{#3}}%
\def\root       #1#2#3{\rootr  {#1\;}{\;}{#2}{#3}}%
\def\rroot    #1#2#3#4{\rrootr {#1\;}{\;}{#2}{#3}{#4}}%

\newbox\stembox
\def\stemaux #1#2#3{\setbox\derboxtwo=\hbox{\unhbox\derboxone}%
   \setbox\derboxthree=\hbox{$#1$}   
   \setbox\derboxfour =\hbox{$#2$}   
      {\global\setbox\derboxone=\hbox{$#3$}%
      \global\derldim=0pt                         
      \global\dercdim=\wd\derboxone               
      \global\derrdim=0pt                         
      }
   \derdim=\dercdim\advance\derdim by-\derccdim\divide\derdim by2 
   \global\derldim=\dercldim\global\advance\derldim by-\derdim
   \global\derrdim=\dercrdim\global\advance\derrdim by-\derdim
   \deropen=\fontdimen5\textfont2\divide\deropen by3
   \setbox\derboxone=\hbox{\vbox{\offinterlineskip
         \hbox{\ifdim\derldim<0pt\kern-\derldim\fi
               \box\derboxtwo
               \ifdim\derrdim<0pt\kern-\derrdim\fi}%
         \kern-\deropen\kern-\ht\strutbox\kern-\dp\strutbox
         \hbox{\ifdim\dercldim>\derldim
                  \ifdim\derldim>0pt\kern\derldim\fi
                  \else\kern\dercldim\fi
               \hbox to0pt{\hss\copy\derboxthree}%
               \vbox{\ifdim\derccdim>\dercdim\hsize=\derccdim
                                        \else\hsize=\dercdim \fi
                    \hbox{$\vcenter{
                    \vbox{\offinterlineskip
                       \hbox{$\copy\stembox$}\kern-1\dp\stembox
                       \hbox{$\copy\stembox$}}}$}}%
               \hbox to0pt{\copy\derboxfour\hss}%
               \ifdim\dercrdim>\derrdim
                  \ifdim\derrdim>0pt\kern\derrdim\fi
                  \else\kern\dercrdim\fi}%
         \kern-\deropen
         \hbox{\ifdim\derldim>0pt\kern\derldim\fi
               \box\derboxone
               \ifdim\derrdim>0pt\kern\derrdim\fi}}}%
   \ifdim\derldim<0pt\global\derldim=0pt\fi       
   \ifdim\derrdim<0pt\global\derrdim=0pt\fi       
   \derdldim=\wd\derboxthree\advance\derdldim by-\dercldim
   \derdrdim=\wd\derboxfour \advance\derdrdim by-\dercrdim
   \ifdim\derdim<0pt
      \ifdim\derdldim<0pt
         \derdldim=0pt                            
      \fi
      \ifdim\derdrdim<0pt
         \derdrdim=0pt                            
      \fi
   \else
      \ifdim\derldim>0pt
         \ifdim\derdldim>-\derdim
            \advance\derdldim by\derdim           
         \else			                  
            \derdldim=0pt                         
         \fi                                      
      \else
         \advance\derdldim by\dercldim            
      \fi
      \ifdim\derrdim>0pt
         \ifdim\derdrdim>-\derdim
            \advance\derdrdim by\derdim           
         \else			                  
            \derdrdim=0pt                         
         \fi                                      
      \else
         \advance\derdrdim by\dercrdim            
      \fi
   \fi
   \global\setbox\derboxone=\hbox
      {\kern\derdldim\unhbox\derboxone\kern\derdrdim}%
   \global\advance\derldim by\derdldim            
   \global\advance\derrdim by\derdrdim            
   }%
\def\stemauxx #1#2#3{\setbox\derboxtwo=\hbox{\unhbox\derboxone}%
   \setbox\derboxthree=\hbox{$#1$}   
   \setbox\derboxfour =\hbox{$#2$}   
   \leaf{#3}
   \derdim=\dercdim\advance\derdim by-\derccdim\divide\derdim by2 
   \global\derldim=\dercldim\global\advance\derldim by-\derdim
   \global\derrdim=\dercrdim\global\advance\derrdim by-\derdim
   \deropen=\fontdimen5\textfont2\divide\deropen by3
   \setbox\derboxone=\hbox{\vbox{\offinterlineskip
         \hbox{\ifdim\derldim<0pt\kern-\derldim\fi
               \box\derboxtwo
               \ifdim\derrdim<0pt\kern-\derrdim\fi}%
         \kern\deropen
         \hbox{\ifdim\dercldim>\derldim
                  \ifdim\derldim>0pt\kern\derldim\fi
                  \else\kern\dercldim\fi
               \hbox to0pt{\hss\copy\derboxthree}%
               \vbox{\ifdim\derccdim>\dercdim\hsize=\derccdim
                                        \else\hsize=\dercdim \fi
                    \hbox{\hfil}}%
               \hbox to0pt{\copy\derboxfour\hss}%
               \ifdim\dercrdim>\derrdim
                  \ifdim\derrdim>0pt\kern\derrdim\fi
                  \else\kern\dercrdim\fi}%
         \kern\deropen
         \hbox{\ifdim\derldim>0pt\kern\derldim\fi
               \box\derboxone
               \ifdim\derrdim>0pt\kern\derrdim\fi}}}%
   \ifdim\derldim<0pt\global\derldim=0pt\fi       
   \ifdim\derrdim<0pt\global\derrdim=0pt\fi       
   \derdldim=\wd\derboxthree\advance\derdldim by-\dercldim
   \derdrdim=\wd\derboxfour \advance\derdrdim by-\dercrdim
   \ifdim\derdim<0pt
      \ifdim\derdldim<0pt
         \derdldim=0pt                            
      \fi
      \ifdim\derdrdim<0pt
         \derdrdim=0pt                            
      \fi
   \else
      \ifdim\derldim>0pt
         \ifdim\derdldim>-\derdim
            \advance\derdldim by\derdim           
         \else			                  
            \derdldim=0pt                         
         \fi                                      
      \else
         \advance\derdldim by\dercldim            
      \fi
      \ifdim\derrdim>0pt
         \ifdim\derdrdim>-\derdim
            \advance\derdrdim by\derdim           
         \else			                  
            \derdrdim=0pt                         
         \fi                                      
      \else
         \advance\derdrdim by\dercrdim            
      \fi
   \fi
   \global\setbox\derboxone=\hbox
      {\kern\derdldim\unhbox\derboxone\kern\derdrdim}%
   \global\advance\derldim by\derdldim            
   \global\advance\derrdim by\derdrdim            
   }%
\def\stemr #1#2#3#4{{#4}%
   \dercldim=\derldim
   \derccdim=\dercdim
   \dercrdim=\derrdim
   \stemaux{#1}{#2}{#3}}%
\def\stemrr #1#2#3#4{{#4}%
   \dercldim=\derldim
   \derccdim=\dercdim
   \dercrdim=\derrdim
   \stemauxx{#1}{#2}{#3}}%
\def\stem #1#2#3#4{\setbox\stembox=\hbox{$\|$}%
   \stemrr{  }{  }{#3              }  {
   \stemr {#1\;}{\;#2}{\kern\wd\stembox} {
   \stemrr{  }{  }{\kern\wd\stembox}{
   #4                             }}}}%
\def\stempr #1#2#3{\setbox\stembox=\hbox{$\|$}%
   \stemrr{  }{  }{#3              }  {
   \stemr {#1\;}{\;#2}{\kern\wd\stembox} {
   \stemrr{  }{  }{\kern\wd\stembox}{
      {\global\setbox\derboxone=\hbox{%
         \vbox to0pt{\vss\hbox{$-$}\vss\kern-2\deropen}}%
      \global\derldim=0pt                            
      \global\dercdim=\wd\derboxone                  
      \global\derrdim=0pt                            
      }                             }}}}%

\def\deraux {\derldim=0pt\dercdim=0pt\derrdim=0pt}%
\def\der       #1#2#3{\deraux\root  {#1}{#2}{#3}        \box\derboxone}%
\def\dder    #1#2#3#4{\deraux\rroot {#1}{#2}{#3}{#4}    \box\derboxone}%
\def\dernote       #1#2#3#4{\deraux\rootr  {#1\;}{\;#2}{#3}{#4}\box\derboxone}%
\def\ddernote    #1#2#3#4#5{\deraux\rrootr {#1\;}{\;#2}{#3}{#4}{#5}\box
                                                                   \derboxone}%
\def\inf       #1#2#3{\der  {#1}{#2}{\leaf{#3}}}%
\def\iinf    #1#2#3#4{\dder {#1}{#2}{\leaf{#3}}{\leaf{#4}}}%

\newcommand{\vcinf}[3]
{\mathchoice
   {\vcenter{\inf{#1}{#2}{#3}}}
   {\hskip.9ex\vcenter{\inf{#1}{#2}{#3}}\hskip.9ex}
   {}{}
}

\newcommand{\vcdernote}[4]
{\vcenter{\dernote{#1}{#2}{#3}{#4}}}

\newcommand{\vciinf}[4]
{\mathchoice
   {\vcenter{\iinf{#1}{#2}{#3}{#4}}}
   {\hskip.9ex\vcenter{\iinf{#1}{#2}{#3}{#4}}\hskip.9ex}
   {}{}
}

\def\sqnsmallderi#1#2{\DerivationFactors{\smash{\phantom{X}}}%
                  {\smash{\phantom{X}}}%
                  {\smash{\phantom{X}}}%
                  {\sqn{#1}}%
                  {#2}{2}{2}}

\def\quand {\quad\mbox{and}\quad}%
\def\qquand {\qquad\mbox{and}\qquad}%
\def\qqquand {\quad\qquad\mbox{and}\qquad\quad}%
\def\qqqquand {\qquad\qquad\mbox{and}\qquad\qquad}%

\newbox\DerivOneBox
\newbox\DerivTwoBox
\newbox\DerivThreeBox
\newbox\DerivFourBox
\newdimen\DerivOneDimen
\newdimen\DerivTwoDimen
\newdimen\DerivThreeDimen
\newdimen\DerivFourDimen
\def\Derivationleaf #1#2#3#4#5{\global\setbox\derboxone=\hbox{\strut
                                    $\DerivationFactors{#1}{#2}{#3}{#4}{#5}11$}}%
\def\DerivationFactors #1#2#3#4#5#6#7{%
   \setbox\DerivOneBox=\hbox{$#1\strut$}%
      \DerivOneDimen=\wd\DerivOneBox\divide\DerivOneDimen by2
   \setbox\DerivThreeBox=\hbox{$#3\strut$}%
      \DerivThreeDimen=\wd\DerivThreeBox\divide\DerivThreeDimen by2
   \setbox\DerivTwoBox=\hbox{\box\DerivOneBox\hbox{$#2$}\box\DerivThreeBox}%
      \DerivTwoDimen=\wd\DerivTwoBox
   \setbox\DerivFourBox=\hbox{$#4\strut$}%
      \DerivFourDimen=\wd\DerivFourBox
   \ifdim\DerivFourDimen>\DerivTwoDimen
      \global\dercdim=\DerivFourDimen                
      \global\derldim=0pt                            
      \global\derrdim=0pt                            
      \advance\DerivFourDimen by-\DerivTwoDimen
      \divide \DerivFourDimen by2
      \advance\DerivTwoDimen  by-\DerivOneDimen
      \advance\DerivTwoDimen  by-\DerivThreeDimen
      \divide \DerivTwoDimen  by 2
   \else
      \global\dercdim=\DerivFourDimen                
      \DerivFourDimen=0pt
      \advance\DerivTwoDimen  by-\DerivOneDimen
      \advance\DerivTwoDimen  by-\DerivThreeDimen
      \global\derldim=\DerivTwoDimen
         \global\advance\derldim by-\dercdim
         \global\divide\derldim by2
         \global\advance\derldim by\DerivOneDimen    
      \global\derrdim=\DerivTwoDimen
         \global\advance\derrdim by-\dercdim
         \global\divide\derrdim by2
         \global\advance\derrdim by\DerivThreeDimen  
      \divide \DerivTwoDimen  by 2
   \fi
   \vbox{\offinterlineskip\hbox{\kern\DerivFourDimen\box\DerivTwoBox}%
         \hbox{\kern\DerivFourDimen\kern\DerivOneDimen
               \kern\DerivTwoDimen\kern-#6\DerivTwoDimen\hbox{$\xy
               0;<#6\DerivTwoDimen,0pt>:<0pt,#7\DerivTwoDimen>::
               (0,1);(2,1)**\crv{(1.25,1.1875)&(0.75,0.8125)};
               (1,0)**@{-};(0,1)**@{-};
               (1,0.625)*{\scriptstyle #5}
               \endxy$}}%
         \hbox{\kern\DerivFourDimen\kern\DerivOneDimen\kern\DerivTwoDimen
               \hbox to0pt{\hss\box\DerivFourBox\hss}%
               \kern\DerivFourDimen\kern\DerivOneDimen\kern\DerivTwoDimen}}}%

\def\clap#1{\hbox to 0pt{\hss#1\hss}}
\def\qlap#1{\hbox to 1em{\hss#1\hss}}
\def\qqlap#1{\hbox to 2em{\hss#1\hss}}
\def\qqqlap#1{\hbox to 3em{\hss#1\hss}}
\def\qqqqlap#1{\hbox to 4em{\hss#1\hss}}
\def\qqqqqlap#1{\hbox to 5em{\hss#1\hss}}
\def\qqqqqqlap#1{\hbox to 6em{\hss#1\hss}}
\def\qqqqqqqlap#1{\hbox to 7em{\hss#1\hss}}
\def\qqqqqqqqlap#1{\hbox to 8em{\hss#1\hss}}
\def\qqqqqqqqqlap#1{\hbox to 9em{\hss#1\hss}}

\def\qlapm#1{\qlap{$#1$}}

\def\qqqlapm#1{\qqqlap{$#1$}}

\def\qqqqqlapm#1{\qqqqqlap{$#1$}}
\def\qqqqqqlapm#1{\qqqqqqlap{$#1$}}

\def\rlapm#1{\hbox to 0pt{$#1$\hss}}
\def\llapm#1{\hbox to 0pt{\hss$#1$}}

\def\qqquad{\quad\qquad}

\def\set#1{\{#1\}}
\def\fcomp{\mathbin{\circ}}

\def\cneg#1{\bar{#1}}
\def\widecneg#1{\overline{#1}}
\def\wcneg#1{\overline{#1}}
\def\cimp{\mathop{\Rightarrow}}
\def\fneg{\widecneg{(-)}}

\def\ctrue{{\mathbf t}}
\def\cfalse{{\mathbf f}}
\def\cand{\wedge}
\def\cor{\vee}
\def\candor{\mathchoice%
  {\mathbin{\rlap{$\vee$}\wedge}}
  {\mathbin{\rlap{$\vee$}\wedge}}
  {\mathbin{\rlap{$\scriptstyle\vee$}\wedge}}
  {\mathbin{\rlap{$\scriptscriptstyle\vee$}\wedge}}
}

\def\lone{1}
\def\lbot{\bot}
\def\ltens{\mathop\varotimes}
\def\lpar{\mathop\bindnasrepma}

\def\cons#1{\{#1\}}
\def\conhole      {\cons{\enspace}}%

\def\sqn  #1{{\vdash #1}}%
\def\ssqn  #1#2{{#1 \vdash #2}}%

\def\cutr{\mathsf{cut}}

\def\conr{\mathsf{cont}}
\def\rr{\mathsf{r}}
\def\swir{\mathsf{s}}
\def\medr{\mathsf{m}}

\def\cA{{\mathcal A}}
\def\cB{{\mathcal B}}
\def\cC{{\mathcal C}}
\def\cE{{\mathcal E}}
\def\cF{{\mathcal F}}
\def\cL{{\mathcal L}}

\def\Nat{\mathbb{N}}

\newcommand{\Hom}[1][]{\mathrm{Hom}_{#1}}
\def\isom{\cong}

\newcommand{\sto}[1][]{\stackrel{#1}{\to}}

\newcommand{\solto}[1][]{\mathrel{\!\xy\ar@{->}^-{#1}(5,0)\endxy\!}}
\newcommand{\longsolto}[1][]{\mathrel{\!\xy\ar@{->}^-{#1}(11,0)\endxy\!}}

\def\name#1{\hat{#1}}

\def\widename#1{\widehat{#1}}

\def\quadfs {\rlap{\rm\quad.}}%

\newbox\cutbox
\newdimen\cutwd
\newdimen\cutht
\newdimen\cutdp
\def\ccut{%
  \setbox\cutbox\hbox{$\lozenge$}
  \cutwd=\wd\cutbox
  \cutht=\ht\cutbox
  \cutdp=\dp\cutbox
  \setbox\cutbox\hbox to\cutwd{\hss\vrule width.3pt height\cutht depth\cutdp\hss}
  \mathbin{\lozenge\hskip-\cutwd\copy\cutbox}}
\def\scriptcut{%
  \setbox\cutbox\hbox{$\scriptstyle\lozenge$}
  \cutwd=\wd\cutbox
  \cutht=\ht\cutbox
  \cutdp=\dp\cutbox
  \setbox\cutbox\hbox to\cutwd{\hss\vrule width.3pt height\cutht depth\cutdp\hss}
  \mathord{\lozenge\hskip-\cutwd\copy\cutbox}}

\def\vccut{%
  \setbox\cutbox\hbox{$\lozenge$}
  \cutwd=\wd\cutbox
  \cutht=\ht\cutbox
  \cutdp=\dp\cutbox
  \setbox\cutbox\hbox to\cutwd{\hss\hskip.3pt\vrule width.3pt height\cutht depth\cutdp\hss}
  \mathbin{\lozenge\hskip-\cutwd\copy\cutbox}}

\def\grammareq {\mathrel{\raise.4pt\hbox{::}{=}}}%

\def\SKS{\mathsf{SKS}}
\def\KS{\mathsf{KS}}
\def\LK{$\mathsf{LK}$}
\def\Bo{$\mathsf{B0}$}
\def\Bi{$\mathsf{B1}$}
\def\Bii{$\mathsf{B2}$}
\def\Biii{$\mathsf{B3}$}
\def\Biv{$\mathsf{B4}$}

\def\Bv{$\mathsf{B5}$}

\def\cCplus{\cC^{\candor}}
\def\SNet{\mathbf{SNet}}
\def\ENet{\mathbf{ENet}}
\def\prfnetsymbol{\mathbin{\vartriangleright}}
\def\prfnet#1#2{#1\prfnetsymbol#2}
\def\seqsep{\quad,\quad}

\newgray{prunegrayline}{.91}
\newgray{prunegraytext}{.86}

\def\prune#1{{\prunegraytext\psset{linecolor=prunegrayline}#1}}
\newbox\wholebox
\setbox\wholebox\hbox{$\scriptscriptstyle\talloblong$}
\newdimen\wdp
\wdp=\dp\wholebox
\def\whole{{\mathord{\raise\wdp\hbox{$\scriptscriptstyle\talloblong$}}}}

\def\fbm#1{\hbox{%
    \rlap{$#1$}\hskip.115pt%
    \rlap{$#1$}\hskip.115pt%
    \rlap{$#1$}\hskip.115pt$#1$}}

\def\fbm#1{#1}

\newcommand{\assoc}[1][]{{\hat\alpha_{#1}}}
\newcommand{\twist}[1][]{{\hat\sigma_{#1}}}
\newcommand{\runit}[1][]{{\hat\varrho_{#1}}}
\newcommand{\lunit}[1][]{{\hat\lambda_{#1}}}
\newcommand{\nid}[1][]{{\check{\hbox{\sf\i}}_{#1}}}
\newcommand{\tens}[1][]{{\hat{\mathsf{t}}_{#1}}}
\newcommand{\diag}[1][]{{{\mathchar"7001}_{#1}}}
\newcommand{\diagsq}[1][]{{{\mathchar"7001}^\mathsf{2}_{#1}}}
\newcommand{\proj}[1][]{{{\mathchar"7005}^{#1}}}

\newcommand{\projl}[2]{{{\mathchar"7005}^{#2}_{#1\whole}}}
\newcommand{\projr}[2]{{{\mathchar"7005}^{#1}_{\whole#2}}}
\def\bprod#1{\left\langle\,#1\,\right\rangle}
\def\bcoprod#1{\left[\,#1\,\right]}
\def\prod#1{\langle #1\rangle}
\def\coprod#1{[#1]}

\newcommand{\coassoc}[1][]{{\check\alpha_{#1}}}
\newcommand{\cotwist}[1][]{{\check\sigma_{#1}}}
\newcommand{\corunit}[1][]{{\check\varrho_{#1}}}
\newcommand{\colunit}[1][]{{\check\lambda_{#1}}}
\newcommand{\conid}[1][]{{\hat{\hbox{\sf\i}}_{#1}}}
\newcommand{\cotens}[1][]{{\check{\mathsf{t}}_{#1}}}
\newcommand{\codiag}[1][]{{\nabla_{#1}}}
\newcommand{\codiagsq}[1][]{{\nabla^\mathsf{2}_{#1}}}
\newcommand{\coproj}[1][]{{\amalg^{#1}}}

\newcommand{\coprojl}[2]{{\amalg^{#2}_{#1\whole}}}
\newcommand{\coprojr}[2]{{\amalg^{#1}_{\whole#2}}}

\newcommand{\switch}[1][]{{\mathsf{s}_{#1}}}
\newcommand{\medial}[1][]{{\mathsf{m}_{#1}}}
\newcommand{\medialsq}[1][]{{\hat{\mathsf{m}}^\mathsf{2}_{#1}}}
\newcommand{\comedialsq}[1][]{{\check{\mathsf{m}}^\mathsf{2}_{#1}}}
\newcommand{\emap}[1][]{{\mathsf{e}_{#1}}}
\newcommand{\nmedial}{{\check{\mathsf{nm}}}}
\newcommand{\conmedial}{{\hat{\mathsf{nm}}}}
\newcommand{\mix}[1][]{\mathsf{mix}_{#1}}

\newcommand{\eval}[1][]{\mathsf{eval}_{#1}}
\newcommand{\idf}[1][]{{1_{#1}}}
\newcommand{\id}[1][]{#1}

\def\Leaf#1{\cL(#1)}
\def\lab#1{\ell(#1)}
\def\anclab{\ell}

\def\fneg{\overline{(-)}}

\def\fple{\preccurlyeq}

\let\dfn=\emph

\def\interdisplayskip{1.5ex}
\def\biginterdisplayskip{3ex}

\psset{%
  labelsep=3pt,
  nodesep=.5ex
  }

\def\boldvec{\psset{linewidth=2pt}}

\def\normalvecheight{1}
\def\loopvecheight{.9}
\def\normalarrpos{.54}
\def\loopangleA{70}
\def\loopangleB{110}
\def\vertangleA{75}
\def\vertangleB{105}

\newcommand{\vecanglesposheight}[6]{%
  \nccurve[ncurv=#6,ArrowInside=->,ArrowInsidePos=#5,angleA=#3,angleB=#4]
  {#1}{#2}}
\newcommand{\duvecanglesheight}[5]{%
  \nccurve[ncurv=#5,angleA=#3,angleB=#4]{#1}{#2}
  \lput{:0}{\psline{->}(.05,.005)(.1,.005)}
  }
\newcommand{\udvecanglesheight}[5]{%
  \nccurve[ncurv=#5,angleA=#3,angleB=#4]{#1}{#2}
  \lput{:0}{\psline{->}(.05,-.005)(.1,-.005)}
  }

\newcommand{\vecanglespos}[5]{%
  \vecanglesposheight{#1}{#2}{#3}{#4}{#5}{\normalvecheight}}
\newcommand{\vecanglesheight}[5]{%
  \vecanglesposheight{#1}{#2}{#3}{#4}{\normalarrpos}{#5}}
\newcommand{\vecangles}[4]{%
  \vecanglesposheight{#1}{#2}{#3}{#4}{\normalarrpos}{\normalvecheight}}

\newcommand{\uvecpos}[3]{%
  \vecanglesposheight{#1}{#2}{90}{-90}{#3}{\normalvecheight}}

\newcommand{\uvec}[2]{%
  \vecanglesposheight{#1}{#2}{90}{-90}{\normalarrpos}{\normalvecheight}}

\newcommand{\urvecpos}[3]{%
  \vecanglesposheight{#1}{#2}{\vertangleA}{-\vertangleB}
		     {#3}{\normalvecheight}}

\newcommand{\urvec}[2]{%
  \vecanglesposheight{#1}{#2}{\vertangleA}{-\vertangleB}
		     {\normalarrpos}{\normalvecheight}}

\newcommand{\ulvecpos}[3]{%
  \vecanglesposheight{#1}{#2}{\vertangleB}{-\vertangleA}
		     {#3}{\normalvecheight}}

\newcommand{\dvecpos}[3]{%
  \vecanglesposheight{#1}{#2}{-90}{90}{#3}{\normalvecheight}}

\newcommand{\dvec}[2]{%
  \vecanglesposheight{#1}{#2}{-90}{90}{\normalarrpos}{\normalvecheight}}

\newcommand{\drvecpos}[3]{%
  \vecanglesposheight{#1}{#2}{-\vertangleA}{\vertangleB}
		     {#3}{\normalvecheight}}

\newcommand{\drvec}[2]{%
  \vecanglesposheight{#1}{#2}{-\vertangleA}{\vertangleB}
		     {\normalarrpos}{\normalvecheight}}

\newcommand{\dlvecpos}[3]{%
  \vecanglesposheight{#1}{#2}{-\vertangleB}{\vertangleA}
		     {#3}{\normalvecheight}}

\newcommand{\druvec}[2]{%
  \duvecanglesheight{#1}{#2}{-\loopangleA}{-\loopangleB}{\loopvecheight}}
\newcommand{\dluvec}[2]{%
  \duvecanglesheight{#1}{#2}{-\loopangleB}{-\loopangleA}{\loopvecheight}}
\newcommand{\dloop}[1]{%
  \udvecanglesheight{#1}{#1}{-55}{-125}{5}}

\newcommand{\urdvecheight}[3]{%
  \udvecanglesheight{#1}{#2}{\loopangleA}{\loopangleB}{#3}}
\newcommand{\uldvecheight}[3]{%
  \udvecanglesheight{#1}{#2}{\loopangleB}{\loopangleA}{#3}}
\newcommand{\urdvec}[2]{%
  \udvecanglesheight{#1}{#2}{\loopangleA}{\loopangleB}{\loopvecheight}}
\newcommand{\uldvec}[2]{%
  \udvecanglesheight{#1}{#2}{\loopangleB}{\loopangleA}{\loopvecheight}}
\newcommand{\uloop}[1]{%
  \udvecanglesheight{#1}{#1}{125}{55}{5}}

\def\anchor{\bullet}
\def\agap{\hskip2em}
\def\agapplus{\hskip2.2em}
\def\agapdemi{\hskip.9em}
\def\agapmed{\hskip1.5em}

\def\verywidecor{\quad\cor\quad}

\def\widecor{\;\,\cor\,\;}
\def\widecand{\;\,\cand\,\;}
\def\medcor{\;\cor\;}
\def\medcand{\;\cand\;}
\def\klam#1{\left(\,#1\,\right)}

\newdimen\diagcol
\newdimen\diagrow

\def\diagsizenormal{
  \diagcol=4em
  \diagrow=5ex
}

\newcommand{\ndiagtriangleup}[9]{%
  \xymatrix@C=.5\diagcol@R=\diagrow{
    &#4 \ar@{#5}[rd]^{#6}\\
    #1 \ar@{#2}[ru]^{#3} \ar@{#7}[rr]_{#8} 
    & & #9
  }
}

\newcommand{\ndiagtriangledown}[9]{%
  \xymatrix@C=.5\diagcol@R=\diagrow{
    #1 \ar@{#2}[rd]_{#3} \ar@{#7}[rr]^{#8} 
    & & #9\\
    &#4 \ar@{#5}[ru]_{#6}
  }
}

\newcommand{\ndiagtriangleright}[9]{%
  \xymatrix@C=\diagcol@R=.5\diagrow{
    #1 \ar@{#2}[rd]^{#3} \ar@{#7}[dd]_{#8}\\ 
    & #4 \ar@{#5}[ld]^{#6}\\
    #9
  }
}

\newcommand{\ndiagtriangleleft}[9]{%
  \xymatrix@C=\diagcol@R=.5\diagrow{
    &#1 \ar@{#2}[ld]_{#3} \ar@{#7}[dd]^{#8}\\ 
    #4 \ar@{#5}[rd]_{#6}\\
    &#9
  }
}

\newcommand{\ndiagsquare}[8]{%
  \xymatrix@C=\diagcol@R=\diagrow{
    #1 \ar[r]^{#2} \ar[d]_{#5} 
    & #3 \ar[d]^{#4}\\
    #6 \ar[r]_{#7} 
    & #8
  }
}

\newcommand{\noar}[1]{\fbm{#1}}
\newcommand{\arbto}[3][r]{\fbm{#2}\ar[#1]^-{#3}}
\newcommand{\arpto}[3][r]{\fbm{#2}\ar[#1]_-{#3}}
\newcommand{\arbfrom}[3][r]{\fbm{#3}\ar@{<-}[#1]^-{#2}}
\newcommand{\arpfrom}[3][r]{\fbm{#3}\ar@{<-}[#1]_-{#2}}

\newcommand{\arbtofrom}[3][r]{#2\ar@{<->}[#1]^-{#3}}
\newcommand{\arptofrom}[3][r]{#2\ar@{<->}[#1]_-{#3}}
\newcommand{\arbfromto}[3][r]{#3\ar@{<->}[#1]^-{#2}}
\newcommand{\arpfromto}[3][r]{#3\ar@{<->}[#1]_-{#2}}

\newcommand{\arbtoid}[3][r]{\fbm{#2}\ar@{=}[#1]^-{#3}}
\newcommand{\arptoid}[3][r]{\fbm{#2}\ar@{=}[#1]_-{#3}}
\newcommand{\arbfromid}[3][r]{\fbm{#3}\ar@{=}[#1]^-{#2}}
\newcommand{\arpfromid}[3][r]{\fbm{#3}\ar@{=}[#1]_-{#2}}

\newcommand{\labupleft}[3]{\strut\smash{\raise#1pt\hbox{%
      $\scriptstyle#3\hskip#2ex$}}}
\newcommand{\labdownleft}[3]{\strut\smash{\lower#1pt\hbox{%
      $\scriptstyle#3\hskip#2ex$}}}
\newcommand{\labupright}[3]{\strut\smash{\raise#1pt\hbox{%
      $\scriptstyle\hskip#2ex#3$}}}
\newcommand{\labdownright}[3]{\strut\smash{\lower#1pt\hbox{%
      $\scriptstyle\hskip#2ex#3$}}}

\newcommand{\diagram}[1]{\xymatrix@C=\diagcol@R=\diagrow{#1}}
\newcommand{\diagramhc}[1]{\xymatrix@C=.5\diagcol@R=\diagrow{#1}}
\newcommand{\diagramhhc}[1]{\xymatrix@C=.25\diagcol@R=\diagrow{#1}}
\newcommand{\diagrammhc}[1]{\xymatrix@C=-.5\diagcol@R=\diagrow{#1}}
\newcommand{\diagrammtqc}[1]{\xymatrix@C=-.65\diagcol@R=\diagrow{#1}}
\newcommand{\diagramdc}[1]{\xymatrix@C=2\diagcol@R=\diagrow{#1}}
\newcommand{\diagramhr}[1]{\xymatrix@C=\diagcol@R=.5\diagrow{#1}}
\newcommand{\diagramdr}[1]{\xymatrix@C=\diagcol@R=2\diagrow{#1}}
\newcommand{\diagramhchr}[1]{\xymatrix@C=.5\diagcol@R=.5\diagrow{#1}}
\newcommand{\diagramzchr}[1]{\xymatrix@C=0.0\diagcol@R=.5\diagrow{#1}}
\newcommand{\diagrammqchr}[1]{\xymatrix@C=-.25\diagcol@R=.5\diagrow{#1}}
\newcommand{\diagramdchr}[1]{\xymatrix@C=2\diagcol@R=.5\diagrow{#1}}
\newcommand{\diagramdcdr}[1]{\xymatrix@C=2\diagcol@R=2\diagrow{#1}}
\newcommand{\diagramscsr}[1]{\xymatrix@C=1.5\diagcol@R=1.5\diagrow{#1}}

\newcommand{\vcdiagram}[1]{\vcenter{\diagram{#1}}}
\newcommand{\vcdiagramhc}[1]{\vcenter{\diagramhc{#1}}}
\newcommand{\vcdiagramdc}[1]{\vcenter{\diagramdc{#1}}}

\newcommand{\vcdiagrammtqc}[1]{\vcenter{\diagrammtqc{#1}}}
\newcommand{\vcdiagramhr}[1]{\vcenter{\diagramhr{#1}}}

\newcommand{\vcdiagramdchr}[1]{\vcenter{\diagramdchr{#1}}}

\newcommand{\diagtriangleup}[3]{%
  \xymatrix@C=.5\diagcol@R=\diagrow{
    &\arpto[dl]#1 \\
    \arpto[rr]#2 && \arpfrom[ul]#3 
    }
  }

\newcommand{\diagtriangledown}[3]{%
  \xymatrix@C=.5\diagcol@R=\diagrow{
    \arbto[rr]#1 &&\arbto[dl]#2\\
    &\arbfrom[ul]#3 
  }
}

\newcommand{\diagtriangleright}[3]{%
  \xymatrix@C=.5\diagcol@R=.5\diagrow{
    \arpto[dd]#1 \\
    &\arpfrom[ul]#3 \\
    \arpto[ur]#2
  }
}

\newcommand{\diagtriangleleft}[3]{%
  \xymatrix@C=.5\diagcol@R=.5\diagrow{
    &\arbto[dd]#2 \\
    \arbto[ur]#1 \\
    &\arbfrom[ul]#3
  }
}

\newcommand{\diagsquare}[4]{%
  \xymatrix@C=\diagcol@R=\diagrow{
    \arbto[r]#1   & \arbto[d]#2\\ 
    \arbfrom[u]#3 & \arbfrom[l]#4
  }
}

\newcommand{\diagsquaredown}[4]{%
  \xymatrix@C=\diagcol@R=\diagrow{
    \arpto[d]#1   & \arpfrom[l]#3\\ 
    \arpto[r]#2 & \arpfrom[u]#4
  }
}

\newcommand{\ddiagsquare}[5]{%
  \xymatrix@C=.7\diagcol@R=.7\diagrow{
    \arbto[rr]#1   && \arbto[dd]#2\\ 
    #3\\
    \arbfrom[uu]#4 && \arbfrom[ll]#5
  }
}

\newcommand{\diagpentagondown}[5]{%
  \xymatrix@C=.5\diagcol@R=\diagrow{
    \arbto[rr]#1   && \arbto[d]#2\\ 
    \arbfrom[u]#4 && \arbto[dl]#3\\
    &\arbfrom[ul]#5
  }
}

\newcommand{\diagpentagonright}[5]{%
  \xymatrix@C=.5\diagcol@R=.5\diagrow{
    \arpto[dd]#1   && \arpfrom[ll]#4 \\
    &&&\arpfrom[ul]#5 \\
    \arpto[rr]#2  && \arpto[ur]#3
  }
}

\newcommand{\diagpentagonleft}[5]{%
  \xymatrix@C=.5\diagcol@R=.5\diagrow{
    &\arbto[rr]#2   && \arbto[dd]#3 \\
    \arbto[ur]#1 \\
    &\arbfrom[ul]#4  && \arbfrom[ll]#5
  }
}

\newcommand{\diaghexagondown}[6]{%
  \xymatrix@C=\diagcol@R=\diagrow{
    \arpto[d]#1   & \arpfrom[l]#4\\
    \arpto[d]#2 & \arpfrom[u]#5\\
    \arpto[r]#3 & \arpfrom[u]#6
  }
}

\newcommand{\diaghexagonhexa}[6]{%
  \xymatrix@C=\diagcol@R=\diagrow{
    & \arpto[dl]#1 \\
    \arpto[d]#2 && \arpfrom[ul]#4\\
    \arpto[dr]#3 && \arpfrom[u]#5\\
     & \arpfrom[ur]#6
  }
}

\newcommand{\diaghexagondowndr}[6]{%
  \xymatrix@C=\diagcol@R=2\diagrow{
    \arpto[d]#1   & \arpfrom[l]#4\\
    \arpto[d]#2 & \arpfrom[u]#5\\
    \arpto[r]#3 & \arpfrom[u]#6
  }
}

\newcommand{\diaghexagonright}[6]{%
  \xymatrix@C=\diagcol@R=\diagrow{
    \arbto[r]#1   & \arbto[r]#2 & \arbto[d]#3\\
    \arbfrom[u]#4 & \arbfrom[l]#5 & \arbfrom[l]#6
  }
}

\newcommand{\diagoctagon}[8]{%
  \xymatrix@C=\diagcol@R=\diagrow{
    \arbto[r]#1   & \arbto[r]#2 & \arbto[d]#3\\
    \arbfrom[u]#5 && \arbto[d]#4\\
    \arbfrom[u]#6 & \arbfrom[l]#7 & \arbfrom[l]#8
  }
}

\newcommand{\diagoctagonslim}[8]{%
  \xymatrix@C=\diagcol@R=\diagrow{
    \arbto[r]#1   & \arbto[d]#2 \\
    \arbfrom[u]#5 & \arbto[d]#3 \\
    \arbfrom[u]#6 & \arbto[d]#4 \\
    \arbfrom[u]#7 & \arbfrom[l]#8
  }
}

\newcommand{\diagdecagonhexa}[9]{%
  \xymatrix@C=\diagcol@R=\diagrow{
    & \arbto[dr]#1 \\
    \arbfrom[ur]#6 && \arbto[d]#2 \\
    \arbfrom[u]#7 && \arbto[d]#3 \\
    \arbfrom[u]#8 && \arbto[d]#4 \\
    \arbfrom[u]#9 && \arbto[dl]#5 \\
    & \arbfrom[ul]{\lastarga}{\lastargb}
  }
}

\newcommand{\vcdiagtriangleup}[3]{\vcenter{\diagtriangleup{#1}{#2}{#3}}}%
\newcommand{\vcdiagtriangledown}[3]{\vcenter{\diagtriangledown{#1}{#2}{#3}}}%
\newcommand{\vcdiagtriangleleft}[3]{\vcenter{\diagtriangleleft{#1}{#2}{#3}}}%
\newcommand{\vcdiagtriangleright}[3]{\vcenter{\diagtriangleright{#1}{#2}{#3}}}%

\newcommand{\vcdiagsquare}[4]{%
  \vcenter{\diagsquare{#1}{#2}{#3}{#4}}}%

\newcommand{\vcdiagsquaredown}[4]{%
  \vcenter{\diagsquaredown{#1}{#2}{#3}{#4}}}%

\newcommand{\vcddiagsquare}[5]{%
  \vcenter{\ddiagsquare{#1}{#2}{#3}{#4}{#5}}}%

\newcommand{\vcdiagpentagondown}[5]{%
  \vcenter{\diagpentagondown{#1}{#2}{#3}{#4}{#5}}}%

\newcommand{\vcdiagpentagonleft}[5]{%
  \vcenter{\diagpentagonleft{#1}{#2}{#3}{#4}{#5}}}%

\newcommand{\vcdiagpentagonright}[5]{%
  \vcenter{\diagpentagonright{#1}{#2}{#3}{#4}{#5}}}%

\newcommand{\vcdiaghexagonright}[6]{%
  \vcenter{\diaghexagonright{#1}{#2}{#3}{#4}{#5}{#6}}}%

\newcommand{\vcdiaghexagondown}[6]{%
  \vcenter{\diaghexagondown{#1}{#2}{#3}{#4}{#5}{#6}}}%

\newcommand{\vcdiaghexagonhexa}[6]{%
  \vcenter{\diaghexagonhexa{#1}{#2}{#3}{#4}{#5}{#6}}}%

\newcommand{\vcdiaghexagondowndr}[6]{%
  \vcenter{\diaghexagondowndr{#1}{#2}{#3}{#4}{#5}{#6}}}%

\newcommand{\vcdiagoctagon}[8]{%
  \vcenter{\diagoctagon{#1}{#2}{#3}{#4}{#5}{#6}{#7}{#8}}}%

\newcommand{\vcdiagoctagonslim}[8]{%
  \vcenter{\diagoctagonslim{#1}{#2}{#3}{#4}{#5}{#6}{#7}{#8}}}%

\newcommand{\vcdiagdecagonhexa}[9]{%
  \vcenter{\diagdecagonhexa{#1}{#2}{#3}{#4}{#5}{#6}{#7}{#8}{#9}}}%

\begin{document}


\title{On the Axiomatisation of Boolean Categories\\
  with and without Medial}

\author{Lutz Stra\ss burger}
\address{Lutz Stra\ss burger\\
INRIA Futurs \& \'Ecole Polytechnique\\
LIX\\
Rue de Saclay\\
91128 Palaiseau Cedex\\
France\\
\tt\url{http://www.lix.polytechnique.fr/Labo/Lutz.Strassburger/}}

\copyrightyear{2007}

\maketitle

\begin{abstract}
  The term ``Boolean category'' should be used for describing an
  object that is to categories what a Boolean algebra is to
  posets. More specifically, a Boolean category should provide the
  abstract algebraic structure underlying the proofs in Boolean Logic,
  in the same sense as a Cartesian closed category captures the proofs
  in intuitionistic logic and a *-autonomous category captures the
  proofs in linear logic. However, recent work has shown that there is
  no canonical axiomatisation of a Boolean category. In this work, we
  will see a series (with increasing strength) of possible such
  axiomatisations, all based on the notion of *-autonomous
  category. We will particularly focus on the medial map, which has
  its origin in an inference rule in KS, a cut-free deductive system
  for Boolean logic in the calculus of structures. Finally, we will
  present a category of proof nets as a particularly well-behaved
  example of a Boolean category.
\end{abstract}

\diagsizenormal

\section{Introduction}

The questions \emph{``What is a proof?''} and \emph{``When are two
proofs the same?''} are fundamental for proof theory. But for the most
prominent logic, Boolean (or classical) propositional logic, we still
have no satisfactory answers.

This is not only embarrassing for proof theory itself, but also for
computer science, where Boolean propositional logic plays a major role
in automated reasoning and logic programming. Also the design and
verification of hardware is based on Boolean logic.  Every area in
which proof search is employed can benefit from a better understanding
of the concept of proof in Boolean logic, and the famous
NP-versus-coNP problem can be reduced to the question whether there is
a short (i.e., polynomial size) proof for every Boolean tautology
\cite{cook:reckhow:79}.

Usually proofs are studied as syntactic objects within some deductive
system (e.g., tableaux, sequent calculus, resolution, \ldots). This
paper takes the point of view that these syntactic objects (also known
as proof trees) should be considered as concrete representations of
certain abstract proof objects, and that such an abstract proof object
can be represented by a resolution proof tree and a sequent calculus
proof tree, or even by several different sequent calculus proof trees.

From this point of view the motivation for this work is to provide an
abstract algebraic theory of proofs. Already Lambek
\cite{lambek:68,lambek:69} observed that such an algebraic treatment
can be provided by category theory.  For this, it is necessary to
accept the following postulates about proofs:
\begin{itemize}
\item for every proof $f$ of conclusion $B$ from hypothesis $A$ (denoted
  by $f\colon A\to B$) and every proof $g$ of conclusion $C$ from
  hypothesis $B$ (denoted by $g\colon B\to C$) there is a uniquely
  defined composite proof $g\fcomp f$ of conclusion $C$ from
  hypothesis $A$ (denoted by $g\fcomp f\colon A\to C$),
\item this composition of proofs is associative,
\item for each formula $A$ there is an identity proof
  $\idf[A]\colon A\to A$ such that for $f\colon A\to B$ we have
  $f\fcomp\idf[A]=f=\idf[B]\fcomp f$.
\end{itemize}
Under these assumptions the proofs are the arrows in a category whose
objects are the formulae of the logic. What remains is to provide the
right axioms for the ``category of proofs''.

It seems that finding these axioms is particularly difficult for the
case of Boolean logic. For intuitionistic logic, Prawitz
\cite{prawitz:71} proposed the notion of \emph{proof normalization}
for identifying proofs. It was soon discovered that this notion of
identity coincides with the notion of identity that results from the
axioms of a Cartesian closed category (see, e.g.,
\cite{lambek:scott:86}). In fact, one can say that the proofs of
intuitionistic logic are the arrows in the free (bi-)Cartesian closed
category generated by the set of propositional variables. An
alternative way of representing the arrows in that category is via
terms in the simply-typed $\lambda$-calculus: arrow composition is
normalization of terms. This observation is well-known as the
Curry-Howard-correspondence \cite{howard:80}.

In the case of linear logic, the relation to *-autonomous categories
\cite{barr:79} was noticed immediately after its discovery
\cite{lafont:thesis,seely:89}. In the sequent calculus linear logic
proofs are identified when they can be transformed into each other via
``trivial'' rule permutations \cite{lafont:95}. For multiplicative
linear logic this coincides with the proof identifications induced by
the axioms of a *-autonomous category \cite{blute:93,str:lam:04:CSL}.
Therefore, we can safely say that a proof in multiplicative linear
logic is an arrow in the free *-autonomous category generated by the
propositional variables \cite{BCST,lam:str:freestar,hughes:freestar}.

But for classical logic no such well-accepted category of proofs
exists.  We can distinguish two main reasons. First, if we start from
a Cartesian closed category and add an involutive
negation\footnote{i.e., a natural isomorphism between $A$ and the
double-negation of $A$ (in this paper denoted by $\cneg{\cneg A}$)},
we get the collapse into a Boolean algebra, i.e., any two proofs
$f,g\colon A\to B$ are identified. For every formula there would be at
most one proof (see, e.g., \cite[p.67]{lambek:scott:86} or the appendix of
\cite{girard:LC} for details).  Alternatively, starting from a
*-autonomous category and adding natural transformations $A\to A\cand
A$ and $A\to\ctrue$, i.e., the proofs for weakening and contraction,
yields the same collapse.\footnote{Since we are dealing with Boolean
logic, we will use the symbols $\cand$ and $\ctrue$ for the tensor
operation (usually~$\ltens$) and the unit (usually {\bf 1} or~{\bf I})
in a *-autonomous category.}

The second reason is that cut elimination in the sequent calculus for
classical logic is not confluent. Since cut elimination is the usual
way of composing proofs, this means that there is no canonical way of
composing two proofs, let alone associativity of composition.

Consequently, for avoiding these two problems, we have to accept that (i)
Cartesian closed categories do not provide an abstract algebraic
  axiomatisation for proofs in classical logic, and that (ii)
the sequent calculus is not the right framework for
  investigating the identity of proofs in classical logic.

There have already been several accounts for a proof theory for
classical logic based on the axioms of Cartesian closed
categories. The first were probably Parigot's
$\lambda\mu$-calculus~\cite{parigot:92} and Girard's
LC~\cite{girard:LC}. The work on polarized proof nets by
Laurent~\cite{laurent:99,laurent:03} shows that there is in fact not
much difference between the two. Later, the category-theoretic
axiomatisations underlying this proof theory has been investigated and
the close relationship to
continuations~\cite{thielecke:phd,streicher:reus:98} has been
established, culminating in Selinger's \emph{control
categories}~\cite{selinger:01}. However, by sticking to the axioms of
Cartesian closed categories, one has to sacrifice the perfect symmetry
of Boolean logic.

In this paper, we will go the opposite way. In the attempt of going
from a Boolean algebra to a Boolean category we insist on keeping the
symmetry between $\cand$ and $\cor$. By doing this we have to leave
the realm of Cartesian closed categories. That this is very well
possible has recently been shown by several authors
\cite{dosen:petric:coherence-book,fuhrmann:pym:oecm,lam:str:05:freebool}.
However, the fact that all three proposals considerably differ from
each other suggests that there might be no canonical way of giving a
categorical axiomatisation for proofs in classical logic.

We will provide a series of possible such axiomatisations with
increasing strength. They will all build on the structure of a
*-autonomous category in which every object has a monoid (and a
comonoid) structure. In this respect it will closely follow the work
of \cite{fuhrmann:pym:oecm} and \cite{lam:str:05:freebool}, but will
differ from \cite{dosen:petric:coherence-book}.

The approach that we take here is mainly motivated by the
investigation in the complexity of proofs. Eventually, a good theory
of proof identification should never identify two proofs if one is
exponentially bigger than the other. 

The main proof-theoretic inspiration for this work comes from the system
$\SKS$ \cite{brunnler:tiu:01}, which is a deductive system for Boolean
logic within the formalism of the calculus of structures
\cite{gug:SIS,gug:str:01,brunnler:tiu:01}. A
remarkable feature of the cut-free version of $\SKS$, which is called
$\KS$, is that it can (cut-free) polynomially simulate not only
sequent calculus and tableaux systems but also resolution and
Frege-Hilbert systems \cite{guglielmi:frogs:speedup,bru:gug:PC-DI}.
This means that if a tautology has a polynomial size proof in any of
these systems, then it has a cut-free polynomial size proof in
$\KS$. This ability of $\KS$ is a consequence of two features:
\begin{enumerate}
\item \emph{Deep inference}: Instead of decomposing the formulae along
  their root connectives into subformulae during the construction of a
  proof, in $\KS$ inference rules are applied deep inside formulae in
  the same way as we know it from term rewriting.
\item The two inference rules \emph{switch} and \emph{medial}, which
  look as follows:
  \begin{equation}\label{eq:swi-med-rule}
    \vcinf{\swir}{F\cons{A\cor (B\cand C)}}
          {F\cons{(A\cor B)\cand C}}
    \qquand
    \vcinf{\medr}{F\cons{(A\cor C)\cand(B\cor D)}}
          {F\cons{(A\cand B)\cor(C\cand D)}}
    \quad,
  \end{equation}
  where $F\conhole$ stands for an arbitrary (positive) formula context and
  $A$, $B$, $C$, and $D$ are formula variables.
\end{enumerate}

\subsection*{From deep inference to algebra}

Deep inference allows us to establish the relationship between proof
theory and algebra in a much cleaner way than this is possible with
shallow inference formalisms like the sequent calculus. The reason is
that from a derivation in a deep inference formalism one can directly
``read off the morphisms''.  Take for example the following derivation
in system $\KS$:
\begin{equation}\label{eq:KSmed1}
  \vcdernote{\medr}{}{(A\cor C)\cand(B\cor D)}{
    \root{\rr}{(A\cand B)\cor(C\cand D)}{
      \leaf{(A'\cand B)\cor(C\cand D)}}}
\end{equation}
where $A$, $A'$, $B$, $C$, and $D$ are arbitrary formulae, and $\rr$
is any inference rule taking $A'$ to $A$. In category-theoretic
language this would be written as a composition of maps: {\diagcol=5em
$$
\vcdiagram{
  \arbto[r]{(A'\cand B)\cor(C\cand D)}{(\rr\cand B)\cor(C\cand D)}
  &
  \arbto[r]{(A\cand B)\cor(C\cand D)}{\medial[A,B,C,D]}
  &
  \noar{(A\cor C)\cand(B\cor D)}
  }
$$}%
where $\medial[A,B,C,D]\colon (A\cand B)\cor(C\cand D)\to (A\cor
C)\cand(B\cor D)$ is called the \emph{medial map}, and $\rr\colon A'\to A$ is
the map corresponding to the rule $\rr$. System $\KS$ also allows the
derivation
\begin{equation}\label{eq:KSmed2}
  \vcdernote{\rr}{}{(A\cor C)\cand(B\cor D)}{
    \root{\medr}{(A'\cor C)\cand(B\cor D)}{
      \leaf{(A'\cand B)\cor(C\cand D)}}}
\end{equation}
From the proof-theoretic point of view it makes perfect sense to
identify the two derivations in \eqref{eq:KSmed1}
and~\eqref{eq:KSmed2} because they do ``essentially'' the same. This
is what Guglielmi calls \dfn{bureaucracy of type B}
\cite{guglielmi:B}. In the language of category theory, the
identification of \eqref{eq:KSmed1} and~\eqref{eq:KSmed2} is saying
that the diagram
\begin{equation}\label{eq:natmedial}
  \vcdiagsquare{{(A'\cand B)\cor(C\cand D)}{\medial[A',B,C,D]}}
  {{(A'\cor C)\cand(B\cor D)}{(\rr\cor C)\cand(B\cor D)}}
  {{(\rr\cand B)\cor(C\cand D)}{(A\cand B)\cor(C\cand D)}}
  {{\medial[A,B,C,D]}{(A\cor C)\cand(B\cor D)}}
\end{equation}
has to commute, which exactly means that the medial map has to be natural.

For deep inference, Guglielmi also introduces the notion of
  \dfn{bureaucracy of type A} \cite{guglielmi:A}, which is the formal
  distinction between the derivations
\begin{equation}\label{eq:KSA}
  \vcdernote{\rr_1}{}{A\cand B}{
    \root{\rr_2}{A'\cand B}{
      \leaf{A'\cand B'}}}
  \qquand
  \vcdernote{\rr_2}{}{A\cand B}{
    \root{\rr_1}{A\cand B'}{
      \leaf{A'\cand B'}}}
\end{equation}
where rule $\rr_1$ takes $A'$ to $A$, and rule $\rr_2$ takes $B'$ to
$B$.  Proof-theoretically, the two derivations in \eqref{eq:KSA} are
``essentially'' the same, so it makes sense to identify
them. Translating this into category theory means to say that the
operation~$\cand$ is a bifunctor.

However, it is not always the case that the demands of algebra and
proof theory coincide so nicely. Sometimes they contradict each other,
which causes ``creative tensions'' \cite{lam:str:freestar}. One example
is the treatment of units. Proof-theoretically it might be desirable
to distinguish between the following two proofs in the sequent
calculus (here $\ctrue$ stands for ``truth'' and $\cfalse$ for ``falsum''):
\begin{equation}
  \label{eq:unit-proof}
  \vcenter{\hbox{
  \dernote{\mathsf{weakening}}{\qqquand}{\sqn{\ctrue,\cfalse}}{
    \root{\mathsf{axiom(true)}}{\sqn{\ctrue}}{
      \leaf{}}}
  \dernote{\mathsf{axiom(identity)}}{}{\sqn{\ctrue,\cfalse}}{
    \leaf{}}}}
\end{equation}
This distinction is made, for example, by the proof nets presented in
\cite{lam:str:05:naming}.  From the algebraic point of view, this causes
certain difficulties: In \cite{lam:str:05:freebool} the concept of weak units
has been introduced in order to give a clean algebraic treatment to the
distinction in \eqref{eq:unit-proof}. However, in this paper we will depart
from this and use proper units instead. This is from the algebraic point of
view more reasonable and simplifies the theory considerably. But it forces the
identification of the two proofs in \eqref{eq:unit-proof}.

\subsection*{Some remarks about switch and medial}

The inference rule switch in \eqref{eq:swi-med-rule}, or the
\dfn{switch map} $\switch[A,B,C]\colon (A\cor B)\cand C\to
A\cor(B\cand C)$ has already been well investigated from the viewpoint
of proof theory \cite{gug:SIS}, as well as from the viewpoint of
category theory, where it is also called \emph{weak distributivity}
\cite{hyland:depaiva:fill,cockett:seely:97}, \emph{linear
distributivity}, or \emph{dissociativity}
\cite{dosen:petric:coherence-book}. On the other hand, the medial rule
or \dfn{medial map} $\medial[A,B,C,D]\colon (A\cand B)\cor(C\cand
D)\to (A\cor C)\cand(B\cor D)$ has not yet been so thoroughly
investigated. Only very recently Lamarche \cite{lamarche:gap} started
to study the consequences of the presence of the medial map in a
*-autonomous category, and Do{\v{s}}en and Petri{\'c}
\cite{dosen:petric:intermutation} investigate it under the name
\emph{intermutation} from the viewpoint of coherence (but without
taking the switch map into account).

Seen from the deductive point of view, the two rules switch and medial
have certain similarities:
\begin{itemize}
\item switch allows the reduction of the identity rule and the cut
  rule to atomic form, and medial allows the reduction of the
  contraction rule (and the cocontraction rule) to atomic form (see
  \cite{brunnler:tiu:01} for details),
\item switch and medial are both self-dual, and
\item they look similar, as can bee seen in
  \eqref{eq:swi-med-rule}. In fact, recent work shows that they can
  both be seen as instance of a single more general inference rule
  \cite{guglielmi:AG8,guglielmi:AG16}.
\end{itemize}
However, from the algebraic point of view, they are quite different:
Switch is a consequence of more primitive properties, namely the
associativity of $\cand$ and $\cor$ and the de Morgan duality between
the two operations\footnote{Nonetheless it has been investigated in
\cite{cockett:seely:97} from the category theoretic viewpoint under
the assumption that negation (and therefore the de Morgan duality) is
absent.}, whereas medial has to be put as additional primitive, if we
want it in the category.\footnote{This fact raises an open problem:
can we find simple primitives from which medial arises naturally, in
the same way as switch arises naturally from associativity and
duality?}

\subsection*{Outline of the paper}

In this work we will present a series of axioms that seem reasonable
(from the proof-theoretic as well as from the algebraic points of
view) to have in a Boolean category. While introducing axioms, we will
also show their consequences. Some of the axioms presented here
coincide with axioms given in the accompanying paper
\cite{lamarche:gap} which has been written at the same time as this
paper and appears in the same issue of this journal. This overlap is
certainly not surprising. However, there are two main differences
between the two papers. First, while \cite{lamarche:gap} works in the
minimal setting of a *-autonomous category with medial (or with
``linear logic plus medial''), we assume from the beginning full
classical propositional logic, i.e., the presence of weakening and
contraction.  Second (and more importantly) we are staying in the
realm of syntax, whereas \cite{lamarche:gap} is primarily concerned
with the construction of concrete models for classical~proofs.

It is in fact a problem of the subject in general that there are only
very few concrete examples of (symmetric) models of classical proofs.
One of them is the category {\bf Rel} of sets and relations
\cite{hyland:04}, but it has the common problem that it identifies
disjunction and conjunction. From the proof-theoretic point of view
this kind of degenerate model is not very interesting. In fact, the
investigation of medial is pointless in this setting.\footnote{For
this reason, we will leave it as an exercise to the reader to verify
that {\bf Rel} fulfills all the axioms presented in this paper.} Here
the work in \cite{lamarche:gap} provides some breakthroughs towards
new kind of models in which disjunction and conjunction do not
coincide.

In the end of this paper, we will also give a concrete example of a
Boolean category, namely a variation of the proof nets of
\cite{lam:str:05:naming}. Although this example might be considered
``only syntactic'', it nonetheless shows that the axioms presented
here do not lead to the collapse into a Boolean algebra. Furthermore,
this last section can be read independently by the reader interested
only in proof nets and not in category theory.

This paper is another attempt to be accessible to both the category
theorist \emph{and} the proof theorist.  Since it is mainly about
algebra, we use here the language of category theory. Nonetheless, the
seasoned proof theorist might find it easier to understand if he
substitutes everywhere \emph{``object''} by \emph{``formula''} and
\emph{``map''/``morphism''/``arrow''} by \emph{``proof''}. Every
commuting diagram in the paper is nothing but an equation between
proofs written in a deep inference formalism. In order to make the
paper easier accessible to proof theorists, all statements are proved
in more detail than the seasoned category theorist might find
appropriate.

\section{What is a Boolean Category ?}

Recall the analogy mentioned in the abstract: A Boolean category
should be for categories, what a Boolean algebra is for posets. This
leads to the following definition:

\begin{definition}\label{def:B0}
  We say a category $\cC$ is a \dfn{\Bo-category} if there is a
  Boolean algebra $\cB$ and a mapping $F\colon\cC\to\cB$ from objects
  of $\cC$ to elements of $\cB$, such that for all objects $A$ and $B$
  in $\cC$, we have $F(A)\le F(B)$ in $\cB$ if and only if there is an
  arrow $f\colon A\to B$ in $\cC$.
\end{definition}

In other words, a \Bo-category is a category whose image under the
forgetful functor from the category of categories to the category of
posets is a Boolean algebra.  From the proof-theoretic point of view
one should have that there is a proof from $A$ to $B$ if and only if
$A\cimp B$ is a valid implication. However, from the algebraic point
of view there are many models, including the category {\bf Rel} of
sets and relations, as well as the models constructed in the in the
accompanying paper \cite{lamarche:gap}, which have a map between any
two objects $A$ and $B$. Note that these models are not ruled out by
Definition~\ref{def:B0} because there is the trivial one-element
Boolean algebra. In any case, we
can make the following (trivial) observation.

\begin{observation}\label{obs:B0}
  In a \Bo-category, we can for any pair of objects $A$ and $B$,
  provide objects $A\cand B$ and $A\cor B$ and $\cneg A$, and there are
  objects $\ctrue$ and $\cfalse$, such that there are maps
  \begin{equation}\label{eq:B0}
    \qlapm{
    \begin{array}{c@{\qquad}c}
      \assoc[A,B,C] \colon A\cand(B\cand C)\to (A\cand B)\cand
      C&
      \coassoc[A,B,C] \colon A\cor(B\cor C)\to (A\cor B)\cor C
      \\[\interdisplayskip]
      \twist[A,B] \colon A\cand B\to B\cand A
      &
      \cotwist[A,B] \colon A\cor B\to B\cor A
      \\[\interdisplayskip]
      \runit[A] \colon  A\cand\ctrue\to A
      &
      \corunit[A] \colon  A\cor\cfalse\to A
      \\[\interdisplayskip]
      \lunit[A] \colon  \ctrue\cand A\to A
      &
      \colunit[A] \colon  \cfalse\cor A\to A
      \\[\biginterdisplayskip]
      \conid[A] \colon  A\cand \cneg A\to\cfalse
      &
      \nid[A] \colon \ctrue\to \cneg A\cor A
      \\[\biginterdisplayskip]
      \multicolumn{2}{c}{\switch[A,B,C]\colon 
        (A\cor B)\cand C\to A\cor(B\cand C)}
      \\[\interdisplayskip]
      \multicolumn{2}{c}{\medial[A,B,C,D]\colon (A\cand
        B)\cor(C\cand D)\to (A\cor C)\cand(B\cor D)}
      \\[\biginterdisplayskip]   
      \diag[A] \colon  A\to A\cand A
      &
      \codiag[A] \colon  A\cor A\to A
      \\[\interdisplayskip]
      \proj[A] \colon  A\to \ctrue
      &
      \coproj[A] \colon  \cfalse\to A
    \end{array}}
  \end{equation}
  for all objects $A$, $B$, and $C$.  This can easily be shown by verifying
  that all of them correspond to valid implications in Boolean logic.
  Conversely, a category in which every arrow can be given as a composite of
  the ones given above by using only the operations of $\cand$, $\cor$, and
  the usual arrow composition, is a \Bo-category. This is a consequence of the
  completeness of system $\SKS$ \cite{brunnler:tiu:01}, which is a deep
  inference deductive system for Boolean logic incorporating the maps in
  (\ref{eq:B0}) as inference rules.
\end{observation}

Note that Definition~\ref{def:B0} is neither enlightening nor useful.
It is necessary to add some additional structure in order to obtain a
``nicely behaved'' theory of Boolean categories. However, as already
mentioned in the introduction, the naive approach of adding structure,
namely adding the structure of a bi-Cartesian closed category (also
called Heyting category) with an involutive negation leads to
collapse: Every Boolean category in that strong sense is a Boolean
algebra. The hom-sets are either singletons or empty. This observation
has first been made by Andr\'e Joyal, and the proof can be found, for
example, in \cite{lambek:scott:86}, page~67. For the sake of
completeness, we repeat the argument here: First, recall that in a
Cartesian closed category, we have, among other properties, (i) binary
products, that we (following the notation of this paper) denote by
$\cand$, (ii) a terminal object $\ctrue$ with the property that
$\ctrue\cand A\isom A$ for all objects $A$, and (iii) a natural
bijection between the maps $f\colon A\cand B\to C$ and $f^\ast\colon
A\to B\cimp C$, where $B\cimp C$ denotes the exponential of $B$
and~$C$. Going from $f$ to $f^\ast$ is also known as
\emph{currying}. Adding an involutive negation means adding a
contravariant endofunctor $\fneg$ such that there is a natural
bijection between maps $f\colon A\to B$ and $\cneg f \colon\cneg
B\to\cneg A$. It also means that there is an initial object
$\cfalse=\cneg\ctrue$. Hence, we have in particular for all objects
$A$ and $B$, that
\begin{equation}
  \label{eq:ccc}
  \Hom(A,B)\isom\Hom(\ctrue\cand A,B)\isom\Hom(\ctrue,A\cimp B)
  \isom\Hom(\wcneg{A\cimp B},\cfalse)
  \quadfs
\end{equation}
Now observe that whenever we have an object $X$ such that the two
projections $$\pi_1,\pi_2\colon X\cand X\to X$$ are equal, then for
all objects $Y$, any two maps $f,g\colon Y\to X$ are equal, because
\begin{equation}
  \label{eq:eq-prod}
  f=\pi_1\circ\prod{f,g}=\pi_2\circ\prod{f,g}=g\quadfs
\end{equation}
Now note that since
$\cfalse$ is initial, there is exactly one map
$\cfalse\to\cfalse\cimp\cfalse$, hence, by uncurrying there is exactly
one map $\cfalse\cand\cfalse\to\cfalse$. Therefore, for every $Y$,
there is at most one map $Y\to\cfalse$. By \eqref{eq:ccc}, for all
$A$, $B$, we have $\Hom(A,B)$ is either singleton or empty.

Recapitulating the situation, we have here two extremes of Boolean
categories: no structure and too much structure. Neither of them is
very interesting, neither for proof theory nor for category
theory. But there is a whole universe between the two, which we will
start to investigate now. On our path, we will stick to \eqref{eq:ccc}
and carefully avoid to have \eqref{eq:eq-prod}. This is what makes our
approach different from control categories \cite{selinger:01}, in
which the equation $f=\pi_1\circ\prod{f,g}$ holds, but the rightmost
bijection in \eqref{eq:ccc} is absent.

\section{*-Autonomous categories}\label{sec:star}

Let us stress the fact that in a plain \Bo-category there is no
relation between the maps listed in \eqref{eq:B0}. In particular,
there is no functoriality of $\cor$ and $\cand$, no naturality of
$\assoc$, $\twist$, \ldots, and no de Morgan duality. Adding this
structure means exactly adding the structure of a *-autonomous
category \cite{barr:79}.

Since we are working in classical logic, we will here use the symbols
$\cand,\cor,\ctrue,\cfalse$ for the usual $\ltens,\lpar,\lone,\lbot$.
\let\ltens=\cand
\let\lpar=\cor
\let\lone=\ctrue
\let\lbot=\cfalse

\begin{definition}\label{def:monoidal}
  A \Bo-category $\cC$ is \dfn{symmetric $\cand$-monoidal} if the
  operation $-\ltens-:\cC\times\cC\to\cC$ is a bifunctor and the maps
  $\assoc[A,B,C],\twist[A,B],\runit[A],\lunit[A]$ in \eqref{eq:B0} are
  natural isomorphisms that obey the following equations:
  $$
  \diagpentagondown
      {{A\ltens(B\ltens (C\ltens D))}{\id[A]\ltens \assoc[B,C,D]}}
      {{A\ltens ((B\ltens C)\ltens D)}{\assoc[A,B\ltens C,D]}}
      {{(A\ltens (B\ltens C))\ltens D}{\;\assoc[A,B,C]\ltens \id[D]}}
      {{\assoc[A,B,C\ltens D]}{(A\ltens B)\ltens (C\ltens D)}}
      {{\assoc[A\ltens B,C,D]}{\qqqlapm{((A\ltens B)\ltens C)\ltens D}}}
  $$
  $$
  \diaghexagondown
      {{A\ltens(B\ltens C)}{\assoc[A,B,C]}}
      {{(A\ltens B)\ltens C}{\twist[A\ltens B,C]}}
      {{C\ltens (A\ltens B)}{\assoc[C,A,B]}}
      {{\id[A]\ltens \twist[B,C]}{A\ltens (C\ltens B)}}
      {{\assoc[A,C,B]}{(A\ltens C)\ltens B}}
      {{\twist[A,C]\ltens\id[B]}{(C\ltens A)\ltens B}}
  $$
  $$
  \diagtriangledown
      {{A\ltens(\lone\ltens B)}{\assoc[A,\lone,B]}}
      {{(A\ltens\lone)\ltens B}{\runit[A]\ltens \id[B]}}
      {{\id[A]\ltens \lunit[B]}{A\ltens B}}
  $$
  $$
  \diagtriangledown
      {{A\ltens B}{\twist[A,B]}}
      {{B\ltens A}{\twist[B,A]}}
      {{\idf[A\ltens B]}{A\ltens B}}
  $$
  The notion of \dfn{symmetric $\cor$-monoidal} is defined in a
  similar way.
\end{definition}

An important property of symmetric monoidal categories is the
coherence theorem \cite{maclane:63}, which says that every diagram
containing only natural isomorphisms built out of $\assoc$, $\twist$,
$\runit$, $\lunit$, and the identity $\idf$ via $\cand$ and $\fcomp$
must commute (for details, see \cite{maclane:71}
and~\cite{kelly:64}).\footnote{In~\cite{kelly:64}, Kelly provides some
simplifications to MacLane's conditions in~\cite{maclane:63}. For
example, the equations
$\runit[\lone]=\lunit[\lone]\colon\lone\ltens\lone\to\lone$ and
$\runit[A]\fcomp\twist[\lone,A]={\lunit[A]}\colon\lone\ltens A\to A$
follow from the ones in Definition~\ref{def:monoidal}.}

As a consequence of the coherence theorem, we can omit certain
parentheses to ease the reading. For example, we will write $A\cand
B\cand C\cand D$ for $(A\cand B)\cand (C\cand D)$ as well as for
$A\cand((B\cand C)\cand D)$. This can be done because there is a
uniquely defined ``coherence isomorphism'' between any two of these
objects.

\medskip

Let us now turn our attention to a very important feature of Boolean
logic: the duality between $\cand$ and $\cor$. We can safely say that it
is reasonable to ask for this duality also in a Boolean category. That
means, we are asking for $\cneg{\cneg A}\isom A$ and $\widecneg{A\cand
  B}\isom\cneg A\cor \cneg B$.  At the same time we ask for the
possibility of transposition (or currying): The proofs of $A\land B\to
C$ are in one-to-one correspondence with the proofs of $A\to \cneg
B\cor C$. This is exactly what makes a monoidal category *-autonomous.

\begin{definition}\label{def:staraut}
  A \Bo-category $\cC$ is \dfn{*-autonomous} if it is symmetric
  $\cand$-monoidal and is equipped with a contravariant functor
  $\fneg:\cC\to\cC$, such that $\widecneg{\fneg}\colon\cC\to\cC$ is
  a natural isomorphism
  and such that for any
  three objects $A$, $B$, $C$ there is a natural bijection
  \begin{equation}\label{eq:star}
    \Hom[\cC](A\ltens B,C)\quad\cong\quad\Hom[\cC](A,\cneg B\lpar C)
    \quadfs \tag{$\star$}
  \end{equation}
  where the bifunctor $-\lpar-$ is defined via $A\lpar
  B=\widecneg{\cneg B\ltens \cneg A}$.\footnote{Although we live in
    the commutative world, we invert the order of the arguments when
    taking the negation.}  
  We also define $\cfalse=\cneg\ctrue$.
\end{definition}

Clearly, if a \Bo-category $\cC$ is \dfn{*-autonomous}, then it is
also $\cor$-monoidal with
$\coassoc[A,B,C]=\widecneg{\assoc[\cneg C,\cneg B,\cneg A]}$,
$\cotwist[A,B]=\widecneg{\twist[\cneg B,\cneg A]}$,
$\corunit[A]=\widecneg{\lunit[\cneg A]}$,
$\colunit[A]=\widecneg{\runit[\cneg A]}$.

Note that our definition is not the original one, but it is not
difficult to show the equivalence, and this was already done in
\cite{barr:79}. For further information, see also
\cite{barr:wells:99,barr:91,hughes:freestar,lam:str:freestar}.

Let us continue with stating some well-known facts about *-autonomous
categories (for proofs of these facts, see e.g. \cite{lam:str:freestar}). Via
the bijection~\eqref{eq:star} we can assign to every map $f\colon A\to B\lpar
C$ a map $g\colon A\ltens\cneg B\to C$, and vice versa.  We say that \(f\) and
\(g\) are \dfn{transposes} of each other if they determine each other
via~\eqref{eq:star}. We will use the term ``transpose'' in a very general
sense: given objects $A$, $B$, $C$, $D$, $E$ such that $D\isom A\cand B$ and
$E\isom \cneg B\cor C$, then any $f\colon D\to C$ uniquely determines a
$g\colon A\to E$, and vice versa. Also in that general case we will say that
$f$ and $g$ are transposes of each other. For example,
$\lunit[A]\colon\lone\ltens A\to A$ and $\corunit[A]\colon A\to A\lpar\lbot$
are transposes of each other, and another way of transposing them yields the
maps
$$
\nid[A]\colon\lone\to \cneg A\lpar A
\qquand
\conid[A]\colon A\ltens \cneg A\to\lbot
\quadfs
$$
If we have $f\colon A\to B\lpar C$ and $b\colon B'\to B$, then 
\begin{equation}\label{eq:transpose-neg}
  \hskip-1em
  \diagramhc{
    \arbto[r]{A\cand B'}{A\cand b}&
    \arbto[r]{A\cand B}{f}&
    \noar{C}
    }
  \quad\mbox{is transpose of}\quad
    \diagramhc{
    \arbto[r]{A}{g}&\arbto[r]{\cneg{B}\cor C}{\cneg b\cor C}&
    \noar{\widecneg{B'}\cor C}
    }
\end{equation}
where $g$ is transpose of $f$.

Let us now transpose the identity $\idf[B\lpar C]\colon B\lpar C\to B\lpar
C$. This yields the \dfn{evaluation map} $\eval\colon(B\lpar C)\ltens \cneg
C\to B$. Taking the $\ltens$ of this with $\idf[A]\colon A\to A$ and
transposing back determines a map $\switch[A,B,C]\colon A\ltens(B\lpar
C)\to(A\ltens B)\lpar C$ that is natural in all three arguments, and that we
call the \dfn{switch map} \cite{gug:SIS,brunnler:tiu:01}\footnote{To
category theorists it is probably better known under the names \emph{weak
distributivity} \cite{hyland:depaiva:fill,cockett:seely:97} or \emph{linear
distributivity}. However, strictly speaking, it is not a form of
distributivity. An alternative is the name \emph{dissociativity}
\cite{dosen:petric:coherence-book}.}. In a similar fashion we obtain maps $(A\lpar
B)\ltens C\to A\lpar(B\ltens C)$ and $A\ltens(B\lpar C)\to B\lpar(A\ltens C)$
and $(A\lpar B)\ltens C\to (A\ltens C)\lpar B$. Alternatively these maps can
be obtained from $\switch$ by composing with $\twist$ and $\cotwist$. For this
reason we will use the term ``switch'' for all of them, and denote them by
$\switch[A,B,C]$ if it is clear from context which one is meant, as for
example in the two diagrams
\begin{equation}
  \label{eq:tens}
  \vcdiagsquare{{(A\cor B)\cand(C\cor D)}{\switch[A,B,C\cor D]}}
  {{A\cor(B\cand(C\cor D))}{A\cor\switch[B,C,D]}}
  {{\switch[A\cor B,C,D]}{((A\cor B)\cand C)\cor D}}
  {{\switch[A,B,C]\cor D}{A\cor(B\cand C)\cor D}}
\end{equation}
and
\begin{equation}
  \label{eq:cotens}
  \vcdiagsquare{{A\cand (B\cor C)\cand D}{A\cand\switch[B,C,D]}}
  {{A\cand(B\cor(C\cand D))}{\switch[A,B,C\cand D]}}
  {{\switch[A,B,C]\cand D}{((A\cand B)\cor C)\cand D}}
  {{\switch[A\cand B,C,D]}{(A\cand B)\cor (C\cand D)}}
\end{equation}
which commute in any *-autonomous category.  Sometimes we will denote the map
defined by \eqref{eq:tens} by $\tens[A,B,C,D]\colon(A\lpar B)\ltens(C\lpar
D)\to A\lpar(B\ltens C)\lpar D$, called the \emph{tensor map}\footnote{This
map describes precisely the tensor rule in the sequent system for linear
logic.} and the one of \eqref{eq:cotens} by $\cotens[A,B,C,D]\colon
A\ltens(B\lpar C)\ltens D\to (A\ltens B)\lpar(C\ltens D)$, called the
\emph{cotensor map}.

Note that the switch map is self-dual, while the two maps $\tens$ and
$\cotens$ are dual to each other, i.e.,
\begin{equation}
  \label{eq:dual-switch}
  \vcdiagsquare{{\widecneg{(A\cand B)\cor C}}{\widecneg{\switch[A,B,C]}}}
  {{\widecneg{A\cand (B\cor C)}}{\isom}}
  {{\isom}{\cneg C\cand(\cneg B\cor \cneg A)}}
  {{\switch[\cneg C,\cneg B,\cneg A]}{(\cneg C\cand\cneg B)\cor \cneg A}}
\end{equation}
and
\begin{equation}
  \label{eq:dual-tens}
  \vcdiagsquare{{\widecneg{(A\cand B)\cor(C\cand D)}}
    {\widecneg{\cotens[A,B,C,D]}}}
  {{\widecneg{A\cand (B\cor C)\cand D}}{\isom}}
  {{\isom}{(\cneg D\cor\cneg C)\cand(\cneg B\cor \cneg A)}}
  {{\tens[\cneg D,\cneg C,\cneg B,\cneg A]}
    {\cneg D\cor(\cneg C\cand\cneg B)\cor \cneg A}}
\end{equation}
where the vertical maps are the canonical isomorphisms determined by the
*-autonomous structure.  Another property of switch that we will use
later is the commutativity of the following diagrams:
\begin{equation}
  \label{eq:swi-unit}
  \vcdiagtriangleright{{(A\cor B)\cand\ctrue}{\switch[A,B,\ctrue]}}
  {{A\cor(B\cand\ctrue)}{A\cor\runit[B]}}
  {{\runit[A\cor B]}{A\cor B}}
  \qquad
  \vcdiagtriangleleft{{A\cand B}{\colunit[A]^{-1}\cand B}}
  {{(\cfalse\cor A)\cand B}{\switch[\cfalse,A,B]}}
  {{\colunit[A\cor B]^{-1}}{\cfalse\cor (A\cand B)}}
\end{equation}

\section{Some remarks on mix}\label{sec:mix}

In this section we will recall what it means for a *-autonomous
category to have mix. Although most of the material of this section
can also be found in \cite{cockett:seely:97:mix},
\cite{fuhrmann:pym:dj}, \cite{dosen:petric:coherence-book}, and
\cite{lamarche:gap}, we give here a complete survey since the main
result, Corollary~\ref{cor:mix}, is rather crucial for the following
sections. This corollary essentially says that the mix-rule in the
sequent calculus
$$\vciinf{\mix}{\sqn{\Gamma,\Delta}}{\sqn{\Gamma}}{\sqn{\Delta}}$$ is
a consequence of the fact that false implies true. Although this is
not a very deep result, it might be surprising for logicians that a
property of sequents (if two sequents can be proved independently,
then they can be proved together) which does not involve any units
comes out of an algebraic property concerning only the units.

\begin{theorem}\label{thm:mix}
  Let $\cC$ be a *-autonomous category and $e\colon\cfalse\to\ctrue$
  be a map in~$\cC$. Then 
  \begin{equation}\label{eq:bot-mix}
    \vcdiagsquare{{\cfalse\cand\cfalse}{e\cand\cfalse}}
    {{\ctrue\cand\cfalse}{\lunit[\cfalse]}}
    {{\cfalse\cand e}{\cfalse\cand\ctrue}}
    {{\runit[\cfalse]}{\cfalse}}
  \end{equation}
  if and only if
  \begin{equation}\label{eq:one-mix}
    \vcdiagsquare{{\ctrue}{\colunit[\ctrue]^{-1}}}
    {{\cfalse\cor\ctrue}{e\cor\ctrue}}
    {{\corunit[\ctrue]^{-1}}{\ctrue\cor\cfalse}}
    {{\ctrue\cor e}{\ctrue\cor\ctrue}}
  \end{equation}
  if and only if
  \begin{equation}\label{eq:sbot-mix}
    \vcdiagoctagon{{A\cand B}{A\cand\colunit[B]^{-1}}}
    {{A\cand(\cfalse\cor B)}{\switch[A,\cfalse,B]}}
    {{(A\cand\cfalse)\cor B}{(A\cand e)\cor B}}
    {{(A\cand\ctrue)\cor B}{\runit[A]\cor B}}
    {{\corunit[A]^{-1}\cand B}{(A\cor\cfalse)\cand B}}
    {{\switch[A,\cfalse,B]}{A\cor(\cfalse\cand B)}}
    {{A\cor (e\cand B)}{A\cor(\ctrue\cand B)}}
    {{A\cor\lunit[B]}{A\cor B}}
  \end{equation}
  for all objects $A$ and $B$.
\end{theorem}

\begin{proof}
  First we show that \eqref{eq:bot-mix} implies \eqref{eq:sbot-mix}.
  For this, chase {\small
    \begin{equation}
      \label{eq:prf-mix}
        \vcdiagrammtqc{
          &&&
          \arbto[dr]{A\cand B}{\colunit[B]^{-1}}
          \arpto[dl]{}{\corunit[A]^{-1}}
          \\ &&
          \arpto[dl]{(A\cor\cfalse)\cand B}{\switch}
          \arbto[dr]{}{\colunit[B]^{-1}}
          &&
          \arpto[dl]{A\cand(\cfalse\cor B)}{\corunit[A]^{-1}}
          \arbto[dr]{}{\switch}
          \\
          &
          \arpto[dl]{A\cor(\cfalse\cand B)}{e}
          \arbto[dr]{}{\colunit[B]^{-1}}
          &&
          \arpto[dl]{(A\cor\cfalse)\cand(\cfalse\cor B)}
          {\switch}
          \arbto[dr]{}{\switch}
          &&
          \arpto[dl]{(A\cand\cfalse)\cor B}{\corunit[A]^{-1}}
          \arbto[dr]{}{e}
          \\
          \arbto[dr]{\quad A\cor(\ctrue\cand B)\quad}{\colunit[B]^{-1}}
          &&
          \arpto[dl]{A\cor(\cfalse\cand(\cfalse\cor B))}{e}
          \arbto[dr]{}{\switch}
          &&
          \arpto[dl]{((A\cor\cfalse)\cand\cfalse)\cor B}
          {\switch}
          \arbto[dr]{}{e}
          &&
          \arpto[dl]{\quad(A\cand\ctrue)\cor B\quad}{\corunit[A]^{-1}}
          \\
          &
          \arbto[dr]{A\cor(\ctrue\cand(\cfalse\cor B))}
          {\switch}
          &&
          \arpto[dl]{A\cor(\cfalse\cand\cfalse)\cor B}{e}
          \arbto[dr]{}{e}
          &&
          \arpto[dl]{((A\cor\cfalse)\cand\ctrue)\cor B}
          {\switch}
          \\
          &&
          \arpto[dl]{A\cor(\ctrue\cand\cfalse)\cor B}{\lunit[\cfalse]}
          &&
          \arbto[dr]{A\cor(\cfalse\cand\ctrue)\cor B}{\runit[\cfalse]}
          \\
          &
          \arbfrom[uu]{\lunit[\cfalse\cor B]}{A\cor\cfalse\cor B}
          \arbtoid[rrrr]{}{}
          \arbto[drr]{}{\colunit[B]}
          &&&&
          \arpfrom[uu]{\runit[A\cor\cfalse]}{A\cor\cfalse\cor B}
          \arpto[dll]{}{\corunit[A]}
          \\
          \arbfrom[uuuu]{\lunit[B]}{A\cor B}
          \arbto[ur]{}{\colunit[B]^{-1}}
          \arbtoid[rrr]{}{}
          &&&
          \arbtoid[rrr]{A\cor B}{}
          &&&
          \arpfrom[uuuu]{\runit[A]}{A\cor B}
          \arpto[ul]{}{\corunit[A]^{-1}}
          }
    \end{equation}}%
  The big triangle at the center is an application of
  \eqref{eq:bot-mix}. The two little triangles next to it are
  (variations of) \eqref{eq:swi-unit}, and the triangles at the bottom
  are trivial.  The topmost square is functoriality of $\cand$, the
  square in the center is \eqref{eq:tens}, and all other squares
  commute because of naturality of $\swir$, $\lunit$, $\runit$,
  $\colunit$, and $\corunit$. Now observe that \eqref{eq:sbot-mix}
  commutes if and only if
  \begin{equation}\label{eq:sone-mix}
    \vcdiagoctagon{{A\cand B}{A\cand\colunit[B]^{-1}}}
    {{A\cand(\cfalse\cor B)}{A\cand(e\cor B)}}
    {{A\cand(\ctrue\cor B)}{\switch[A,\ctrue,B]}}
    {{(A\cand\ctrue)\cor B}{\runit[A]\cor B}}
    {{\corunit[A]^{-1}\cand B}{(A\cor\cfalse)\cand B}}
    {{(A\cor e)\cand B}{(A\cor\ctrue)\cand B}}
    {{\switch[A,\ctrue,B]}{A\cor(\ctrue\cand B)}}
    {{A\cor\lunit[B]}{A\cor B}}
  \end{equation}
  commutes (because of naturality of switch), and that the diagonals
  of \eqref{eq:sbot-mix} and \eqref{eq:sone-mix} are the same map
  $\mix[A,B]\colon A\cand B\to A\cor B$.  Note that by the dual of
  \eqref{eq:prf-mix} we get that \eqref{eq:one-mix} implies
  \eqref{eq:sone-mix}. Therefore we also get that \eqref{eq:one-mix}
  implies \eqref{eq:sbot-mix}.
  Now we show that \eqref{eq:sone-mix} implies
  \eqref{eq:one-mix}.  We will do this by showing that 
  \begin{equation}\label{eq:one-mix-a}
    \vcdiagramdc{
      &
      \arpto[dl]{\ctrue}{\corunit[\ctrue]^{-1}}
      \arbto[d]{}{\runit[\ctrue]^{-1}=\lunit[\ctrue]^{-1}}
      \arbto[dr]{}{\colunit[\ctrue]^{-1}}
      \\
      \arpto[dr]{\ctrue\cor\cfalse}{\ctrue\cor e}
      &
      \arbto[d]{\ctrue\cand\ctrue}{\mix[\ctrue,\ctrue]}
      &
      \arbto[dl]{\cfalse\cor\ctrue}{e\cor\ctrue}
      \\
      &
      \noar{\ctrue\cor\ctrue}
      }
  \end{equation}
  commutes. For this, consider
  \begin{equation}\label{eq:one-mix-b}
    \vcdiagramhr{
      \arbto[dr]{\ctrue}{\runit[\ctrue]^{-1}}
      \\
      &
      \arbto[dl]{\ctrue\cand\ctrue}{\runit[\ctrue]}
      \arbto[dd]{}{\corunit[\ctrue]^{-1}\cand\ctrue}
      \\
      \arbfromid[uu]{}{\ctrue}
      \\
      &
      \arbto[dl]{(\ctrue\cor\cfalse)\cand\ctrue}{\runit[\ctrue\cor\cfalse]}
      \arbto[dd]{}{(\ctrue\cor e)\cand\ctrue}
      \\
      \arbfrom[uu]{\corunit[\ctrue]^{-1}}{\ctrue\cor\cfalse}
      \\
      &
      \arbto[dl]{(\ctrue\cor\ctrue)\cand\ctrue}{\runit[\ctrue\cor\ctrue]}
      \arbto[dd]{}{\switch[\ctrue,\ctrue,\ctrue]}
      \\
      \arbfrom[uu]{\ctrue\cor e}{\ctrue\cor\ctrue}
      \\
      &
      \arbto[dl]{\ctrue\cor(\ctrue\cand\ctrue)}{\ctrue\cor\lunit[\ctrue]}
      \arbto[ul]{}{\ctrue\cor\runit[\ctrue]}
      \\
      \arbfromid[uu]{}{\ctrue\cor\ctrue}
      }
  \end{equation}
  which says that the left triangle in \eqref{eq:one-mix-a} commutes
  because the right down path in \eqref{eq:one-mix-b} is exactly the
  lower left path in \eqref{eq:sone-mix}.  Similarly we obtain the
  commutativity of the right triangle in \eqref{eq:one-mix-a}.  In the
  same way we show that \eqref{eq:sbot-mix} implies
  \eqref{eq:bot-mix}, which completes the proof.   
\end{proof}

Therefore, in a *-autonomous category every map
$e\colon\cfalse\to\ctrue$ obeying \eqref{eq:bot-mix} uniquely
determines a map $\mix[A,B]\colon A\cand B\to A\cor B$ which is natural in
$A$ and $B$.  It can be shown that this \emph{mix map} goes well with
the twist, associativity, and switch maps:

\begin{proposition}
  The map $\mix[A,B]\colon A\cand B\to A\cor B$ obtained from
  \eqref{eq:sbot-mix} is natural in both arguments and obeys the
  equations
  \begin{equation}
    \label{eq:mix-twist}
    \vcdiagsquare{{A\cand B}{\mix[A,B]}}{{A\cor B}{\cotwist[A,B]}}
    {{\twist[A,B]}{B\cand A}}{{\mix[B,A]}{B\cor A}}
    \tag{$\mix$-$\twist$}
  \end{equation}
  and
  \begin{equation}
    \label{eq:mix-assoc}
    \vcdiaghexagonright{{A\cand(B\cand C)}{A\cand\mix[B,C]}}
    {{A\cand(B\cor C)}{\mix[A,B\cor C]}\arbto[d]{}{\switch[A,B,C]}}
    {{A\cor(B\cor C)}{\coassoc[A,B,C]}}
    {{\assoc[A,B,C]}{(A\cand B)\cand C}}
    {{\mix[A\cand B,C]}{(A\cand B)\cor C}}
    {{\mix[A,B]\cor C}{(A\cor B)\cor C}}
    \tag{$\mix$-$\assoc$}
  \end{equation}
\end{proposition}

\begin{proof}
  Naturality of mix follows immediately from the naturality of switch.
  Equation \eqref{eq:mix-twist} follows immediately from the
  definition of switch, and \eqref{eq:mix-assoc} can be shown with a
  similar diagram as \eqref{eq:prf-mix}.
\end{proof}

\begin{corollary}\label{cor:mix}
  In a *-autonomous category there is a one-to-one correspondence
  between the maps $e\colon\cfalse\to\ctrue$
  obeying~\eqref{eq:bot-mix} and the natural transformations
  $\mix[A,B]\colon A\cand B\to A\cor B$ obeying \eqref{eq:mix-twist}
  and~\eqref{eq:mix-assoc}.
\end{corollary}

\begin{proof}
  Whenever we have a map $\mix[A,B]\colon A\cand B\to A\cor B$ for all $A$ and
  $B$, we can form the map
  \begin{equation}
    \label{eq:mix-e}
    \diagram{
      \arbto[r]{e\colon\cfalse}{\runit[\cfalse]^{-1}}&
      \arbto[r]{\cfalse\cand\ctrue}{\mix[\cfalse,\ctrue]}&
      \arbto[r]{\cfalse\cor\ctrue}{\colunit[\ctrue]}&
      \noar{\ctrue}
      }
  \end{equation}
  One can now easily show that naturality of mix, as well as
  \eqref{eq:mix-twist} and~\eqref{eq:mix-assoc} are exactly what is needed to
  let the map $e\colon\cfalse\to\ctrue$ defined in \eqref{eq:mix-e} obey
  equation~\eqref{eq:bot-mix}. We leave the details to the reader. Hint: Show
  that both maps of \eqref{eq:bot-mix} are equal to 
  $$
  \diagram{
    \arbto[r]{\cfalse\cand\cfalse}{\mix[\cfalse,\cfalse]}&
    \arbto[r]{\cfalse\cor\cfalse}{\colunit[\cfalse]=\corunit[\cfalse]}&
    \noar{\cfalse}
  }
  \quadfs
  $$ 
  It remains to show that plugging the map of \eqref{eq:mix-e} into
  \eqref{eq:sbot-mix} gives back the same natural transformation
  $\mix[A,B]\colon A\cand B\to A\cor B$ we started from. Similarly, plugging
  in the the mix defined via \eqref{eq:sbot-mix} into \eqref{eq:mix-e} gives
  back the same map $e\colon\cfalse\to\ctrue$ that has been plugged into
  \eqref{eq:sbot-mix}. Again, we leave the details to the reader.
\end{proof}

Note that a *-autonomous category can have many different
maps $e\colon\cfalse\to\ctrue$ with the property of
Theorem~\ref{thm:mix}, each of them defining its own natural mix obeying
\eqref{eq:mix-twist} and \eqref{eq:mix-assoc}.

\section{$\cor$-Monoids and $\cand$-comonoids}\label{sec:monoids}

The structure investigated so far is exactly the same as for proofs in linear
logic (with or without mix). For classical logic, we need to provide algebraic
structure for the maps $\codiag[A]\colon A\cor A\to A$ and
$\coproj[A]\colon\cfalse\to A$, as well as $\diag[A]\colon A\to A\cand A$ and
$\proj[A]\colon A\to\ctrue$, which are listed in \eqref{eq:B0}.  This is done
via monoids and comonoids.

\begin{definition}\label{def:monoid-comonoid}
  A \Bo-category has \dfn{commutative $\cor$-monoids} if it is
  symmetric $\cor$-mo\-noidal and for every object $A$, the maps
  $\codiag[A]$ and $\coproj[A]$ obey the equations
  \begin{equation}
    \label{eq:monoid}
      \vcdiagpentagonright
	  {{A\cor(A\cor A)}{\coassoc[A,A,A]}}
	  {{(A\cor A)\cor A}{\codiag[A]\cor\id[A]}}
	  {{A\cor A}{\codiag[A]}}
	  {{\id{A}\cor\codiag[A]}{A\cor A}}
	  {{\codiag[A]}{A}}
      \qquad\!
      \vcdiagtriangleright
	  {{A\cor A}{\cotwist[A,A]}}
	  {{A\cor A}{\codiag[A]}}
	  {{\codiag[A]}{A}}
      \qquad\!
      \vcdiagtriangleright
	  {{A\cor\cfalse}{\id[A]\cor\coproj[A]}}
	  {{A\cor A}{\codiag[A]}}
	  {{\corunit[A]}{A}}
  \end{equation}
  Dually, we say that a \Bo-category has \dfn{cocommutative
  $\cand$-comonoids} if  it is
  symmetric $\cand$-monoidal and for every object $A$, the maps
  $\diag[A]$ and $\proj[A]$ obey the equations
  \begin{equation}
    \label{eq:comonoid}
      \vcdiagpentagonleft
	  {{A}{\diag[A]}}
	  {{A\cand A}{\diag[A]\cand\id[A]}}
	  {{(A\cand A)\cand A}{\assoc[A,A,A]^{-1}}}
	  {{\diag[A]}{A\cand A}}
	  {{\id{A}\cand\diag[A]}{A\cand(A\cand A)}}
      \qquad\!
      \vcdiagtriangleleft
	  {{A}{\diag[A]}}
	  {{A\cand A}{\twist[A,A]^{-1}}}
	  {{\diag[A]}{A\cand A}}
      \qquad\!
      \vcdiagtriangleleft
	  {{A}{\diag[A]}}
	  {{A\cand A}{\id{A}\cand\proj[A]}}
	  {{\runit[A]^{-1}}{A\cand\ctrue}}
  \end{equation}
\end{definition}

Translated into the language of the sequent calculus
(cf.~\cite{fuhrmann:pym:oecm}), having the structure of a $\cor$-monoid, i.e.,
the equations in \eqref{eq:monoid}, means
\begin{enumerate}[\rm(i)]
\item to force the identification of the two possible proofs of the shape
  $$
  \vcdernote{\mathsf{contraction}}{}{\ssqn{\Lambda}{A,\Gamma}}{
    \root{\mathsf{contraction}}{\ssqn{\Lambda}{A,A,\Gamma}}{
      \leaf{\ssqn{\Lambda}{A,A,A,\Gamma}}}}
  $$
\item to identify the two proofs
  $$
  \vcenter{\hbox{
  \dernote{\mathsf{contraction}}{\qqquand}{\ssqn{\Lambda}{A,\Gamma}}{
    \root{\mathsf{exchange}}{\ssqn{\Lambda}{A,A,\Gamma}}{
      \leaf{\ssqn{\Lambda}{A,A,\Gamma}}}}
  \dernote{\mathsf{contraction}}{}{\ssqn{\Lambda}{A,\Gamma}}{
    \leaf{\ssqn{\Lambda}{A,A,\Gamma}}}}}
  $$
\item to say that the derivation 
  $$
  \vcdernote{\mathsf{contraction}}{}{\ssqn{\Lambda}{A,\Gamma}}{
    \root{\mathsf{weakening}}{\ssqn{\Lambda}{A,A,\Gamma}}{
      \leaf{\ssqn{\Lambda}{A,\Gamma}}}}
  $$
  is the same as doing nothing (i.e., the identity).
\end{enumerate}
The equations in \eqref{eq:comonoid}, i.e., the structure of an
$\cand$-comonoid, forces the same identification on the left-hand side of the
turnstile. See \cite{fuhrmann:pym:oecm} for a detailed discussion of this
correspondence.

\begin{remark}
  The (co)associativity of the maps $\diag[A]$ and $\codiag[A]$ allows us to
  use the notation $\diagsq[A]\colon A\to A\cand A\cand A$ and 
  $\codiagsq[A]\colon A\cor A\cor A\to A$.
\end{remark}

\begin{proposition}\label{prop:unique-unit}
  Let $\cC$ be a category with commutative
  $\cor$-monoids, and let 
  $$
  \diagtriangleright
      {{A\cor\cfalse}{\id[A]\cor f}}
      {{A\cor A}{\codiag[A]}}
      {{\corunit[A]}{A}}
  $$
  commute for some $f\colon\cfalse\to A$. Then $f=\coproj[A]$.
\end{proposition}

\begin{proof}
  This is a well-known fact from algebra: in a monoid the unit is
  uniquely defined. Written as diagram, the standard proof looks as
  follows:
  $$
  \vcdiagram{
    &&
    \arbto[d]{\cfalse}{\colunit[\cfalse]^{-1}=\corunit[\cfalse]^{-1}}
    \arbtoid[dll]{}{}
    \arbtoid[drr]{}{}
    \\
    \arbtoid[dd]{\cfalse}{}
    &&
    \arbto[ll]{\cfalse\cor\cfalse}{\colunit[\cfalse]}
    \arpto[rr]{}{\corunit[\cfalse]}
    \arpto[dl]{}{\cfalse\cor f} 
    \arbto[dr]{}{\coproj[A]\cor\cfalse}
    &&
    \arbtoid[dd]{\cfalse}{}
    \\
    &
    \arbto[r]{\cfalse\cor A}{\coproj[A]\cor A}
    \arpto[dr]{}{\colunit[A]}
    &
    \arbto[d]{A\cor A}{\codiag[A]}
    &
    \arpto[l]{A\cor\cfalse}{A\cor f} 
    \arbto[dl]{}{\corunit[A]}
    \\
    \arpto[rr]{\cfalse}{f} 
    &&
    \noar{A}
    &&
    \arbto[ll]{\cfalse}{\coproj[A]}
    }
  $$
  Note that in the same way it follows that the counit in a comonoid
  is uniquely defined.  
\end{proof}

Although the operations $\cand$ and $\cor$ are \emph{not} the product and
coproduct in the category-theoretic sense, we use the notation:
\begin{equation}\label{eq:prod}
  \prod{f,g}=(f\cand g)\fcomp\diag[A]\colon A\to C\cand D
  \quand
  \coprod{f,h}=\codiag[C]\fcomp(f\cor h)\colon A\cor B\to C
  \quad
\end{equation}
where $f\colon A\to C$ and $g\colon A\to D$ and $h\colon B\to C$ are
arbitrary maps.

Another helpful notation (see \cite{lam:str:05:freebool}) is the
following:
{
\begin{equation}\label{eq:projlr}
  \begin{array}{l@{\qquad}l@{\quad}}
    \projl{A}{B}=\runit[A]\fcomp(A\cand\proj[B])\colon A\cand B\to A&
    \projr{A}{B}=\lunit[B]\fcomp(\proj[A]\cand B)\colon A\cand B\to B
    \\[\interdisplayskip]
    \coprojl{A}{B}=(A\cor\coproj[B])\fcomp\corunit[A]^{-1}\colon A\to A\cor B&
    \coprojr{A}{B}=(\coproj[A]\cor B)\fcomp\colunit[B]^{-1}\colon B\to A\cor B
  \end{array}
\end{equation}
}%
Note that 
\begin{equation}\label{eq:diag-proj-id}
  \codiag[A]\fcomp\coprojr{A}{A}=\idf[A]=\codiag[A]\fcomp\coprojl{A}{A}
  \qquand
  \projr{A}{A}\fcomp\diag[A]=\idf[A]=\projl{A}{A}\fcomp\diag[A]
\end{equation}

\begin{definition}\label{def:strong}
  Let $f\colon A\to B$ be a map in a \Bo-category with commutative
  $\cor$-monoids and cocommutative $\cand$-comonoids.  Consider the
  following four diagrams:\strut{\small
  $$
    \diagsquare{{A\cor A}{f\cor f}}{{B\cor B}{\codiag[B]}}
    {{\codiag[A]}{A}}{{f}{B}}
    \quad\!
    \diagtriangleup{{\cfalse}{\coproj[A]}}{{A}{f}}{{\coproj[B]}{B}}
    \quad\!
    \diagtriangledown{{A}{f}}{{B}{\proj[B]}}{{\proj[A]}{\ctrue}}
    \quad\!
    \diagsquare{{A}{f}}{{B}{\diag[B]}}
    {{\diag[A]}{A\cand A}}{{f\cand f}{B\cand B}}
  $$}%
  We say that 
  \begin{itemize}
  \item $f$ \dfn{preserves the $\cor$-multiplication} if the left square
    commutes,
  \item $f$ \dfn{preserves the $\cor$-unit} if the left triangle commutes,
  \item $f$ \dfn{preserves the $\cand$-counit} if the right triangle
    commutes,
  \item $f$ \dfn{preserves the $\cand$-comultiplication} if the right square
    commutes,
  \item $f$ is a \dfn{$\cor$-monoid morphism} if the two left diagrams
    commute,
  \item $f$ is a \dfn{$\cand$-comonoid morphism} if the two right diagrams
    commute,
  \item $f$ is a \dfn{quasientropy} if both triangles commute,
  \item $f$ is \dfn{clonable} if both squares commute,
  \item $f$ is \dfn{strong} if all four diagrams commute.
  \end{itemize}
\end{definition}

\begin{definition}\label{def:B1}
  A \dfn{\Bi-category} is a \Bo-category that is *-autonomous and has
  cocommutative $\cand$-comonoids.
\end{definition}

Clearly, a \Bi-category also has commutative $\cor$-monoids with
$\codiag$ dual to~$\diag$, and $\coproj$ dual to~$\proj$.

\begin{remark}\label{rem:iso1}
  Definition~\ref{def:B1} exhibits another ``creative tension'' between
  algebra and proof theory. From the algebraic point of view one should add
  the phrase ``and all isomorphisms preserve the $\cand$-comonoid structure''
  because in a semantics of proofs this will probably be inevitable. But here
  we do not assume it from the beginning, but systematically give conditions
  that will ensure it in the end (cf. Theorem~\ref{thm:B4} and
  Remark~\ref{rem:B4}). From the proof-theoretic view point this is more
  interesting because when seen syntactically, these conditions are more
  primitive. The reason is that in syntax the morphisms (i.e., proofs) come
  after the objects (i.e., formulae), and the formulae can always be
  decomposed into subformulae, whereas in semantics we have no access to the
  outermost connective.  Furthermore, forcing all isomorphisms to preserve the
  $\cand$-comonoid structure can cause identifications of proofs that might
  not necessarily be wanted by every proof theorist (see, e.g.,
  Proposition~\ref{prop:med-assoc}).
\end{remark}

\begin{remark}\label{rem:closed-comp}
  For each object $A$ in a \Bi-category $\cC$, the identity map
  $\idf[A]\colon A\to A$ is strong, and all kinds of maps defined in
  Definition~\ref{def:strong} are closed under composition. Therefore,
  each kind defines a wide subcategory (i.e., a subcategory that has all
  objects) of $\cC$, e.g., the wide
  subcategory of quasientropies, or the wide subcategory of
  $\cor$-monoid morphisms.
\end{remark}

In a \Bi-category we have two canonical maps $\cfalse\to\ctrue$,
namely $\proj[\cfalse]$ and $\coproj[\ctrue]$. Because of the
$\cand$-comonoid structure on $\cfalse$ and the $\cor$-monoid
structure on $\ctrue$, we have
$$
\qlapm{
\vcdiagram{
  \arbto[r]{\cfalse\cor\ctrue}{\coproj[\ctrue]\cor\ctrue}
  &
  \arbto[d]{\ctrue\cor\ctrue}{\codiag[\ctrue]}
  &
  \arpto[l]{\ctrue\cor\cfalse}{\ctrue\cor\coproj[\ctrue]}
  \\ &
  \arbfrom[ul]{\colunit[\ctrue]}{\ctrue}
  \arpfrom[ur]{\corunit[\ctrue]}{}
  }
\quand
\vcdiagram{
  \arbfrom[r]{\proj[\cfalse]\cand\cfalse}{\ctrue\cand\cfalse}
  &
  \arbfrom[d]{\diag[\cfalse]}{\cfalse\cand\cfalse}
  &
  \arpfrom[l]{\cfalse\cand\proj[\cfalse]}{\cfalse\cand\ctrue}
  \\ &
  \arbto[ul]{\cfalse}{\lunit[\cfalse]^{-1}}
  \arpto[ur]{}{\runit[\cfalse]^{-1}}
  }
}
$$
(which even hold if the (co)monoids are not (co)commutative.) Since
$\colunit[\ctrue]$, $\corunit[\ctrue]$, $\lunit[\cfalse]$, and
$\runit[\cfalse]$ are isomorphisms, we immediately can conclude that 
the following two diagrams commute (cf.~\cite{fuhrmann:pym:dj}):
$$
  \vcdiagsquare{{\ctrue}{\colunit[\ctrue]^{-1}}}
  {{\cfalse\cor\ctrue}{\coproj[\ctrue]\cor\ctrue}}
  {{\corunit[\ctrue]^{-1}}{\ctrue\cor\cfalse}}
  {{\ctrue\cor\coproj[\ctrue]}{\ctrue\cor\ctrue}}
  \qquand
  \vcdiagsquare{{\cfalse\cand\cfalse}{\cfalse\cand\proj[\cfalse]}}
  {{\cfalse\cand\ctrue}{\runit[\cfalse]}}
  {{\proj[\cfalse]\cand\cfalse}{\ctrue\cand\cfalse}}
  {{\lunit[\cfalse]}{\cfalse}}
$$
By Section~\ref{sec:mix}, this gives us two different mix maps $A\cand
B\to A\cor B$, and motivates the following definition:

\begin{definition}\label{def:single-mixed}
  A \Bi-category is called \dfn{single-mixed} if
  $\proj[\cfalse]=\coproj[\ctrue]$.
\end{definition}

In a single-mixed \Bi-category we have, as the name says, a single
canonical mix map $\mix[A,B]\colon A\cand B\to A\cor B$ obeying
\eqref{eq:mix-twist} and \eqref{eq:mix-assoc}. The naturality of mix,
i.e., the commutativity of
\begin{equation}
  \label{eq:mix-nat}
  \vcdiagsquare{{A\cand B}{\mix[A,B]}}{{A\cor B}{f\cor g}}
  {{f\cand g}{C\cand D}}{{\mix[C,D]}{C\cor D}}
\end{equation}
for all maps $f\colon A\to C$ and $g\colon B\to D$, uniquely
determines a map $f\candor g\colon A\cand B\to C\cor D$. Then, for
every $f,g \colon A\to B$ we can define
$$f+g=\codiag_B\fcomp(f\candor g)\fcomp\diag_A\colon A\to B \quadfs$$
It follows from (co)-associativity and (co)-commutativity of $\diag$
and $\codiag$, along with naturality of $\mix$, that the operation $+$
on maps is associative and commutative.  This gives us for $\Hom(A,B)$
a commutative semigroup structure.

Note that in general the semigroup structure on the Hom-sets is not an
enrichment, e.g., $(f+g)h$ is in general not the same as $fh+gh$.

\begin{definition}\label{def:idempotent}
  Let $\cC$ be a single-mixed \Bi-category. Then $\cC$ is called
  \emph{idempotent} if for every $A$ and $B$, the semigroup on
  $\Hom(A,B)$ is idempotent, i.e., for every $f\colon A\to B$ we have
  $f+f=f$.
\end{definition}

In an idempotent \Bi-category the semigroup structure on $\Hom(A,B)$
is in fact a sup-semilattice structure, given by $f\le g$ iff $f+g=g$.

\bigskip

One can argue that the structure of \Bi-categories is in some sense
the minimum of algebraic structure that a Boolean category should
have: *-autonomous categories provide the right structure for linear
logic proofs, and the $\cor$-monoids and $\cand$-comonoids seem to be
exactly what is needed to ``model contraction and weakening'' in
classical logic. There are certainly reasons to argue against that,
since it is by no means God-given that the proofs in classical logic
obey the bijection \eqref{eq:star} nor that ``contraction is
associative''.  But let us, for the time being, assume that proofs in
classical logic form a \Bi-category. Then it is desirable that there
is some more structure. This can be, for example, an agreement between
the $\cand$-monoidal structure (Definition~\ref{def:monoidal}) and the
$\cand$-comonoid structure (Definition~\ref{def:monoid-comonoid}), or,
a more sophisticated condition like the commutativity of the diagram
\begin{equation}\label{eq:switchy}
  \vcdiagoctagonslim
      {{((A\cand B)\cor(A\cand B))\cand(A\cor B)}
	{\codiag[A\cand B]\cand(A\cor B)}}
      {{A\cand B\cand (A\cor B)}{\isom}}
      {{A\cand(B\cor A)\cand B}{\cotens[A,B,A,B]}}
      {{(A\cand B)\cor(A\cand B)}{\codiag[A\cand B]}}
      {{\switch[A\cand B,A\cand B,A\cor B]}
	{(A\cand B)\cor(A\cand B\cand (A\cor B))}}
      {{\isom}{(A\cand B)\cor(A\cand(B\cor A)\cand B)}}
      {{(A\cand B)\cor\cotens[A,B,A,B]}
	{(A\cand B)\cor(A\cand B)\cor(A\cand B)}}
      {{\codiagsq[A\cand B]}{A\cand B}}
\end{equation}
for all objects $A$ and $B$.
We now start to add the axioms for this.

\begin{proposition}\label{prop:B2a}
  Let $\cC$ be a \Bi-category in which the equation
  \begin{equation}\label{eq:B2a}
    \proj[\ctrue]=\idf[\ctrue]\colon\ctrue\to\ctrue
    \tag{\sf B2a} 
  \end{equation}
  holds. Then we have that 
  \begin{enumerate}[\rm(i)]
  \item\label{l:diag-unit}
    $\diag[\ctrue]=\runit[\ctrue]^{-1}\colon\ctrue\to\ctrue\cand\ctrue$
  \item For all objects $A$, the map $\proj[A]$ is a $\cand$-comonoid
    morphism.
  \end{enumerate}
\end{proposition}

\begin{proof}
  The equation $\diag[\ctrue]=\runit[\ctrue]^{-1}$ follows immediately
  from $\proj[\ctrue]=\idf[\ctrue]$ and the definition of
  $\cand$-comonoids. That $\proj[A]$ preserves the $\cand$-counit is
  trivial and that it preserves the $\cand$-comultiplication follows
  from
  \begin{equation}
    \vcddiagsquare{{A}{\proj[A]}}{{\ctrue}{\diag[\ctrue]=\runit[\ctrue]^{-1}}}
    {&\arbfrom[ul]{\runit[A]^{-1}}{A\cand\ctrue}
      \arpfrom[dl]{A\cand\proj[A]}{}
      \arbto[dr]{}{\proj[A]\cand\ctrue}}
    {{\diag[A]}{A\cand A}}{{\proj[A]\cand\proj[A]}{\ctrue\cand\ctrue}}
    \tag*{}
  \end{equation}
  where the left triangle is the definition of $\cand$-comonoids, the
  lower triangle is functoriality of $\cand$ and the big ``triangle''
  is naturality of $\runit$.  
\end{proof}

\begin{lemma}\label{lem:tf-idem}
  If a \Bi-category is single-mixed and obeys \eqref{eq:B2a}, then
  \begin{equation}\label{eq:tf-idem}
    \idf[\ctrue]+\idf[\ctrue]=\idf[\ctrue]
    \qquand
    \idf[\cfalse]+\idf[\cfalse]=\idf[\cfalse]
  \end{equation}  
\end{lemma}

\begin{proof}
  First, we show that 
  \begin{equation}\label{eq:pi-nabla-mix}
    \projr{\cfalse}{\cfalse}=\codiag[\cfalse]\fcomp\mix[\cfalse,\cfalse]
    \colon\cfalse\cand\cfalse\to\cfalse
  \end{equation}  
  This is done by chasing the diagram
  \begin{equation}\label{eq:pi-nabla-mix-dia}
    \vcdiagram{
      \arbto[r]{\cfalse\cand\cfalse}{\colunit[\cfalse]^{-1}\cand\cfalse}
      &
      \arbto[d]{(\cfalse\cor\cfalse)\cand\cfalse}
      {(\cfalse\cand\proj[\cfalse])\cand\cfalse}
      \\
      \arbfrom[u]{\proj[\cfalse]\cand\cfalse}{\ctrue\cand\cfalse}
      \arpto[dr]{}{\colunit[\ctrue\cand\cfalse]^{-1}}
      &
      \arpfrom[l]{\colunit[\ctrue]^{-1}\cand\cfalse}
      {(\cfalse\cor\ctrue)\cand\cfalse}
      \arbto[d]{}{\switch[\cfalse,\ctrue,\cfalse]}
      \\ &
      \arbto[d]{\cfalse\cor(\ctrue\cand\cfalse)}
      {\cfalse\cor\lunit[\cfalse]}
      \\
      \arbfromid[uu]{}{\ctrue\cand\cfalse}
      \arbfrom[ur]{\colunit[\ctrue\cand\cfalse]}{}
      &
      \arbto[d]{\cfalse\cor\cfalse}{\codiag[\cfalse]}
      \\
      \arbfrom[u]{\lunit[\cfalse]}{\cfalse}
      \arbfrom[ur]{\colunit[\cfalse]}{}
      &
      \arbfromid[l]{}{\cfalse}
      }
  \end{equation}  
  The right-down path is $\codiag[\cfalse]\fcomp\mix[\cfalse,\cfalse]$
  and the left down path is $\projr{\cfalse}{\cfalse}$. The two
  squares commute because of naturality of $\colunit$, the upper
  triangle holds because \eqref{eq:swi-unit}, the big triangle in the
  center is trivial, and that the lower triangle commutes follows from
  (the dual of) Proposition~\ref{prop:B2a}~\eqref{l:diag-unit}.  Now
  we can proceed:
  $$
  \idf[\cfalse]=\projr{\cfalse}{\cfalse}\fcomp\diag[\cfalse]
  =\codiag[\cfalse]\fcomp\mix[\cfalse,\cfalse]\fcomp\diag[\cfalse]
  =\idf[\cfalse]+\idf[\cfalse]
  $$
  The equation $\idf[\ctrue]=\idf[\ctrue]+\idf[\ctrue]$ follows by
  duality.
\end{proof}

Note that Lemma~\ref{lem:tf-idem} is a consequence of having proper units. In
the case of weak units (see \cite{lam:str:05:naming,lam:str:05:freebool}) it
does not hold.

\begin{proposition}\label{prop:f-plus-pi}
  In a \Bi-category that is single-mixed and obeys \eqref{eq:B2a}, we have
  \begin{equation}
    \label{eq:f-plus-pi}
    f+\proj[A]\;=\;f
  \end{equation}
  for all maps $f\colon A\to\ctrue$. Dually, we have  
  \begin{equation}
    \label{eq:co-f-plus-pi}
    g+\coproj[B]\;=\;g
  \end{equation}
  for all maps $g\colon\cfalse\to B$. 
\end{proposition}

\begin{proof}
  Chase the diagram
  \begin{equation}\label{eq:f-plus-pi-dia}
    \vcdiagram{
      \arbto[r]{A}{\diag[A]}
      &
      \arbto[d]{A\cand A}{A\cand\proj[A]}
      \\
      \arbfromid[u]{}{A}
      &
      \arpto[l]{A\cand\ctrue}{\runit[A]}
      \arbto[d]{}{f\cand\ctrue}
      \\
      \arbfrom[u]{f}{\ctrue}
      &
      \arpto[l]{\ctrue\cand\ctrue}{\runit[\ctrue]}
      \arbtoid[d]{}{}
      \\&
      \arbfrom[ul]{\diag[\ctrue]}{\ctrue\cand\ctrue}
      \arbto[d]{}{\mix[\ctrue,\ctrue]}
      \\
      \arbfromid[uu]{}{\ctrue}
      &
      \arbto[l]{\ctrue\cor\ctrue}{\codiag[\ctrue]}
      }
  \end{equation}  
  The first square is the comonoid equation, the second one is
  naturality of $\runit$, the triangle commutes because of
  Proposition~\ref{prop:B2a}~\eqref{l:diag-unit}, and the lower
  quadrangle is~\eqref{eq:tf-idem}.   
\end{proof}

\begin{proposition}\label{prop:B2b}
  In a \Bi-category obeying \eqref{eq:B2a}, the equation
  \begin{equation}\label{eq:B2b}
    \vcdiagtriangleup{{A\cand B}{\proj[A]\cand\proj[B]}}
    {{\ctrue\cand\ctrue}{\runit[\ctrue]}}{{\proj[A\cand B]}{\ctrue}}
    \tag{{\sf B2b}} 
  \end{equation}
  holds if and only if
  \begin{enumerate}[\rm(i)]
  \item\label{l:B2b1} $\proj[\ctrue\cand\ctrue]=
    \runit[\ctrue]\colon\ctrue\cand\ctrue\to\ctrue$ and
  \item\label{l:B2b2} the maps that preserve the $\cand$-counit are closed
    under $\cand$.
  \end{enumerate}
\end{proposition}

\begin{proof}
  We see that \eqref{l:B2b1} follows from \eqref{eq:B2a} and
  \eqref{eq:B2b} by plugging in $\ctrue$ for $A$ and $B$ in
  \eqref{eq:B2b}. That \eqref{l:B2b2} holds follows from
  \begin{equation}\label{eq:proj-and}
    \vcdiagramdchr{
      \arbto[rr]{A\cand B}{f\cand g}&&
      \arbto[lddd]{C\cand D}{\proj[C\cand D]}\\ &
      \arbto[dd]{\ctrue\cand\ctrue}{\runit[\ctrue]}
      \arbfrom[lu]{\proj[A]\cand\proj[B]}{}
      \arpfrom[ru]{\proj[C]\cand\proj[D]}{}\\ \\ &
      \arbfrom[luuu]{\proj[A\cand B]}{\ctrue}
      }
  \end{equation}
  where $f\colon A\to B$ and $g\colon C\to D$ are maps that preserve
  the $\cand$-counit. Conversely, it follows from \eqref{l:B2b2} and
  Proposition~\ref{prop:B2a} that $\proj[A]\cand\proj[B]$ preserves the
  $\cand$-counit. With \eqref{l:B2b1} this yields \eqref{eq:B2b}.
\end{proof}

\begin{proposition}\label{prop:assoc-counit}
  In a \Bi-category obeying \eqref{eq:B2a} and \eqref{eq:B2b} the maps
  $\assoc[A,B,C]$, $\twist[A,B]$, $\runit[A]$, $\lunit[A]$,
  $\proj[A]$, $\projl{B}{A}$, and $\projr{A}{B}$ all preserve the
  $\cand$-counit. And dually, the maps $\coassoc[A,B,C]$,
  $\cotwist[A,B]$, $\corunit[A]$, $\colunit[A]$, $\coproj[A]$,
  $\coprojl{B}{A}$, and $\coprojr{A}{B}$ all preserve the $\cor$-unit.
\end{proposition}

\begin{proof}
  We show the case for $\twist[A,B]$:
  $$
  \diagramhchr{
    \arbto[rrrrrr]{A\cand B}{\twist[A,B]}&&&&&&
    \arbto[lllddd]{B\cand A}{\proj[B\cand A]}\\ &&
    \arbto[ddr]{\ctrue\cand\ctrue}{\runit[\ctrue]}
    \arbto[rr]{}{\twist[\ctrue,\ctrue]}    
    \arpfrom[llu]{\proj[A]\cand\proj[B]}{}&&
    \arpto[ddl]{\ctrue\cand\ctrue}{\runit[\ctrue]}
    \arbfrom[rru]{\proj[B]\cand\proj[A]}{}\\ \\ &&&
    \arbfrom[llluuu]{\proj[A\cand B]}{\ctrue}
    }
  $$
  The quadrangle in naturality of $\twist$ and the commutativity of
  triangle in the center is a consequence of the coherence theorem for
  monoidal categories. The two slim triangles are just \eqref{eq:B2b}.
  The cases for $\assoc[A,B,C]$, $\runit[A]$, $\lunit[A]$ are similar.
  For $\proj[A]$, it follows directly from \eqref{eq:B2a} and for
  $\projl{B}{A}$ and $\projr{A}{B}$ from Proposition
  \ref{prop:B2b}~\eqref{l:B2b2} and from~\eqref{eq:projlr}.  
\end{proof}

\begin{proposition}\label{prop:B2c}
  If a \Bi-category obeys \eqref{eq:B2a} and the equation
  \begin{equation}\label{eq:B2c}
    \vcdiagtriangleup{{A\cand B}{\diag[A]\cand\diag[B]}}
    {{A\cand A\cand B\cand B}{A\cand\twist[A,B]\cand B}}
    {{\diag[A\cand B]}{A\cand B\cand A\cand B}}
    \tag{{\sf B2c}} 
  \end{equation}
  then 
  \begin{enumerate}[\rm(i)]
  \item\label{l:B2c1} also the equation \eqref{eq:B2b} holds,
  \item\label{l:B2c2} for every $A$, the map $\diag[A]$ is a
    $\cand$-comonoid morphism, and
  \item\label{l:B2c3} the maps that preserve the
    $\cand$-comultiplication are closed under $\cand$.
  \end{enumerate}
\end{proposition}

\begin{proof}
  \eqref{l:B2c1} For showing that \eqref{eq:B2b} holds, consider the
  diagram 
    $$
    \qlapm{\vcdiagram{
        \arbtoid[rrr]{A\cand B}{} 
        &&& 
        \arbto[dd]{A\cand B}{\runit[A\cand B]^{-1}}
        \\ &
        \arbto[r]{A\cand A\cand B\cand B}{\;\;\;
          {A\cand\proj[A]\cand B\cand\proj[B]}}
        \arbfrom[ul]{\diag[A]\cand\diag[B]}{}
        \arbto[dl]{}{\labupright{5}{3}
          {A\cand \twist[A,B]\cand B}}
        &
        \arbto[d]{A\cand\ctrue\cand B\cand\ctrue}
        {A\cand\twist[\ctrue,B]\cand\ctrue} 
        \arbfrom[ur]{\labdownleft{5}{5}
          {\runit[A]^{-1}\cand\runit[B]^{-1}}}{} 
        \\
        \arbfrom[uu]{\diag[A\cand B]}{A\cand B\cand A\cand B} 
        &&
        \arbfrom[ll]{A\cand B\cand\proj[A]\cand\proj[B]}
        {A\cand B\cand\ctrue\cand\ctrue} 
        &
        \arbfrom[l]{A\cand B\cand\runit[\ctrue]}
        {A\cand B\cand\ctrue} 
        }}
    $$
  The triangle on the left is \eqref{eq:B2c}, the upper quadrangle is
  the comonoid equation, the lower quadrangle is naturality of
  $\twist$ and the quadrangle on the right commutes because of the
  coherence in monoidal categories. The outer square says that
  $\runit[\ctrue]\fcomp(\proj[A]\cand\proj[B])$ is $\cand$-counit for
  $\diag[A\cand B]$. By Proposition~\ref{prop:unique-unit} (uniqueness
  of units) it must therefore be equal to $\proj[A\cand B]$. 
  \eqref{l:B2c2} That
  $\diag[A]$ preserves the $\cand$-comultiplication follows from
  \begin{equation}\label{eq:diag-diag}
    \vcddiagsquare{{A}{\diag[A]}}{{A\cand A}{\diag[A\cand A]}}
    {&\arbto[dr]{A\cand A\cand A\cand A}{A\cand \twist[A,A]\cand A}
      \arbfrom[ur]{\diag[A]\cand\diag[A]}{}}
    {{\diag[A]}{A\cand A}}{{\diag[A]\cand\diag[A]}{A\cand A\cand A\cand A}}
  \end{equation}
  where the pentagon commutes because of the coassociativity and
  cocommutativity of $\diag[A]\colon A\to A\cand A$.  For showing that
  $\diag[A]$ preserves the $\cand$-counit, consider the diagram
  \begin{equation}
    \diagram{
      \arbto[rr]{A}{\diag[A]}&&
      \arbto[dd]{A\cand A}{\proj[A\cand A]}\\
      &\arpfrom[ur]{\proj[A]\cand\proj[A]}{\ctrue\cand\ctrue}
      \arpfrom[dl]{\diag[\ctrue]}{}
      \arbto[dr]{}{\runit[\ctrue]}\\
      \arbfrom[uu]{\proj[A]}{\ctrue} &&
      \arpfromid[ll]{}{\ctrue}
    }
    \tag*{}
  \end{equation}
  The big and the lower triangle commute by Proposition~\ref{prop:B2a}, 
  and the left triangle is \eqref{eq:B2b} which has been shown before.
  For \eqref{l:B2c3} chase 
    $$
      \qlapm{\vcdiagramhc{
        \arbto[rrrr]{A\cand B}{f\cand g} &&&& 
        \arbto[dd]{C\cand D}{\diag[C\cand D]}\\ &
        \arbto[rr]{A\cand A\cand B\cand B}{f\cand f\cand g\cand g}
        \arbfrom[ul]{\diag[A]\cand\diag[B]}{}
        \arbto[dl]{}{\labupright{5}{4}
          {A\cand \twist[A,B]\cand B}} 
        &&
        \arpto[dr]{C\cand C\cand D\cand D}{\labupleft{5}{4}
          {C\cand \twist[C,D]\cand D} }
        \arpfrom[ur]{\diag[C]\cand\diag[D]}{} \\
        \arbfrom[uu]{\diag[A\cand B]}{A\cand B\cand A\cand B} &&&&
        \arbfrom[llll]{f\cand g\cand f\cand g}{C\cand D\cand C\cand D}
        }}
    $$
  where $f\colon A\to B$ and $g\colon C\to D$ are maps preserving the
  $\cand$-comultiplication.
\end{proof}

\begin{proposition}\label{prop:assoc-comult}
  In a \Bi-category obeying \eqref{eq:B2a} and \eqref{eq:B2c} the maps
  $\assoc[A,B,C]$, $\twist[A,B]$, $\runit[A]$, $\lunit[A]$,
  $\proj[A]$, $\projl{A}{B}$, and $\projr{A}{B}$ all preserve the
  $\cand$-comultiplication. Dually,  the maps
  $\coassoc[A,B,C]$, $\cotwist[A,B]$, $\corunit[A]$, $\colunit[A]$,
  $\coproj[A]$, $\coprojl{A}{B}$, and $\coprojr{A}{B}$ all preserve the
  $\cor$-multiplication. 
\end{proposition}

\begin{proof}
  Again, we show the case only for $\twist$:
  $$
  \qlapm{\diagramhc{
    \arbto[rrrr]{A\cand B}{\twist[A,B]} &&&& 
    \arbto[dd]{B\cand A}{\diag[B\cand A]}\\ &
    \arbto[rr]{A\cand A\cand B\cand B}{\twist[A\cand A,B\cand B]}
    \arbfrom[ul]{\diag[A]\cand\diag[B]}{}
    \arbto[dl]{}{\labupright{5}{4}
      {A\cand \twist[A,B]\cand B}}
      &&
    \arpto[dr]{B\cand B\cand A\cand A}{\labupleft{5}{4}
      {B\cand \twist[B,A]\cand A}} 
    \arpfrom[ur]{\diag[B]\cand\diag[A]}{} \\
    \arbfrom[uu]{\diag[A\cand B]}{A\cand B\cand A\cand B} &&&&
    \arbfrom[llll]{\twist[A,B]\cand\twist[A,B]}{B\cand A\cand B\cand A}
    }}
  $$
  The two triangles are \eqref{eq:B2c}, the upper square is
  naturality of $\twist$ and the lower square commutes because of
  coherence in monoidal categories. For $\assoc$, $\runit$, and
  $\lunit$ the situation is similar. For $\proj[A]$ it has been shown
  already in Proposition~\ref{prop:B2a}, and for $\projl{A}{B}$, and
  $\projr{A}{B}$ it follows from Proposition~\ref{prop:B2c}.  
\end{proof}

Propositions~\ref{prop:B2a}--\ref{prop:B2c} give
rise to the following definition:
 
\begin{definition}\label{def:B2}
  A \dfn{\Bii-category} is a \Bi-category which obeys equations
  \eqref{eq:B2a} and \eqref{eq:B2c} for all objects $A$ and $B$.
\end{definition}

The following theorem summarizes the properties of \Bii-categories.

\begin{theorem}\label{thm:B2}
  In a \Bii-category, the maps $\assoc[A,B,C]$, $\twist[A,B]$,
  $\runit[A]$, $\lunit[A]$, $\diag[A]$, $\proj[A]$, $\projl{A}{B}$,
  and $\projr{A}{B}$, all are $\cand$-comonoid morphisms, and the
  $\cand$-comonoid morphisms are closed under $\cand$.
  Dually, the maps $\coassoc[A,B,C]$, $\cotwist[A,B]$, $\corunit[A]$,
  $\colunit[A]$, $\codiag[A]$, $\coproj[A]$, $\coprojl{A}{B}$, and
  $\coprojr{A}{B}$, all are $\cor$-monoid morphisms, and the
  $\cor$-monoid morphisms are closed under $\cor$.
\end{theorem}

\begin{proof}
  Propositions \ref{prop:B2a}, \ref{prop:B2b},
  \ref{prop:assoc-counit}, \ref{prop:B2c},  and \ref{prop:assoc-comult}.
\end{proof}

\begin{proposition}\label{prop:prod}
  Let $f\colon A\to C$ and $g\colon A\to D$ and $h\colon B\to C$ and
  $a\colon A'\to A$ and $b\colon B'\to B$ and $c\colon C\to C'$ and
  $d\colon D\to D'$ be maps for some objects $A$, $B$, $C$, $D$, $A'$,
  $B'$, $C'$, $D'$ in
  a \Bii-category. Diagrammatically:
  $$
  \diagramhc{
    \arbto[r]{A'}{a}&
    \arbto[r]{A}{f}&
    \noar{C}&
    \arpfrom[l]{c}{C'}\\
    \arbto[r]{B'}{b}&
    \arbto[ur]{B}{g\;\;\;}&
    \arbfrom[ul]{h\;\;\;}{D}&
    \arpfrom[l]{d}{D'}
    }
  $$
  Then we have:
  \begin{enumerate}[\rm (i)]
  \item\label{l:a} $(c\cand d)\fcomp\prod{f,g}=\prod{c\fcomp f,d\fcomp g}$.
  \item\label{l:b} If $a$ preserves the comultiplication, then
    $\prod{f,g}\fcomp a=\prod{f\fcomp a,g\fcomp a}$.
  \item\label{l:c} If $g$ preserves the counit, then
    $\projr{C}{D}\fcomp\prod{f,g}=f$.\\ 
    If $f$ preserves the counit, then
    $\projl{C}{D}\fcomp\prod{f,g}=g$.
  \item\label{l:d} $\prod{\projl{C}{D},\projr{C}{D}}=\idf[C\cand D]$.
  \end{enumerate}
  Dually, we also have:
  \begin{enumerate}[\rm (i)]
  \item\label{l:aa} $\coprod{f,h}\fcomp(a\cor b)=\coprod{f\fcomp a,h\fcomp
  b}$.
  \item\label{l:bb} If $c$ preserves the multiplication, then
    $c\fcomp\coprod{f,h}=\coprod{c\fcomp f,c\fcomp h}$.
  \item\label{l:cc} If
    $h$ preserves the unit, then $\coprod{f,h}\fcomp\coprojl{A}{B}=f$.\\
    If $f$ preserves the unit, then $\coprod{f,h}\fcomp\coprojr{A}{B}=h$.
  \item\label{l:dd} $\coprod{\coprojl{A}{B},\coprojr{A}{B}}=\idf[A\cor
    B]$.
  \end{enumerate}
\end{proposition}

\begin{proof}
  Straightforward calculation. Note that \eqref{l:a}--\eqref{l:c}
  hold already in a \Bi-category, only for \eqref{l:d} is the equation
  \eqref{eq:B2c} needed.   
\end{proof}

As observed before, if a \Bi-category is single-mixed then $\Hom(A,B)$
carries a semigroup structure. If we additionally have the structure
of a \Bii-category, then the bijection~\eqref{eq:star} of
Definition~\ref{def:staraut} preserves this semigroup structure:

\begin{proposition}\label{prop:semigroup-iso}
  In a single-mixed \Bii-category the bijection \eqref{eq:star} is a
  semigroup isomorphism.
\end{proposition}

\begin{proof}
  Let $f,g\colon A\cand B\to C$ be two maps for some objects $A$, $B$,
  and $C$, and let $f',g'\colon A\to \cneg B\cor C$ be their
  transposes. We have to show that $f'+g'$ is the transpose of $f+g$.
  First note, that in any *-autonomous category the map
  $$
  \vcdiagram{
    \arbto[rr]{A\cand A\cand B\cand B}{A\cand\twist[A,B]\cand B}
    &&
    \arbto[r]{A\cand B\cand A\cand B}{f\cand g}
    &
    \noar{C\cand C}
    }
  $$
  is a transpose of
  $$
  \vcdiagram{
    \arbto[r]{A\cand A}{f'\cand g'}
    &
    \arbto[r]{(\cneg B\cor C)\cand(\cneg B\cor C)}{\tens}
    &
    \noar{\cneg B\cor\cneg B\cor(C\cand C)}
    }
  $$
  where $\tens$ is the canonical map obtained from two switches,
  cf.~\eqref{eq:tens}.   Now, by definition, $f+g$ is the map
  $$
  \vcdiagram{
    \arbto[r]{A\cand B}{\diag[A\cand B]}
    \arpto[dr]{}{\diag[A]\cand\diag[B]}
    &
    \arbto[r]{A\cand B\cand A\cand B}{f\cand g}
    &
    \arbto[r]{C\cand C}{\mix[C,C]}
    &
    \arbto[r]{C\cor C}{\codiag[C]}
    &
    \noar{C}
    \\&
    \arpto[u]{A\cand A\cand B\cand B}{A\cand\twist[A,B]\cand B}
    }
  $$
  By \eqref{eq:transpose-neg} and what has been said above, the
  transpose of the lower path is the outermost path of the following:
  $$
  \vcdiagram{
    \arbto[r]{A}{\diag[A]}
    &
    \arbto[r]{A\cand A}{f'\cand g'}
    &
    \arbto[r]{(\cneg B\cor C)\cand(\cneg B\cor C)}{\tens}
    \arpto[d]{}{\mix[\cneg B\cor C,\cneg B\cor C]}
    &
    \arbto[d]{\cneg B\cor\cneg B\cor(C\cand C)}
    {\cneg B\cor\cneg B\cor\mix[C,C]}
    \\&&
    \arbto[r]{\cneg B\cor C\cor\cneg B\cor C}
    {\cneg B\cor\twist[C,\cneg B]\cor C}
    &
    \arbto[d]{\cneg B\cor\cneg B\cor C\cor C}{\codiag[\cneg B]\cor\codiag[C]}
    \\&&&
    \arbfrom[ul]{\codiag[\cneg B\cor C]}{\cneg B\cor C}
    }
  $$
  The innermost path is by definition $f'+g'$. The square commutes
  because of \eqref{eq:mix-twist} and \eqref{eq:mix-assoc}, and the
  triangle is the dual of \eqref{eq:B2c}.
\end{proof}

\section{Order enrichment}

In \cite{fuhrmann:pym:oecm}, F\"uhrman and Pym equipped
\Bii-categories with an order enrichment, such that the proof
identifications induced by the axioms are exactly the same as the
proof identifications made by Gentzen's sequent calculus
\LK~\cite{gentzen:34}, modulo ``trivial rule permutations'' (see
\cite{lafont:95,robinson:03}), and such that $f\fple g$ if $g$ is
obtained from $f$ via cut elimination (which is not confluent in \LK).

\begin{definition}\label{def:LK}
  A \Bii-category is called an \dfn{\LK-category} if for every $A$, $B$, the
  set $\Hom(A,B)$ is equipped with a partial order structure $\fple$ such that
  \begin{enumerate}[\rm(i)]
  \item\label{l:LK-monoton} the arrow composition $\fcomp$, as well as the
    bifunctors $\cand$ and $\cor$ are monotonic in both arguments,
  \item for every map $f\colon A\to B$ we have 
    \begin{align*}
      \label{eq:LK1}
      \proj[B]\fcomp f&\, \fple\;\proj[A] \tag{\sf LK-$\proj$}\\[1ex]
      \label{eq:LK2}
      \diag[B]\fcomp f&\,\fple\;(f\cand f)\fcomp\diag[A]\tag{\sf LK-$\diag$}
    \end{align*}
  \item\label{l:LK-iso} and the bijection \eqref{eq:star} of
    Definition~\ref{def:staraut} is an order isomorphism for~$\fple$.
  \end{enumerate}
\end{definition}

Although in \cite{fuhrmann:pym:oecm,fuhrmann:pym:04} F\"uhrmann and
Pym use the term ``classical category'', We use here the term
\LK-categories because---as worked out in detail in
\cite{fuhrmann:pym:oecm}---they provide a category-theoretic
axiomatisation of sequent calculus proofs in Gentzen's system
\LK~\cite{gentzen:34}. However, it should be clear that \LK-categories
are only one particular example of a wide range of possible
category-theoretic axiomatisations of proofs in classical logic.

\begin{remark}\label{rem:LK}
  Note that in \cite{fuhrmann:pym:oecm} F\"uhrmann and Pym give a
  different definition for \LK-categories. Since they start from a
  weakly distributive category \cite{cockett:seely:97} instead of a
  *-autonomous one, they do not have immediate access to
  transposition. For this reason, they have to give a larger set of
  inequalities, defining the order~$\fple$: {\small
    \begin{equation}
      \label{eq:FP}
      \begin{array}{r@{\;\fple\;}l@{\qqquad}r@{\;\fple\;}l}
        \diag[B]\fcomp f&(f\cand f)\fcomp\diag[A]&
        f\fcomp\codiag[A]&\codiag[B]\fcomp(f\cor f)\\[\interdisplayskip]
        \proj[B]\fcomp f&\proj[A]&
        f\fcomp\coproj[A]&\coproj[B]\\[\interdisplayskip]
        A\cor\diag[B]&
        (\codiag[A]\cor(B\cand B))\fcomp\tens\fcomp\diag[A\cor B]&
        A\cand\codiag[B]&
        \diag[A\cand B]\fcomp\cotens\fcomp(\diag[A]\cand(B\cor B))
        \\[\interdisplayskip]
        A\cor\proj[B]&
        (\coproj[A]\cor\ctrue)\fcomp\colunit[\ctrue]^{-1}\fcomp\proj[A\cor B]&
        A\cand\coproj[B]&
        \coproj[A\cand B]\fcomp\lunit[\cfalse]\fcomp(\proj[A]\cand\cfalse)
      \end{array}
      \tag{\sf FP}
    \end{equation}}%
  where $f\colon A\to B$ is an arbitrary map and $\tens\colon(A\cor
  B)\cand(A\cor B)\to A\cor A\cor (B\cand B)$ and $\cotens\colon
  A\cand A\cand(B\cor B)\to (A\cand B)\cor(A\cand B)$ are the tensor and
  cotensor map,
  cf.~\eqref{eq:tens} and~\eqref{eq:cotens}.  One can now easily show
  that both definitions are equivalent: Clearly the inequations on the
  right in \eqref{eq:FP} are just transposes of the ones on the left.
  The two top ones on the left are just \eqref{eq:LK1}
  and~\eqref{eq:LK2}, and the two bottom ones follow as follows.  If
  we transpose $A\cor B\longsolto[{A\cor\diag[B]}]A\cor(B\cand B)$ we
  get the map
  $$
  \noar{\cneg A\cand(A\cor B)}
  \longsolto[\eval]
  \noar{B}
  \longsolto[{\diag[B]}]
  \noar{B\cand B}
  $$
  By \eqref{eq:LK2}, this is smaller or
  equal to
  $$
  \noar{\cneg A\cand(A\cor B)}
  \longsolto[{\diag[\cneg A\cand(A\cor B)]}]
  \noar{\cneg A\cand(A\cor B)\cand\cneg A\cand(A\cor B)}
  \longsolto[\eval\cand\eval]
  \noar{B\cand B}
  $$
  By \eqref{eq:B2c} this is the same map as
  $$
  \noar{\cneg A\cand(A\cor B)}
  \longsolto[{\diag[\cneg A]\cand\diag[A\cor B]}]
  \noar{\cneg A\cand\cneg A\cand(A\cor B)\cand(A\cor B)}
  \solto[\isom]
  \noar{\cneg A\cand(A\cor B)\cand\cneg A\cand(A\cor B)}
  \longsolto[\eval\cand\eval]
  \noar{B\cand B}
  $$
  Transposing back yields
  $$
  \noar{A\cor B}
  \longsolto[{\diag[A\cor B]}]
  \noar{(A\cor B)\cand (A\cor B)}
  \solto[\tens] 
  \noar{A\cor A\cor (B\cand B)}
  \longsolto[{\codiag[A]}]
  \noar{A\cor(B\cand B)}
  $$
  This shows the third inequation on the left in
  \eqref{eq:FP}. For the last one, we proceed similarly: Transposing
  $A\cor B\longsolto[{A\cor\proj[B]}]A\cor\ctrue$ yields
  $$
  \noar{\cneg A\cand(A\cor B)}
  \longsolto[\eval]
  \noar{B}
  \longsolto[{\proj[B]}]
  \noar{\ctrue}
  $$
  which is by
  \eqref{eq:LK1} smaller or equal to
  $$
  \noar{\cneg A\cand(A\cor B)}
  \longsolto[{\proj[\cneg A\cand(A\cor B)]}]
  \noar{\ctrue}
  $$
  which is by \eqref{eq:B2b} and~\eqref{eq:projlr} the same as
  $$
  \noar{\cneg A\cand(A\cor B)}
  \longsolto[{\projr{\cneg A}{A\cor B}}] 
  \noar{A\cor B}
  \longsolto[{\proj[A\cor B]}]
  \noar{\ctrue}
  $$
  If transpose back, we get
  $$
  \noar{A\cor B}
  \longsolto[{\proj[A\cor B]}]
  \noar{\ctrue}
  \longsolto[\coprojr{A}{\ctrue}]
  \noar{A\cor\ctrue}
  $$
  as desired. We do not show here the other direction because it is
  rather tedious: It is almost literally the same as the proof for
  showing that any weakly distributive category with negation is a
  *-autonomous category (see \cite{cockett:seely:97,BCST}).
\end{remark}

The following theorem states the main properties of \LK-categories. It
has first been observed and proved by F\"uhrmann and Pym in
\cite{fuhrmann:pym:dj}.

\begin{theorem}\label{thm:LK}
  Every \LK-category is single-mixed and idempotent. Furthermore, for
  all maps $f,g\colon A\to B$, we have $f\le g$ iff $g\fple f$. 
\end{theorem}

\begin{proof}
  Because of \eqref{eq:B2a} and~\eqref{eq:LK1} we have that
  $\coproj[\ctrue]=\idf[\ctrue]\fcomp\coproj[\ctrue]=
  \proj[\ctrue]\fcomp\coproj[\ctrue]\fple\proj[\cfalse]$.  By duality,
  we also get $\proj[\cfalse]\fple\coproj[\ctrue]$.  Therefore
  $\proj[\cfalse]=\coproj[\ctrue]$, i.e., the category is
  single-mixed.  Next, we show that $f+g\fple f$ for all maps
  $f,g\colon A\to B$.  For this, note that
  $$
  \noar{A\cand B}
  \longsolto[{\mix[A,B]}]
  \noar{A\cor B}
  \quad\fple\quad
  \noar{A\cand B}
  \longsolto[\projr{A}{B}]
  \noar{B}
  \longsolto[\coprojr{A}{B}]
  \noar{A\cor B}
  $$
  because these are the transposes of
  $$
  \noar{A\cand\cneg A}
  \longsolto[{\conid[A]}]
  \noar{\cfalse}
  \longsolto[{\proj[\cfalse]}]
  \noar{\ctrue}
  \longsolto[{\nid[B]}]
  \noar{B\cor\cneg B}
  \quad\fple\quad 
  \noar{A\cand\cneg A}
  \longsolto[{\proj[A\cand\cneg A]}]
  \noar{\ctrue}
  \longsolto[{\nid[B]}]
  \noar{B\cor\cneg B}
  \qquad\qquad\qquad
  $$
  Now we can proceed as follows:
  \begin{eqnarray*}
    f+g&=&\codiag[B]\fcomp(f\cor g)\fcomp\mix[A,A]\fcomp\diag[A]\\
    &\fple&\codiag[B]\fcomp(f\cor g)\fcomp\coprojr{A}{A}\fcomp
    \projr{A}{A}\fcomp\diag[A]\\
    &=&\codiag[B]\fcomp(f\cor g)\fcomp\coprojr{A}{A}
    \fcomp\idf[A]\\
    &=&\codiag[B]\fcomp(f\cor B)\fcomp(A\cor g)\fcomp(A\cor\coproj[A])
    \fcomp\corunit[A]\\
    &=&\codiag[B]\fcomp(f\cor B)\fcomp(A\cor g\fcomp\coproj[A])
    \fcomp\corunit[A]\\
    &\fple&\codiag[B]\fcomp(f\cor B)\fcomp(A\cor\coproj[B])
    \fcomp\corunit[A]\\
    &=&\codiag[B]\fcomp\coprojr{B}{B}\fcomp f\\
    &=&f
  \end{eqnarray*}
  Similarly, we get $f+g\fple g$.  Now we show that $f\fple f+f$ for
  $f\colon A\to B$. Let $\name{f}\colon A\cand\cneg B\to\cfalse$ be
  the transpose of $f$. 
  Then we have
  \begin{eqnarray*}
    \name{f}&=&\idf[\cfalse]\fcomp\name{f}\\
    &=&(\idf[\cfalse]+\idf[\cfalse])\fcomp\name{f}\\
    &=&\codiag[\cfalse]\fcomp\mix[\cfalse,\cfalse]\fcomp\diag[\cfalse]
    \fcomp\name{f}\\
    &\fple&\codiag[\cfalse]\fcomp\mix[\cfalse,\cfalse]\fcomp
    (\name{f}\cand\name{f})\fcomp\diag[A\cand\cneg B]\\
    &=&\name{f}+\name{f}\\
    &=&\widename{f+f}
  \end{eqnarray*}
  The second equation is Lemma~\ref{lem:tf-idem}, the third one is the
  definition of $+$, the fourth one is~\eqref{eq:LK2}, the fifth again
  the definition of $+$, and the last equation uses
  Proposition~\ref{prop:semigroup-iso}.  By transposing back, we get
  $f\fple f+f$.  From this together with $f+f\fple f$ we get
  idempotency.  For showing that $f\le g$ iff $g\fple f$, we need to
  show that $g\fple f$ iff $f+g=g$. Since $f+g\fple f$, we have that
  $f+g=g$ implies $g\fple f$. Now suppose $g\fple f$. Then we have
  $g=g+g\fple f+g$.  This finishes the proof since $f+g\fple g$ has
  been shown already.   
\end{proof}

Note that the converse is not necessarily true. Not every single-mixed
idempotent \Bii-category is an \LK-category. Nonetheless, because of
Proposition~\ref{prop:f-plus-pi}, in every single-mixed idempotent
\Bii-category we have for every $f\colon A\to B$ that $\proj[B]\fcomp
f+\proj[A]=\proj[B]\fcomp f$, and hence $\proj[A]\le\proj[B]\fcomp f$
which is exactly~\eqref{eq:LK1}. However, the
inequality~\eqref{eq:LK2} does not follow from idempotency. One can
easily construct countermodels along the lines of \cite{str:SD05}
(see also Section~\ref{sec:pn}).

\section{The medial map and the nullary medial map}\label{sec:medial}

That \LK-categories are idempotent means that they are already at the
degenerate end of the spectrum of Boolean categories. On the other
hand, \Bii-categories have (apart from Theorem~\ref{thm:B2}) very
little structure. The question that arises now is therefore, how we
can add additional structure to \Bii-categories without getting too
much collapse. In particular, can we extend the structure such that
all the maps mentioned in Theorem~\ref{thm:B2} become $\cor$-monoid
morphisms \emph{and} $\cand$-comonoid morphisms? This is
where medial enters the scene.

\begin{definition}\label{def:medial}
  We say, a \Bii-category $\cC$ \dfn{has medial} if for all objects
  $A$, $B$, $C$, and $D$ there is a map $\medial[A,B,C,D]\colon
  (A\cand B)\cor(C\cand D)\to (A\cor C)\cand(B\cor D)$ with the
  following properties:
  \begin{itemize}
  \item it is natural in $A$, $B$, $C$ and $D$, 
  \item it is self-dual, i.e.,
    \begin{equation}\label{eq:medial-selfdual}
      \vcdiagsquare{{\widecneg{(A\cor C)\cand(B\cor D)}}
        {\widecneg{\medial[A,B,C,D]}}}
      {{\widecneg{(A\cand B)\cor(C\cand D)}}{\isom}}
      {{\isom}{(\cneg D\cand\cneg B)\cor(\cneg C\cand\cneg A)}}
      {{\medial[\cneg D,\cneg B,\cneg C,\cneg A]}
        {(\cneg D\cor\cneg C)\cand(\cneg B\cor\cneg A)}}
    \end{equation}
    commutes, where the vertical maps are the canonical isomorphisms
    induced by Definition~\ref{def:staraut},
  \item and it obeys the equation
    \begin{equation}\label{eq:medial}
      \vcdiagtriangleup{{A\cor B}{\diag[A]\cor\diag[B]\;}}
      {{(A\cand A)\cor(B\cand B)}{\medial[A,A,B,B]}}
      {{\diag[A\cor B]}{(A\cor B)\cand (A\cor B)}}
      \tag{\sf B3c} 
    \end{equation}
    for all objects $A$ and $B$.
  \end{itemize}
\end{definition}

The following equation is a consequence of~\eqref{eq:medial} and 
the self-duality of medial.
\begin{equation}\label{eq:comedial}
  \vcdiagtriangledown
  {{(A\cand B)\cor(A\cand B)}{\medial[A,B,A,B]}}
  {{(A\cor A)\cand (B\cor B)}{\;\;\codiag[A]\cand\codiag[B]}}
  {{\codiag[A\cand B]\!}{A\cand B}}
  \tag{$\mathsf{B3c'}$} 
\end{equation}

\begin{theorem}\label{thm:medial}
  Let $\cC$ be a \Bii-category that has medial. Then 
  \begin{enumerate}[\rm(i)]
  \item\label{l:closed-and-or} The maps that preserve the
    $\cand$-comultiplication are closed under $\cor$, and dually, the
    maps that preserve the $\cor$-multiplication are closed under
    $\cand$.
  \item\label{l:bcom} 
    For all maps $A\sto[f]C$, $A\sto[g]D$, $B\sto[h]C$, and
    $B\sto[k]D$, we have that
    $$
    \coprod{\prod{f,g},\prod{h,k}}=\prod{\coprod{f,h},\coprod{g,k}}\colon
    A\cor B\to C\cand D\quadfs 
    $$
  \item\label{l:med-pp} For all objects $A$, $B$, $C$, and $D$,  
    \begin{eqnarray*}
      \medial[A,B,C,D]
      &=&
      \bcoprod{
        \bprod{
          \coprojl{A}{C}\fcomp\projl{A}{B},
          \coprojl{B}{D}\fcomp\projr{A}{B}},
        \bprod{
          \coprojr{A}{C}\fcomp\projl{C}{D},
          \coprojr{B}{D}\fcomp\projr{C}{D}}}\\[1mm]
      &=&
      \bprod{
        \bcoprod{
          \coprojl{A}{C}\fcomp\projl{A}{B},
          \coprojr{A}{C}\fcomp\projl{C}{D}},
        \bcoprod{
          \coprojl{B}{D}\fcomp\projr{A}{B},
          \coprojr{B}{D}\fcomp\projr{C}{D}}}
    \end{eqnarray*}
  \item\label{l:dppd} For all objects $A$, $B$, $C$, and $D$,
    the following diagram commutes: 
    \begin{equation}\label{eq:medsquare}
      \vcdiagram{
        &
        \arbto[dr]{\qqqqqqlapm{((A\cand B)\cor(C\cand D))\cand
            ((A\cand B)\cor(C\cand D))}}
        {\labdownright{6}{12}{(\projl{A}{B}\cor\projl{C}{D})\cand
          (\projr{A}{B}\cor\projr{C}{D})}}
        \\
        \arbto[ur]{(A\cand B)\cor(C\cand D)}
        {\labdownleft{6}{7}{\diag[(A\cand B)\cor(C\cand D)]}}
        &&
        \arbfrom[dl]{\labupright{6}{7}{\codiag[(A\cor C)\cand(B\cor D)]}}
        {(A\cor C)\cand(B\cor D)}
        \\
        &
        \arbfrom[ul]{\labupleft{6}{12}{(\coprojl{A}{C}\cand\coprojl{B}{D})\cor
          (\coprojr{A}{C}\cand\coprojr{B}{D})}}
        {\qqqqqqlapm{((A\cor C)\cand(B\cor D))\cor((A\cor C)\cand(B\cor D))}}
        }
    \end{equation} 
  \item\label{l:med-dp} The horizontal diagonal of \eqref{eq:medsquare} is
    equal to $\medial[A,B,C,D]$.
  \end{enumerate}
\end{theorem}

\begin{proof}
  For \eqref{l:closed-and-or}, chase the following (compare with the
  proof of Proposition~\ref{prop:B2c}~\eqref{l:B2c3})
  {\small
    $$
    \diagrow=6ex
    \diagcol=2em
    \vcdiagramhc{
        \arbto[rrrrr]{A\cor B}{f\cor g} &&&&&
        \arbto[dd]{C\cor D}{\diag[C\cor D]}\\ &
        \arbto[rrr]{(A\cand A)\cor(B\cand B)}{(f\cand f)\cor(g\cand g)\strut}
        \arpfrom[ul]{\;\;\diag[A]\cor\diag[B]}{}
        \arbto[dl]{}{\;\;\;\;\medial[A,A,B,B]} &&&
        \arpto[dr]{(C\cand C)\cor(D\cand D)}{\medial[C,C,D,D]\;\;\;} 
        \arbfrom[ur]{\diag[C]\cor\diag[D]\;\;\;}{} \\
        \arbfrom[uu]{\diag[A\cor B]}{(A\cor B)\cand (A\cor B)} &&&&&
        \arbfrom[lllll]{(f\cor g)\cand(f\cor g)}{(C\cor D)\cand(C\cor D)}
        }
    $$}
  For \eqref{l:bcom} chase the diagram {\small
  $$
    \diagrow=8ex
    \diagcol=3em
  \diagramhr{
      \arbto[rrrr]{A\cor B}{\coprod{\prod{f,g},\prod{h,k}}}
      \arpto[rrrdd]{}{\prod{f,g}\cor\prod{h,k}\;\;}
      \arbtoid[ddd]{}{}
      &&&&
      \arbtoid[ddd]{C\cand D}{}
      \\ \\ &
      \arpto[rr]{(A\cand A)\cor(B\cand B)}{(f\cand g)\cor(h\cand k)}
      \arpto[dd]{}{\medial[A,A,B,B]}
      &&
      \arbto[dr]{(C\cand D)\cor(C\cand D)}{\codiag[C\cand D]}
      \arbto[dd]{}{\medial[C,D,C,D]}
      \\
      \arbto[ur]{A\cor B}{\diag[A]\cor\diag[B]}
      \arpto[dr]{}{\diag[A\cor B]}
      \arbtoid[ddd]{}{}
      &&&&
      \arbtoid[ddd]{C\cand D}{}
      \\ &
      \arbto[rr]{(A\cor B)\cand(A\cor B)}{(f\cor h)\cand(g\cor k)}
      &&
      \arpto[ur]{(C\cor C)\cand(D\cor D)}{\;\codiag[C]\cand\codiag[D]}
      \\ \\
      \arpto[rrrr]{A\cor B}{\prod{\coprod{f,h},\coprod{g,k}}}
      &&&&
      \arpfrom[llluu]{\;\;\coprod{f,h}\cand\coprod{g,k}}{C\cand D}
      }
  $$}%
  where the square in the center is naturality of medial, the two
  small triangles are \eqref{eq:medial} and~\eqref{eq:comedial}. The
  big triangles are just \eqref{eq:prod}. Note the importance of
  naturality of medial in the two diagrams above.  Let us now continue
  with \eqref{l:dppd} and~\eqref{l:med-dp}, which are proved by {\small
    $$
    \diagrow=8ex
    \diagcol=-2em
    \diagram{
      \arbto[rr]{(A\cand B)\cor(C\cand D)}{\diag[(A\cand B)\cor(C\cand D)]}
      \arpto[dr]{}{\diag[A\cand B]\cor\diag[C\cand D]\;}
      \arpto[dd]{}{(\diag[A]\cand\diag[B])\cor(\diag[C]\cand\diag[D])}
      &&
      \arbto[dd]
      {{((A\cand B)\cor(C\cand D))\cand((A\cand B)\cor(C\cand D))}}
      {(\projl{A}{B}\cor\projl{C}{D})\cand(\projr{A}{B}\cor\projr{C}{D})}
      \\ &
      \arpto[ur]{\qqqqqlapm{(A\cand B\cand A\cand B)\cor
          (C\cand D\cand C\cand D)}}
        {\medial} 
      \arbto[dd]{}
      {\labupleft{22}{0}{
	  (\projl{A}{B}\cand\projr{A}{B})\cor(\projl{C}{D}\cand\projr{C}{D})}}
      \\
      \arpto[ur]
      {{(A\cand A\cand B\cand B)\cor(C\cand C\cand D\cand D)}}
      {\isom} 
      \arpto[dr]{}
      {\labupleft{4}{8}
        {(\projl{A}{A}\cand\projr{B}{B})\cor(\projl{C}{C}\cand\projr{D}{D})}}
      &&
      \noar{(A\cor C)\cand(B\cor D)}
      \\ &
      \arpto[ur]{(A\cand B)\cor(C\cand D)}{\medial} 
      }
    $$}%
  and \eqref{eq:diag-proj-id} and the self-duality of medial. 
  It remains to show \eqref{l:med-pp}. For this consider
  {\small
    $$
    \diagrow=6ex
    \diagcol=3.3em
    \vcdiaghexagondowndr
      {{(A\cand B)\cor(C\cand D)}{\medial[A,B,C,D]}}
      {{(A\cor C)\cand(B\cor D)}
        {(\coprojr{A}{A}\cor\coprojl{C}{C})\cand
          (\coprojr{B}{B}\cor\coprojl{D}{D})}
        \arpfrom[ur]{\labupright{3}{11}
          {(\projl{A}{B}\cor\projl{C}{D})\cand
            (\projr{A}{B}\cor\projr{C}{D})}}{}
        \arpto[r]{}{(\coprojl{A}{C}\cor\coprojr{A}{C})\cand
          (\coprojl{B}{D}\cor\coprojr{B}{D})}}
      {{(A\cor A\cor C\cor C)\cand(B\cor B\cor D\cor D)}
        {(\codiag[A]\cor\codiag[C])\cand(\codiag[B]\cor\codiag[D])}
        \arpto[ur]{}
        {\qquad\qquad(A\cor\cotwist[A,C]\cor C)\cand
          (B\cor\cotwist[B,D]\cor D)}} 
      {{\diag[(A\cand B)\cor(C\cand D)]}
        {((A\cand B)\cor(C\cand D))\cand((A\cand B)\cor(C\cand D))}}
      {{\begin{array}{c}\scriptstyle(\coprojl{A}{C}\fcomp\projl{A}{B}\cor
          \coprojr{A}{C}\fcomp\projl{C}{D})\cand \quad\\\scriptstyle\quad
          (\coprojl{B}{D}\fcomp\projr{A}{B}\cor
          \coprojr{B}{D}\fcomp\projr{C}{D})\end{array}}
        {(A\cor C\cor A\cor C)\cand(B\cor D\cor B\cor D)}}
      {{\codiag[A\cor C]\cand\codiag[B\cor D]}
        {(A\cor C)\cand(B\cor D)}}
    $$}%
  The topmost triangle is \eqref{l:med-dp}, the two middle ones are
  trivial, and the bottommost triangle is \eqref{eq:B2c}. Note that
  the first-right-then-down path is 
  $$
  \bprod{
    \bcoprod{
      \coprojl{A}{C}\fcomp\projl{A}{B},
      \coprojr{A}{C}\fcomp\projl{C}{D}},
    \bcoprod{
      \coprojl{B}{D}\fcomp\projr{A}{B},
      \coprojr{B}{D}\fcomp\projr{C}{D}}}
  $$
  by definition, and the first-down-then-right path is $\medial[A,B,C,D]$
  because of \eqref{eq:diag-proj-id}.
  We get \eqref{l:med-pp} by self-duality of medial.
\end{proof}

\begin{remark}\label{rem:medial}
  Because of \eqref{l:med-pp} and~\eqref{l:med-dp} in
  Theorem~\ref{thm:medial}, we could obtain a \emph{weak medial map}
  by adding \eqref{l:dppd} or~\eqref{l:bcom} as axiom to a
  \Bii-category. This weak medial map would be self-dual. By also
  adding Theorem~\ref{thm:medial}~\eqref{l:closed-and-or} as axiom, we
  could even recover equations \eqref{eq:medial}
  and~\eqref{eq:comedial}, as the following diagram shows: {\small
  $$
    \diagrow=7ex
    \diagcol=2em
  \diagram{
    \arbto[r]{A\cor B}{\diag[A\cor B]}
    &
    \arpto[dr]{(A\cor B)\cand(A\cor B)}
    {\labupleft{6}{14}{(\diag[A]\cor\diag[B])\cand(\diag[A]\cor\diag[B])}}
    \arbtoid[r]{}{}
    &
    \arbfrom[d]
    {(\projl{A}{A}\cor\projl{B}{B})\cand(\projr{A}{A}\cor\projr{B}{B})}
    {(A\cor B)\cand(A\cor B)}
    \\
    \arbfrom[u]{\diag[A]\cor\diag[B]}{(A\cand A)\cor(B\cand B)}
    &&
    \arbfrom[ll]{\diag[(A\cand A)\cor(B\cand B)]}
    {((A\cand A)\cor(B\cand B))\cand((A\cand A)\cor(B\cand B))}
    }
  $$}
  where the left square says that $\diag[A]\cor\diag[B]$ preserves
  the $\cand$-comultiplication.  However, by
  doing this, we would \emph{not} get naturality of medial, which is
  crucial for algebraic as well as for proof-theoretic reasons (see also the
  introduction).
\end{remark}

\begin{definition}\label{def:nmedial}
  A \Bii-category $\cC$ \dfn{has nullary medial} if there is a
  map $\nmedial\colon\ctrue\cor\ctrue\to\ctrue$ (called the
  \dfn{nullary medial map}) such that for all objects $A$, $B$, the
  following holds:
  \begin{equation}\label{eq:nmedial}
    \vcdiagtriangleup{{A\cor B}{\proj[A]\cor\proj[B]}}
    {{\ctrue\cor\ctrue}{\nmedial}}
    {{\proj[A\cor B]}{\ctrue}}
    \tag{\sf B3b} 
  \end{equation}
\end{definition}

Clearly, if a a \Bii-category has nullary medial, then
$\nmedial=\proj[\ctrue\cor\ctrue]$. This can be seen by plugging in
$\ctrue$ for $A$ and $B$ in \eqref{eq:nmedial}.  By duality
$\coproj[\cfalse\cand\cfalse]=\conmedial\colon\cfalse\to\cfalse\cand\cfalse$
(the \dfn{nullary comedial map}) obeys the dual of \eqref{eq:nmedial}.

\begin{proposition}\label{prop:nmedial}
  In a \Bii-category $\cC$ that has nullary medial, we have that 
  \begin{enumerate}[\rm(i)]
  \item The maps that preserve the $\cand$-counit are closed under
    $\cor$, and dually, the maps that preserve the $\cor$-unit are
    closed under $\cand$.
  \item For all objects $A,B,C$, the map $\switch[A,B,C]$ is a
    quasientropy.
  \end{enumerate}
\end{proposition}

\begin{proof}
  For showing the first statement, replace in \eqref{eq:proj-and}
  every $\cand$ by an $\cor$, and $\runit[\ctrue]$ by $\nmedial$.  The
  second statement is shown by
  $$
  \diagramhc{
    \arbto[rr]{(A\cor B)\cand C}{\switch[A,B,C]}
    \arpto[d]{}{(\proj[A]\cor\proj[B])\cand\proj[C]}
    &&
    \arbto[d]{A\cor(B\cand C)}{\proj[A]\cor(\proj[B]\cand\proj[C])}
    \\
    \arbto[rr]{(\ctrue\cor\ctrue)\cand\ctrue}{\switch[\ctrue,\ctrue,\ctrue]}
    \arpto[d]{}{\nmedial\cand\ctrue}
    \arpto[dr]{}{\runit[\ctrue\cor\ctrue]}
    &&
    \arbto[dl]{\ctrue\cor(\ctrue\cand\ctrue)}{\ctrue\cor\runit[\ctrue]}
    \\
    \arpto[dr]{\ctrue\cand\ctrue}{\runit[\ctrue]}
    &
    \arbto[d]{\ctrue\cor\ctrue}{\nmedial}
    \\
    &
    \noar{\ctrue}
    }
  $$
  where the left down-path is $\proj[(A\cor B)\cand C]$ and the
  right down-path is $\proj[A\cor(B\cand C)]$ (because of
  \eqref{eq:B2b} and \eqref{eq:nmedial}). The two squares are
  naturality of $\switch$ and $\runit$, and the triangle at the center
  is just \eqref{eq:swi-unit}. Hence, switch preserves the
  $\cand$-counit, and by duality also the $\cor$-unit.   
\end{proof}

\begin{proposition}\label{prop:med-runit}
  Let $\cC$ be a \Bii-category with medial and nullary medial. Then
  $\cC$ obeys the equation
  \begin{equation}
    \label{eq:med-runit}
    \vcdiagsquare{{(A\cand\ctrue)\cor(B\cand\ctrue)}
      {\medial[A,\ctrue,B,\ctrue]}}
		 {{(A\cor B)\cand(\ctrue\cor\ctrue)}
		   {(A\cor B)\cand\nmedial}}
		 {{\runit[A]\cor\runit[B]}{A\cor B}}
		 {{\runit[A\cor B]^{-1}}{(A\cor B)\cand\ctrue}}
		 \tag{$\medial$-$\runit$}
  \end{equation}%
\end{proposition}

\begin{proof}
  Chase
  {\small
    $$
    \diagrow=5ex
    \diagcol=3.3em
    \diagramhc{
      \arbto[rrr]{(A\cand\ctrue)\cor(B\cand\ctrue)}
      {\medial[A,\ctrue,B,\ctrue]}
      \arbto[dr]{}{\diag[(A\cand\ctrue)\cor(B\cand\ctrue)]}
      &&&
      \arbto[ddd]{(A\cor B)\cand(\ctrue\cor\ctrue)}
      {(A\cor B)\cand\nmedial}
      \\ &
      \arpto[d]{((A\cand\ctrue)\cor(B\cand\ctrue))
        \cand((A\cand\ctrue)\cor(B\cand\ctrue))\qquad}
      {(\runit[A]\cor\runit[B])\cand(\runit[A]\cor\runit[B])}
      \arbto[urr]{}{\labdownleft{3}{10}
        {(\projl{A}{\ctrue}\cor\projl{B}{\ctrue})\cand
          (\projr{A}{\ctrue}\cor\projr{B}{\ctrue})\quad}}
      \\ &
      \arbfrom[dl]{\diag[A\cor B]}{(A\cor B)\cand(A\cor B)}
      \arpto[uurr]{}{\qquad(A\cor B)\cand(\proj[A]\cor\proj[B])}
      \arbto[drr]{}{\qquad(A\cor B)\cand(\proj[A\cor B])}
      \\
      \arbfrom[uuu]{\runit[A]\cor\runit[B]}{A\cor B}
      &&&
      \arbfrom[lll]{\runit[A\cor B]^{-1}}{(A\cor B)\cand\ctrue}
      }
    $$
    }%
  The upper triangle is Theorem~\ref{thm:medial}~\eqref{l:med-dp}, the
  lower triangle is the comonoid equation, the left square says that
  $\runit[A]\cor\runit[B]$ preserves the $\cand$-comultiplication
  (Theorems \ref{thm:B2}
  and~\ref{thm:medial}~\eqref{l:closed-and-or}), the triangle on the
  right is~\eqref{eq:nmedial}, and the triangle in the middle commutes
  because $\projl{A}{\ctrue}=\runit[A]$ and
  $\projr{A}{\ctrue}=\proj[A]\fcomp\runit[A]$, where the latter
  equation holds because of \eqref{eq:projlr} and naturality of
  $\runit$.   
\end{proof}

\begin{proposition}\label{prop:B3a}
  A \Bii-category with medial and nullary medial the following are equivalent:
  \begin{enumerate}[\rm(i)]
  \item We have
    \begin{equation}\label{eq:B3a}
      \proj[\ctrue\cor\ctrue]=\nmedial=\codiag[\ctrue]
      \colon\ctrue\cor\ctrue\to\ctrue
      \tag{\sf B3a}
    \end{equation}
  \item For all objects $A$, the map $\runit[A]$ preserves the
    $\cor$-multiplication.
  \end{enumerate}
\end{proposition}

\begin{proof}
  Chasing the diagram
    $$
    \diagram{
      \arbto[rrr]{(A\cand\ctrue)\cor(A\cand\ctrue)}
      {\runit[A]\cor\runit[A]}
      \arpto[dr]{}{\medial[A,\ctrue,B,\ctrue]}
      &&&
      \arbto[ddd]{A\cor A}{\codiag[A]}
      \\&
      \arbto[r]{(A\cor A)\cand(\ctrue\cor\ctrue)}{(A\cor A)\cand\nmedial}
      \arbto[d]{}{\codiag[A]\cand(\ctrue\cor\ctrue)}
      \arpto[ddl]{}{\codiag[A]\cand\codiag[\ctrue]}
      &
      \arpto[ur]{(A\cor A)\cand\ctrue}{\runit[A\cor A]}
      \arbto[d]{}{\codiag[A]\cand\ctrue}
      \\ &
      \arpto[r]{A\cand(\ctrue\cor\ctrue)}{A\cand\nmedial}
      \arbto[dl]{}{A\cand\codiag[\ctrue]}
      &
      \arbto[dr]{A\cand\ctrue}{\runit[A]}
      \\
      \arbfrom[uuu]{\codiag[A\cand\ctrue]}{A\cand\ctrue}
      &&&
      \arbfrom[lll]{\runit[A]}{A}
      }
    $$
  shows that in the presence of medial, nullary medial, and
  \eqref{eq:B3a} the map $\runit[A]$ preserves the
  $\cor$-multiplication. Note that in that diagram the uppermost
  square is \eqref{eq:med-runit} from the previous proposition. The
  lowermost square commutes because of \eqref{eq:B3a}, and the big
  left triangle is \eqref{eq:comedial}. Conversely, consider
  the diagram {\small
    $$
    \diagrow=6ex
    \diagcol=3.3em
    \diagramhc{
      \arbto[rrrr]{\ctrue\cor\ctrue}{\diag[\ctrue\cor\ctrue]}
      \arbto[dr]{}{\smash{\lower1ex\hbox{$\scriptstyle\qquad
            \diag[\ctrue]\cor\diag[\ctrue]=
            \runit[\ctrue]^{-1}\cor\runit[\ctrue]^{-1}$}}}
      \arpto[dddr]{}{\smash{\lower4ex\hbox{$\scriptstyle
            \coprojl{\ctrue}{\ctrue}\cor\coprojr{\ctrue}{\ctrue}
            \hskip-1em$}}}
      &&&&
      \arbto[dddd]{(\ctrue\cor\ctrue)\cand(\ctrue\cor\ctrue)}
      {\idf[\ctrue\cor\ctrue]\cand\codiag[\ctrue]}
      \\&
      \arpto[rr]{(\ctrue\cand\ctrue)\cor(\ctrue\cand\ctrue)}
	    {p}     
      \arbto[urrr]{}{\medial[\ctrue,\ctrue,\ctrue,\ctrue]}
      \arpto[ddrr]{}{(\coprojl{\ctrue}{\ctrue}\cand\ctrue)\cor
        (\coprojr{\ctrue}{\ctrue}\cand\ctrue)\quad\;\;}
      &&
      \arpto[ur]{((\ctrue\cor\ctrue)\cand(\ctrue\cor\ctrue))\cor
        ((\ctrue\cor\ctrue)\cand(\ctrue\cor\ctrue))}
      {\raise2ex\hbox{$\scriptstyle\qquad
          \codiag[(\ctrue\cor\ctrue)\cand(\ctrue\cor\ctrue)]$}}
      \arbto[dd]{}{(\idf[\ctrue\cor\ctrue]\cand\codiag[\ctrue])\cor
        (\idf[\ctrue\cor\ctrue]\cand\codiag[\ctrue])}
      \\ \\&
      \arpto[dl]{\ctrue\cor\ctrue\cor\ctrue\cor\ctrue}
      {\codiag[\ctrue\cor\ctrue]}
      \arpto[rr]{}{\runit[\ctrue\cor\ctrue]^{-1}\cor
        \runit[\ctrue\cor\ctrue]^{-1}}
      &&
      \arbto[dr]{((\ctrue\cor\ctrue)\cand\ctrue)\cor
        ((\ctrue\cor\ctrue)\cand\ctrue)}
      {\codiag[(\ctrue\cor\ctrue)\cand\ctrue]}
      \\
      \arbfromid[uuuu]{}{\ctrue\cor\ctrue}
      &&&&
      \arbfrom[llll]{\runit[\ctrue\cor\ctrue]^{-1}}
      {(\ctrue\cor\ctrue)\cand\ctrue}
      }
    $$
    }%
  where $p=(\coprojl{\ctrue}{\ctrue}\cor\coprojl{\ctrue}{\ctrue})\cand
  (\coprojr{\ctrue}{\ctrue}\cor\coprojr{\ctrue}{\ctrue})$.  The upper
  two triangles are \eqref{eq:medial} and
  Theorem~\ref{thm:medial}~\eqref{l:med-dp}.  The left triangle
  commutes because of Proposition~\ref{prop:prod}~\eqref{l:dd}, and
  the triangle at the center is the monoid equation. The
  triangle-shaped square is the naturality of $\runit$, and the
  rightmost square commutes because
  $\idf[\ctrue\cor\ctrue]\cand\codiag[\ctrue]$ preserves the
  $\cor$-multiplication, which follows from (the dual of)
  Proposition~\ref{prop:B2c}~\eqref{l:B2c2} and
  Theorem~\ref{thm:medial}~\eqref{l:closed-and-or}. 
  Finally, the lower square commutes because
  we assumed that $\runit[A]$ preserved the $\cor$-multiplication.
  Note that the commutativity of the outer square says that
  $\codiag[\ctrue]$ is unit for $\diag[\ctrue\cor\ctrue]$. Therefore,
  by Proposition~\ref{prop:unique-unit}, we can conclude that
  $\nmedial=\proj[\ctrue\cor\ctrue]=\codiag[\ctrue]$.
\end{proof}

\begin{definition}\label{def:B3}
  A \dfn{\Biii-category} is a \Bii-category that obeys~\eqref{eq:B3a} and has
  medial and nullary medial.
\end{definition}

\begin{corollary}\label{cor:unit-clon}
  In a \Biii-category, the maps $\runit[A]$, $\lunit[A]$,
  $\corunit[A]$, and $\colunit[A]$ are clonable for all objects $A$,
  i.e, they preserve both the $\cor$-multiplication and the
  $\cand$-comultiplication.
\end{corollary}

\begin{proof}
  Theorem~\ref{thm:B2} and Proposition~\ref{prop:B3a} suffice to show
  that $\runit[A]$ is clonable. For $\lunit[A]$ it is similar, and for
  $\corunit[A]$ and $\colunit[A]$ it follows by duality.  
\end{proof}

It has first been observed by Lamarche~\cite{lamarche:gap} that the
presence of a natural and self-dual map $\medial[A,B,C,D]\colon
(A\cand B)\cor(C\cand D)\to (A\cor C)\cand(B\cor D)$ in a *-autonomous
category induces two canonical maps
$\emap_1,\emap_2\colon\cfalse\to\ctrue$, namely
$$
\emap_1\colon\cfalse
\solto[{\colunit[\cfalse]^{-1}}]
\cfalse\cor\cfalse
\longsolto[{\runit[\cfalse]^{-1}\cor\lunit[\cfalse]^{-1}}]
(\cfalse\cand\ctrue)\cor(\ctrue\cand\cfalse)
\longsolto[{\medial[\cfalse,\ctrue,\ctrue,\cfalse]}]
(\cfalse\cor\ctrue)\cand(\ctrue\cor\cfalse)
\longsolto[{\colunit[\ctrue]\cand\corunit[\ctrue]}]
\ctrue\cand\ctrue
\solto[{\runit[\ctrue]}]
\ctrue
$$
and
$$
\emap_2\colon\cfalse
\solto[{\colunit[\cfalse]^{-1}}]
\cfalse\cor\cfalse
\longsolto[{\lunit[\cfalse]^{-1}\cor\runit[\cfalse]^{-1}}]
(\ctrue\cand\cfalse)\cor(\cfalse\cand\ctrue)
\longsolto[{\medial[\ctrue,\cfalse,\cfalse,\ctrue]}]
(\ctrue\cor\cfalse)\cand(\cfalse\cor\ctrue)
\longsolto[{\corunit[\ctrue]\cand\colunit[\ctrue]}]
\ctrue\cand\ctrue
\solto[{\runit[\ctrue]}]
\ctrue
$$
which are both self-dual (while
$\proj[\cfalse]$ and $\coproj[\ctrue]$ are dual to each other).  By
adding sufficient structure one can
enforce that $\emap_1=\emap_2$ and that this map has the properties of
Theorem~\ref{thm:mix}. In~\cite{lamarche:gap}, Lamarche shows how this
can be done without the $\cand$-comonoid and $\cor$-monoid structure
for every object by using equation~\eqref{eq:med-twist} that we will
introduce in Proposition~\ref{prop:med-twist}. 
In our case the structure of a \Bii-category is
sufficient to obtain that $\emap_1=\emap_2$.  But for letting this map
have the properties of Theorem~\ref{thm:mix}, as it is the case with
$\proj[\cfalse]$ and $\coproj[\ctrue]$, we need all the structure of a
\Biii-category. Then we have the following:

\begin{theorem}\label{thm:B3-mix}
  In a \Biii-category we have
  $\proj[\cfalse]=\emap_1=\emap_2=\coproj[\ctrue]$, i.e., every 
  \Biii-category is single-mixed.
\end{theorem}

\begin{proof}
  We will first show that $\proj[\cfalse]=\emap_1$. For this, note
  that {\small
    \begin{eqnarray*}
      &\rlapm{\!\!\!(\colunit[\ctrue]\cand\corunit[\ctrue])\fcomp
      \medial[\cfalse,\ctrue,\ctrue,\cfalse]\fcomp
      (\runit[\cfalse]^{-1}\cor\lunit[\cfalse]^{-1})}\\
      &=&\!
      (\colunit[\ctrue]\cand\corunit[\ctrue])\fcomp
      \bprod{
        \bcoprod{
          \coprojl{\cfalse}{\ctrue}\fcomp\projl{\cfalse}{\ctrue},
          \coprojr{\cfalse}{\ctrue}\fcomp\projl{\ctrue}{\cfalse}},
        \bcoprod{
          \coprojl{\ctrue}{\cfalse}\fcomp\projr{\cfalse}{\ctrue},
          \coprojr{\ctrue}{\cfalse}\fcomp\projr{\ctrue}{\cfalse}}}\fcomp
      (\runit[\cfalse]^{-1}\cor\lunit[\cfalse]^{-1})\\
      &=&\!
      \bprod{
        \bcoprod{
          \colunit[\ctrue]\fcomp\coprojl{\cfalse}{\ctrue}
          \fcomp\projl{\cfalse}{\ctrue}\fcomp\runit[\cfalse]^{-1},
          \colunit[\ctrue]\fcomp\coprojr{\cfalse}{\ctrue}
          \fcomp\projl{\ctrue}{\cfalse}\fcomp\lunit[\cfalse]^{-1}},
        \bcoprod{
          \corunit[\ctrue]\fcomp\coprojl{\ctrue}{\cfalse}
          \fcomp\projr{\cfalse}{\ctrue}\fcomp\runit[\cfalse]^{-1},
          \corunit[\ctrue]\fcomp\coprojr{\ctrue}{\cfalse}
          \fcomp\projr{\ctrue}{\cfalse}\fcomp\lunit[\cfalse]^{-1}}}\\
      &=&\!
      \bprod{
        \bcoprod{\coproj[\ctrue],\proj[\cfalse]},
        \bcoprod{\proj[\cfalse],\coproj[\ctrue]}}
    \end{eqnarray*}
    }%
  The first equation is an application of
  Theorem~\ref{thm:medial}~\eqref{l:med-pp}, the second one uses
  Proposition~\ref{prop:prod} together with the fact that
  $\runit[\cfalse]$ and $\lunit[\cfalse]$ preserve the
  $\cand$-co\-multi\-pli\-ca\-tion (Theorem~\ref{thm:B2}) and that
  these maps are closed under $\cor$
  (Theorem~\ref{thm:medial}~\eqref{l:closed-and-or}). The third
  equation is an easy calculation, involving \eqref{eq:prod} and the
  naturality of $\runit$ and~$\lunit$. Before we proceed, notice that:
  \begin{equation}
    \label{eq:proj-unit}
    \projr{\ctrue}{\ctrue}=\lunit[\ctrue]=\runit[\ctrue]=
    \projl{\ctrue}{\ctrue}\colon\ctrue\cand\ctrue\to\ctrue
    \qquand
    \coprojr{\cfalse}{\cfalse}=\lunit[\cfalse]^{-1}=\runit[\cfalse]^{-1}=
    \coprojl{\cfalse}{\cfalse}\colon\cfalse\to\cfalse\cor\cfalse
  \end{equation}
  Now we have:
    \begin{eqnarray*}
      \emap_1
      &=&\runit[\ctrue]\fcomp(\colunit[\ctrue]\cand\corunit[\ctrue])\fcomp
      \medial[\cfalse,\ctrue,\ctrue,\cfalse]\fcomp
      (\runit[\cfalse]^{-1}\cor\lunit[\cfalse]^{-1})\fcomp
      \colunit[\cfalse]^{-1}\\
      &=&
      \runit[\ctrue]\fcomp
      \bprod{
        \bcoprod{\coproj[\ctrue],\proj[\cfalse]},
        \bcoprod{\proj[\cfalse],\coproj[\ctrue]}}
      \fcomp\colunit[\cfalse]^{-1}\\
      &=&
      \runit[\ctrue]\fcomp
      \bprod{
        \bcoprod{\coproj[\ctrue],\proj[\cfalse]}\fcomp\colunit[\cfalse]^{-1},
        \bcoprod{\proj[\cfalse],\coproj[\ctrue]}\fcomp\colunit[\cfalse]^{-1}}
      \\
      &=&
      \projl{\ctrue}{\ctrue}\fcomp
      \bprod{
        \bcoprod{\coproj[\ctrue],\proj[\cfalse]}
        \fcomp\coprojr{\cfalse}{\cfalse},
        \bcoprod{\proj[\cfalse],\coproj[\ctrue]}
        \fcomp\coprojl{\cfalse}{\cfalse}} \\ 
      &=&
      \projl{\ctrue}{\ctrue}\fcomp
      \bprod{\proj[\cfalse],\proj[\cfalse]}\\
      &=&
      \proj[\cfalse]
    \end{eqnarray*}
  The first two equations are just the definition of $\emap_1$ and the
  previous calculation. The third equation uses
  Proposition~\ref{prop:prod} and the fact that
  $\colunit[\cfalse]=\corunit[\cfalse]$ preserves the
  $\cand$-comultiplication (Corollary~\ref{cor:unit-clon}). The fourth
  equation applies \eqref{eq:proj-unit}, and the last two equations are
  again Proposition~\ref{prop:prod}, together with the fact that
  $\coproj[\ctrue]$ preserves the $\cor$-unit and $\proj[\cfalse]$
  preserves the $\cand$-counit (Theorem~\ref{thm:B2}).  Similarly, we
  show that $\emap_2=\proj[\cfalse]$ and dually, we obtain
  $\emap_1=\emap_2=\coproj[\ctrue]$.  
\end{proof}

\begin{theorem}\label{thm:B3}
  In a \Biii-category, the strong maps (in fact, all types of maps
  defined in Definition~\ref{def:strong}) are closed under $\cand$ and
  $\cor$. Furthermore, the maps $\medial[A,B,C,D]$ and\/ $\nmedial$
  and\/ $\conmedial$ are strong.
\end{theorem}

\begin{proof}
  By Propositions \ref{prop:B2b} and~\ref{prop:B2c}, the
  $\cand$-comonoid morphisms are closed under $\cand$, and by
  Proposition~\ref{prop:nmedial} and Theorem~\ref{thm:medial} they are
  closed under $\cor$. Dually, the $\cor$-monoid morphisms are closed
  under $\cor$ and $\cand$, and therefore also the strong maps have
  this property.  Since by Theorem~\ref{thm:medial}~\eqref{l:med-dp}, 
  medial is
  $((\projl{A}{B}\cor\projl{C}{D})\cand
  (\projr{A}{B}\cor\projr{C}{D}))\fcomp {\diag[(A\cand B)\cor(C\cand
    D)]}$ as well as $\codiag[(A\cor C)\cand(B\cor D)]\fcomp
  ((\coprojl{A}{C}\cand\coprojl{B}{D})\cor
  (\coprojr{A}{C}\cand\coprojr{B}{D}))$, we have by
  Theorem~\ref{thm:B2} that it is a $\cand$-comonoid morphism and a
  $\cor$-monoid morphism, and therefore strong.  Since
  $\nmedial=\proj[\ctrue\cor\ctrue]=\codiag[\ctrue]$, we get again
  from Theorem~\ref{thm:B2} that it is a $\cand$-comonoid morphism and
  a $\cor$-monoid morphism. Similarly for
  $\conmedial=\coproj[\cfalse\cand\cfalse]=\diag[\cfalse]$.   
\end{proof}

\begin{proposition}\label{prop:B3-assoc}
  In a \Biii-category the maps $\coassoc[A,B,C]$, $\cotwist[A,B]$,
  $\colunit[A]$, and $\corunit[A]$ preserve the $\cand$-counit for all
  objects $A,B,C$. Dually, the maps $\assoc[A,B,C]$, $\twist[A,B]$,
  $\lunit[A]$, and $\runit[A]$ all preserve the $\cor$-unit.
\end{proposition}

\begin{proof}
  As before, the cases for $\coassoc[A,B,C]$ and $\cotwist[A,B]$ are
  similar. This time, we show the case for $\coassoc[A,B,C]$:
    $$
    \diagramhchr{
      \arbto[rr]{A\cor(B\cor C)}{\coassoc[A,B,C]}
      \arpto[dd]{}{\proj[A]\cor(\proj[B]\cor\proj[C])}
      &&
      \arbto[dd]{(A\cor B)\cor C}{(\proj[A]\cor\proj[B])\cor\proj[C]}
      \\ \\
      \arpto[rr]{\ctrue\cor(\ctrue\cor\ctrue)}
      {\coassoc[\ctrue,\ctrue,\ctrue]}
      \arpto[dd]{}{\ctrue\cor\codiag[\ctrue]}
      &&
      \arbto[dd]{(\ctrue\cor\ctrue)\cor\ctrue}{\codiag[\ctrue]\cor\ctrue}
      \\ \\
      \arpto[dr]{\ctrue\cor\ctrue}{\codiag[\ctrue]}
      &&
      \arbto[dl]{\ctrue\cor\ctrue}{\codiag[\ctrue]}
      \\&
      \noar{\ctrue}
      }
    $$
  The square is naturality of $\coassoc$, and the pentagon is
  associativity of $\codiag$. The left down path is $\proj[A\cor(B\cor
  C)]$ and the right down path is $\proj[(A\cor B)\cor C]$ (because of
  \eqref{eq:nmedial} and~\eqref{eq:B3a}). 
  For $\corunit[A]$, chase
    $$
    \diagramhr{
      \arbto[rr]{A\cor\cfalse}{\corunit[A]}
      \arbto[dr]{}{\proj[A]\cor\cfalse}
      \arpto[dddr]{}{\proj[A]\cor\proj[\cfalse]\!\!\!}
      &&
      \arbto[dddd]{A}{\proj[A]}
      \\ &
      \arpto[dd]{\ctrue\cor\cfalse}{\smash{\raise1ex\hbox{$\scriptstyle
            \ctrue\cor\coproj[\ctrue]\!\!$}}}
      \arbto[dddr]{}{\!\!\!\corunit[\ctrue]}
      \\ \\ &
      \arpto[dl]{\ctrue\cor\ctrue}{\nmedial}
      \arbto[dr]{}{\!\!\!\!\!\!\!\!\codiag[\ctrue]}
      \\
      \arbfrom[uuuu]{\proj[A\cor\cfalse]}{\ctrue}
      &&
      \arbfromid[ll]{}{\ctrue}
      }
    $$
  The upper right quadrangle is naturality of $\corunit$. The leftmost
  triangle is \eqref{eq:nmedial}. The one in the center next to it
  commutes because of functoriality of $\cor$ and
  $\proj[\cfalse]=\coproj[\ctrue]$ (Theorem~\ref{thm:B3-mix}).
  The lower right triangle the the monoid equation and the triangle at
  the bottom is~\eqref{eq:B3a}. The case for $\colunit[A]$ is similar.
\end{proof}


\begin{proposition}\label{prop:med-twist}
  In a \Biii-category the following are equivalent:
  \begin{enumerate}[\rm(i)]
  \item\label{l:med-twist} The equation 
    \begin{equation}
      \label{eq:med-twist}
      \vcdiagsquaredown{{(A\cand B)\cor(C\cand D)}{\medial[A,B,C,D]}}
      {{(A\cor C)\cand(B\cor D)}{\twist[A\cor C,B\cor D]}}
      {{\twist[A,B]\cor\twist[C,D]}{(B\cand A)\cor(D\cand C)}}
      {{\medial[B,A,D,C]}{(B\cor D)\cand(A\cor C)}}
      \tag{$\medial$-$\twist$}
    \end{equation}
    holds for all objects $A$, $B$, $C$, and $D$.
  \item\label{l:twist-mult} The map $\twist[A,B]\colon A\cand B\to
    B\cand A$ preserves the $\cor$-multiplication.
  \item\label{l:cotwist-comult} The map $\cotwist[A,B]\colon A\cor
    B\to B\cor A$ preserves the $\cand$-comultiplication.
  \item\label{l:med-cotwist} The equation 
    \begin{equation}
      \label{eq:med-cotwist}
      \vcdiagsquaredown{{(A\cand B)\cor(C\cand D)}{\medial[A,B,C,D]}}
      {{(A\cor C)\cand(B\cor D)}{\cotwist[A,C]\cand\cotwist[B,D]}}
      {{\cotwist[A\cand B,C\cand D]}{(C\cand D)\cor(A\cand B)}}
      {{\medial[C,D,A,B]}{(C\cor A)\cand(D\cor B)}}
      \tag{$\medial$-$\cotwist$}
    \end{equation}
    holds for all objects $A$, $B$, $C$, and $D$.
  \end{enumerate}
\end{proposition}

\begin{proof} 
  Suppose
  \eqref{eq:med-twist} does hold. Then we have
  $$
  \vcdiaghexagondown
  {{(A\cand B)\cor(A\cand B)}{\medial[A,B,A,B]}}
  {{(A\cor A)\cand(B\cor B)}{\codiag[A]\cand\codiag[B]}
    \arbto[r]{}{\twist[A\cor A,B\cor B]}}
  {{A\cand B}{\twist[A,B]}}
  {{\twist[A,B]\cor\twist[A,B]}{(B\cand A)\cor(B\cand A)}}
  {{\medial[B,A,B,A]}{(B\cor B)\cand(A\cor A)}}
  {{\codiag[B]\cand\codiag[A]}{B\cand A}}
  $$
  which together with \eqref{eq:comedial} says that
  $\twist[A,B]$ preserves the $\cor$-multiplication. Conversely, we
  have
  {\small
    $$
    \diagrow=6ex
    \diagcol=2.3em
  \vcdiaghexagondown
    {{(A\cand B)\cor(C\cand D)}
      {(\coprojl{A}{C}\cand\coprojl{B}{D})\cor
        (\coprojr{A}{C}\cand\coprojr{B}{D})}}
    {{((A\cor C)\cand(B\cor D))\cor((A\cor C)\cand(B\cor D))}
      {\codiag[(A\cor C)\cand(B\cor D)]}
      \arbto[r]{}{\phantom{\Big\vert}
        \twist[A\cor C,B\cor D]\cor\twist[A\cor C,B\cor D]}}
    {{(A\cor C)\cand(B\cor D)}{\twist[A\cor C,B\cor D]}}
    {{\twist[A,B]\cor\twist[C,D]}{(B\cand A)\cor(D\cand C)}}
    {{(\coprojl{B}{D}\cand\coprojl{A}{C})\cor
        (\coprojr{B}{D}\cand\coprojr{A}{C})}
      {((B\cor D)\cand(A\cor C))\cor((B\cor D)\cand(A\cor C))}}
    {{\codiag[(B\cor D)\cand(A\cor C)]}{(B\cor D)\cand(A\cor C)}}
  $$}%
  The upper square is naturality of $\twist$, and the lower square
  says that $\twist[A,B]$ preserves the $\cor$-multiplication.
  Together with Theorem~\ref{thm:medial}~\eqref{l:med-dp}, this is
  \eqref{eq:med-twist}. Hence \eqref{l:med-twist} and
  \eqref{l:twist-mult} are equivalent. The other equivalences follow
  because of duality.  
\end{proof}

\begin{proposition}\label{prop:med-assoc}
  In a \Biii-category the following are equivalent:
  \begin{enumerate}[\rm(i)]
  \item\label{l:med-assoc} The equation 
    \begin{equation}
      \label{eq:med-assoc}\hskip-3em
      \qlapm{
        \vcdiaghexagondown{{(A\cand (B\cand C))\cor(D\cand (E\cand F))}
        {\medial[A,B\cand C,D,E\cand F]}}
      {{(A\cor D)\cand((B\cand C)\cor (E\cand F))}
        {(A\cor D)\cand\medial[B,C,E,F]}}
      {{(A\cor D)\cand((B\cor E)\cand(C\cor F))}
        {\labdownleft{2}{0}{\assoc[A\cor D,B\cor E,C\cor F]}}}
      {{\labupleft{2}{0}{\assoc[A,B,C]\cor\assoc[D,E,F]}}
        {((A\cand B)\cand C)\cor((D\cand E)\cand F)}}
      {{\medial[A\cand B,C,D\cand E,F]}
        {((A\cand B)\cor(D\cand E))\cand (C\cor F)}}
      {{\medial[A,B,D,E]\cand (C\cor F)}
        {((A\cor D)\cand(B\cor E))\cand(C\cor F)}}}
      \tag{$\medial$-$\assoc$}
    \end{equation}
    holds for all objects $A$, $B$, $C$, $D$, $E$, and $F$.
  \item The map $\assoc[A,B,C]\colon A\cand(B\cand C)\to (A\cand B)\cand C$
    preserves the $\cor$-multiplication.
  \end{enumerate}
\end{proposition}

\begin{proof}
  Similar to the previous proposition. (Here the statements corresponding to
  \eqref{l:cotwist-comult} and~\eqref{l:med-cotwist} in
  Proposition~\ref{prop:med-twist} are omitted to save space, but obviously
  they hold accordingly.)  
\end{proof}

\begin{remark}
  This proposition allows us to speak of uniquely defined maps 
  $$
  \medialsq[A,B,C,D,E,F]\colon 
  (A\cand B\cand C)\cor(D\cand E\cand F)
  \to
  (A\cor D)\cand (B\cor E)\cand(C\cor F)
  $$
  and dually
  $$
  \comedialsq[A,B,C,D,E,F]\colon 
  (A\cand B)\cor (C\cand D)\cor (E\cand F)
  \to
  (A\cor C\cor E)\cand(B\cor D\cor F)
  $$
  A more sophisticated and more general notation for composed variations of
  medial is introduced by Lamarche in \cite{lamarche:gap}. 
\end{remark}

\begin{proposition}\label{prop:med-switch}
  In a a \Biii-category obeying \eqref{eq:med-twist} and
  \eqref{eq:med-assoc} the following are equivalent:
  \begin{enumerate}[\rm(i)]
  \item The equation
    \begin{equation}
      \vcdiagsquaredown{{((A\cand B)\cor(C\cand D))\cand E}
        {\medial[A,B,C,D]\cand E}}
      {{(A\cor C)\cand(B\cor D)\cand E}{(A\cor C)\cand\switch[B,D,E]}}
      {{\switch[A\cand B,C\cand D,E]}{(A\cand B)\cor(C\cand D\cand E)}}
      {{\medial[A,B,C,D\cand E]}{(A\cor C)\cand(B\cor (D\cand E))}}
      \label{eq:med-switch}
      \tag{$\medial$-$\switch$}
    \end{equation}
    holds for all objects $A$, $B$, $C$, $D$, and $E$.
  \item The map $\switch[A,B,C]\colon A\cand(B\cor C)\to (A\cand
    B)\cor C$ preserves the $\cand$-comultiplication.
  \end{enumerate}
\end{proposition}

\begin{proof}
  First note that if the equations \eqref{eq:med-twist},
  \eqref{eq:med-assoc}, and~\eqref{eq:med-switch} hold, we can compose
  them to get the commutativity of diagrams like
  \begin{equation}
    \label{eq:med-swi-swi}
    \vcdiagsquare{{((A\cand B)\cor(C\cand D))\cand E\cand F}{}}
    {{(A\cand B)\cor(C\cand E\cand D\cand F)}{\medial[A,B,C\cand E,D\cand F]}}
    {{\medial[A,B,C,D]\cand E\cand F}{(A\cor C)\cand(B\cor D)\cand E\cand F}}
    {{}{(A\cor(C\cand E))\cand(B\cor(D\cand F))}}
  \end{equation}
  where the horizontal maps are the canonical maps (composed of twist,
  associativity, and switch) that are uniquely determined by the
  *-autonomous structure. Now chase
  $$
  \vcdiagram{
    \arpto[d]{(A\cor B)\cand C}{(\diag[A]\cor\diag[B])\cand\diag[C]}
    \arbto[r]{}{\switch[A,B,C]}
    &
    \arbto[d]{A\cor (B\cand C)}{\diag[A]\cor(\diag[B]\cand\diag[C])}
    \\
    \arpto[d]{((A\cand A)\cor (B\cand B))\cand C\cand C}
    {\medial[A,A,B,B]\cand C\cand C}
    \arbto[r]{}{\switch[A\cand A,B\cand B,C\cand C]}
    \arbto[dr]{}{}
    &
    \arbto[d]{(A\cand A)\cor(B\cand B\cand C\cand C)}
    {(A\cand A)\cor(B\cand\twist[B,C]\cand C)}
    \\
    \arpto[d]{(A\cor B)\cand(A\cor B)\cand C\cand C}
    {(A\cor B)\cand\twist[A\cor B,C]\cand C}
    \arbto[dr]{}{}
    &
    \arbto[d]{(A\cand A)\cor(B\cand C\cand B\cand C)}
    {\medial[A,A,B\cand C,B\cand C]}
    \\
    \arpto[r]{(A\cor B)\cand C\cand(A\cor B)\cand C}
    {\switch[A,B,C]\cand\switch[A,B,C]}
    &
    \noar{(A\cor(B\cand C))\cand(A\cor(B\cand C))}
    }
  $$
  where the parallelogram is just \eqref{eq:med-swi-swi}, the upper
  square is naturality of switch and the two triangles are laws of
  *-autonomous categories. Note that, by \eqref{eq:B2c}
  and~\eqref{eq:medial}, the vertical paths are just $\diag[(A\cor
  B)\cand C]$ and $\diag[A\cor(B\cand C)]$. Therefore switch preserves
  the $\cand$-comultiplication. Conversely, consider the diagram
  {\small
    \def\cor{\mathord\vee}
    \def\cand{\mathord\wedge}
  $$
    \diagrow=6ex
    \diagcol=3em
  \vcdiagram{
    \arpto[d]{((A\cand B)\cor (C\cand D))\cand E}
    {\diag[((A\cand B)\cor (C\cand D))\cand E]}
    \arbto[r]{}{\switch[A\cand B,C\cand D,E]}
    &
    \arbto[d]{(A\cand B)\cor(C\cand D\cand E)}
    {\diag[(A\cand B)\cor(C\cand D\cand E)]}
    \\
    \arpto[d]{((A\cand B)\cor (C\cand D))\cand E\cand
      ((A\cand B)\cor (C\cand D))\cand E}
    {p}
    \arbto[r]{}{\switch\cand\switch}
    &
    \arbto[d]{((A\cand B)\cor(C\cand D\cand E))\cand
      ((A\cand B)\cor(C\cand D\cand E))}
    {q}
    \\
    \arpto[d]{((A\cand\ctrue)\cor (C\cand\ctrue))\cand\ctrue\cand
      ((\ctrue\cand B)\cor (\ctrue\cand D))\cand E}
    {\isom}
    \arbto[r]{}{\switch\cand\switch}
    &
    \arbto[d]{((A\cand\ctrue)\cor(C\cand\ctrue\cand\ctrue))\cand
      ((\ctrue\cand B)\cor(\ctrue\cand D\cand E))}
    {\isom}
    \\
    \arpto[r]{(A\cor C)\cand(B\cor C)\cand E}
    {(A\cor C)\cand\switch[B,D,E]}
    &
    \noar{(A\cor C)\cand(B\cor(D\cand E))}
    }
  $$}%
  where 
  \begin{eqnarray*}
    p&=&((A\cand\proj[B])\cor (C\cand\proj[D]))\cand\proj[E]\cand
      ((\proj[A]\cand B)\cor (\proj[C]\cand D))\cand E\\[1ex]
    q&=&((A\cand\proj[B])\cor (C\cand\proj[D]\cand\proj[E]))\cand
      ((\proj[A]\cand B)\cor (\proj[C]\cand D\cand E))  
  \end{eqnarray*}
  Note that the left vertical map is $\medial[A,B,C,D]\cand\idf[E]$
  while the right vertical map is $\medial[A,B,C,D\cand E]$. The upper
  square commutes because we assumed that switch preserves the
  $\cand$-co\-mul\-ti\-pli\-ca\-tion, the middle one is naturality of
  switch, and the lower one commutes because the category is
  *-autonomous (the isomorphisms are just compositions of $\runit$ and
  $\lunit$).  
\end{proof}

\begin{definition}\label{def:B4}
  A \dfn{\Biv-category} is a \Biii-category that obeys the equations
  \eqref{eq:med-twist}, \eqref{eq:med-assoc}, and~\eqref{eq:med-switch}.
\end{definition}

\begin{remark}
  Equivalently, one can define a \Biv-category as a \Biii-category in which
  $\twist$, $\assoc$, and $\switch$ are strong. We have chosen the form of
  Definition~\eqref{def:B4} to show the resemblance to the work
  \cite{lamarche:gap} where the equations \eqref{eq:med-twist},
  \eqref{eq:med-assoc}, and~\eqref{eq:med-switch} are also considered as
  primitives.
\end{remark}

\begin{theorem}\label{thm:B4}
  In a \Biv-category, the maps $\assoc[A,B,C]$, $\twist[A,B]$,
  $\runit[A]$, $\lunit[A]$ and $\coassoc[A,B,C]$, $\cotwist[A,B]$,
  $\corunit[A]$, $\colunit[A]$, as well as $\switch[A,B,C]$ and
  $\mix[A,B]$ are all strong.
\end{theorem}

\begin{proof}
  That $\assoc[A,B,C]$, $\twist[A,B]$, $\runit[A]$, $\lunit[A]$ and
  $\coassoc[A,B,C]$, $\cotwist[A,B]$, $\corunit[A]$, $\colunit[A]$ are
  quasientropies follows from Theorem~\ref{thm:B2} and
  Proposition~\ref{prop:B3-assoc}. That $\runit[A]$, $\lunit[A]$ and
  $\corunit[A]$, $\colunit[A]$ are clonable has been said already in
  Corollary~\ref{cor:unit-clon}. For $\assoc[A,B,C]$, $\twist[A,B]$ and
  $\coassoc[A,B,C]$, $\cotwist[A,B]$ this follows from
  Theorem~\ref{thm:B2} and Propositions \ref{prop:med-twist}
  and~\ref{prop:med-assoc} (and by duality). Hence, all these maps are
  strong. That $\switch[A,B,C]$ is strong follows from
  Proposition~\ref{prop:nmedial} and Proposition~\ref{prop:med-switch} (and
  self-duality of switch).
  For showing that $\mix[A,B]$ is also strong it suffices to observe that
  mix is a composition of strong maps via $\fcomp$, $\cand$, and
  $\cor$. See \eqref{eq:sbot-mix}, Theorem~\ref{thm:B3-mix}, and
  Theorem~\ref{thm:B3}.
\end{proof}

\begin{remark}\label{rem:B4}
  Theorem~\ref{thm:B4} gives justification to the algebraic concern
  raised in Remark~\ref{rem:iso1}. In a \Biv-category all isomorphisms
  that are imposed by the \Biv-structure do preserve the $\cor$-monoid
  and $\cand$-comonoid structure (and are therefore ``proper
  isomorphisms''). Note that there might still be ``improper
  isomorphisms'' in a \Biv-category. But these live outside the
  \Biv-structure and are therefore not accessible to proof-theoretic
  investigations.
\end{remark}

It has first been observed by Lamarche in \cite{lamarche:gap} that the
equation \eqref{eq:med-mix} (see below) is a consequence of the equations
\eqref{eq:med-assoc}, \eqref{eq:med-twist}, and~\eqref{eq:med-switch}. Due to
the presence of the $\cor$-monoids and $\cand$-comonoids, we can give here a
simpler proof of that fact:
 
\begin{corollary}\label{cor:mix-medial}
  In a \Biv-category, the diagram
    \begin{equation}
      \label{eq:med-mix}
      \vcdiagsquare
      {{A\cand B\cand C\cand D}{A\cand\twist[B,C]\cand D}}
      {{A\cand C\cand B\cand D}{\mix[A,C]\cand\mix[B,D]}}
      {{\mix[A\cand B,C\cand D]}{(A\cand B)\cor(C\cand D)}}
      {{\medial[A,B,C,D]}{(A\cor C)\cand(B\cor D)}}
      \tag{$\medial$-$\mix$}
    \end{equation}
  commutes.
\end{corollary}

\begin{proof}
  Chase {\small
    \diagrow=8ex
    \diagcol=2em
    $$
    \vcdiagram{
        \arbto[rr]{A\cand B\cand C\cand D}{A\cand\twist[B,C]\cand D}
        \arpto[dr]{}
        {\labupleft{6}{8}{\diag[A]\cand\diag[B]\cand\diag[C]\cand\diag[D]}}
        &&
        \arbtoid[ddd]{A\cand C\cand B\cand D}{}
        \arbto[ddl]{}{\quad\diag[A]\cand\diag[C]\cand\diag[B]\cand\diag[D]}
        \\&
        \arpto[d]{A\cand A\cand B\cand B\cand C\cand C\cand D\cand D}
        {A\cand A\cand\twist[B\cand B,C\cand C]\cand D\cand D}
        \\&
        \arpto[d]{A\cand A\cand C\cand C\cand B\cand B\cand D\cand D}
        {\isom}
        \arbto[dr]{}
        {\labdownright{3}{10}
          {\projr{A}{A}\cand\projr{C}{C}\cand\projl{B}{B}\cand\projl{D}{D}}}
        \\
        \arbto[r]{A\cand B\cand C\cand D}{\diag[A\cand B\cand C\cand D]}
        \arbfromid[uuu]{}{}
        &
        \arpto[r]{A\cand B\cand C\cand D\cand A\cand B\cand C\cand D}
        {\labdownright{3}{0}
          {(\projl{A}{B}\cand\projl{C}{D})\cand\projr{A}{B}\cand\projr{C}{D}}}
        \arpto[d]{}{\mix[A\cand B,C\cand D]\cand\mix[A\cand B,C\cand D]}
        &
        \arbto[dd]{A\cand C\cand B\cand D}{\mix\cand\mix}
        \\&
        \arpto[dr]{((A\cand B)\cor(C\cand D))\cand((A\cand B)\cor(C\cand D))}
        {\labupleft{3}{10}
          {(\projl{A}{B}\cor\projl{C}{D})\cand(\projr{A}{B}\cor\projr{C}{D})}}
        \\
        \arbfrom[uu]{\mix}{(A\cand B)\cor(C\cand D)}
        \arbto[ur]{}{\diag[(A\cand B)\cor(C\cand D)]\quad\quad}
        &&
        \arbfrom[ll]{\medial[A,B,C,D]}{(A\cor C)\cand(B\cor D)}
        }
    $$}%
  The topmost quadrangle commutes because of naturality of $\twist$.
  The pentagon below consists of several applications of
  \eqref{eq:B2c}. The two triangles on the right are trivial.  The
  quadrangle on the lower left commutes because mix preserves the
  $\cand$-comultiplication, and the quadrangle on the lower right
  because of naturality of mix. Finally, the triangle on the bottom is
  Theorem~\ref{thm:medial}~\eqref{l:med-dp}.  
\end{proof}

Obviously one can come up with more diagrams like \eqref{eq:med-mix} or
\eqref{eq:med-runit} and ask whether they commute, for example the
following due to McKinley \cite{mckinley:frogs}:
\begin{equation}
  \label{eq:mckinley}
  \vcdiagpentagondown%
  {{(A\cand\cfalse)\cor(B\cand C)}{\medial[A,\cfalse,B,C]}}
  {{(A\cor B)\cand(\cfalse\cor C)}{(A\cor B)\cand\colunit[C]}}
  {{(A\cor B)\cand C}{\switch[A,B,C]}}
  {{(A\cand\proj[\cfalse])\cor(B\cand C)}{(A\cand\ctrue)\cor(B\cand C)}}
  {{\runit[A]\cor(B\cand C)}{A\cor(B\cand C)}}
\end{equation}
It was soon discovered independently by several people that
\eqref{eq:mckinley} is equivalent to:
\begin{equation}\label{eq:mix-med-tens}
  \tag{$\mix$-$\medial$-$\tens$}
  \diagrow=7ex
  \vcdiagtriangleright
      {{(A\cand B)\cor(C\cand D)}{\medial[A,B,C,D]}}
      {{(A\cor C)\cand(B\cor D)}{\qquad\tens[A,C,B,D]}}
      {{\qquad\mix[A,B]\cor(C\cand D)}{A\cor B\cor(C\cand D)}}
\end{equation}
Here are two other examples that do not contain the units:
{\small
\begin{equation}\label{eq:med-tens-swi}
  \tag{$\medial$-$\cotens$-$\switch$}
  \diagcol=6em
  \vcdiaghexagondown
      {{((A\cand B)\cor(C\cand D))\cand(E\cor F)}
	{\switch[A\cand B,C\cand D,E\cand F]}}
      {{(A\cand B)\cor(C\cand D\cand(E\cor F))}
	{(A\cand B)\cor\cotens[C,D,E,F]}}
      {{(A\cand B)\cor(C\cand F)\cor(E\cand D)}
	{\comedialsq[A,B,C,F,E,D]}}
      {{\medial[A,B,C,D]\cand(E\cor F)}
	{(A\cor C)\cand(B\cor D)\cand(E\cor F)}}
      {{\cotens[A\cor C,B\cor D,E,F]}
	{((A\cor C)\cand F)\cor(E\cand(B\cor D))}}
      {{\medial[A\cor C,F,E,B\cor D]}
	{(A\cor C\cor E)\cand(F\cor B\cor D)}}
\end{equation}
\begin{equation}\label{eq:medsq-swi-medsq}
  \tag{$\comedialsq$-$\switch$-$\comedialsq$}
    \def\cor{\mathord\vee}
    \def\cand{\mathord\wedge}
  \diagcol=-7em
  \vcdiaghexagonhexa
      {{{(A'\cor A)\cand(B'\cor B)\cand(C'\cor C)\cand(D'\cor D)}}
	{p}}
      {{((A'\cor B')\cand(C'\cor D'))\cor(D\cand C)\cor (B\cand A)}
	{\comedialsq[A'\cor B',C'\cor D',D,C,B,A]}}
      {{(A'\cor B'\cor B\cor D)\cand(D'\cor C'\cor C\cor A)}
	{\labupleft{8}{12}{\tens[A'\cor B',B\cor D,D'\cor C',C\cor A]}}}
      {{q}
	{((A'\cor A)\cand(B'\cor C))\cor(B\cand D')\cor(D\cor C')}}
      {{\comedialsq[A'\cor A,B'\cor C,B,D',D,C']}
	{(A'\cor A\cor B\cor D)\cand(D'\cor C'\cor B'\cor C)}}
      {{\labupright{8}{13}{\tens[A'\cor A,B\cor D,D'\cor C',B'\cor C]}}
	{{A'\cor B'\cor((B\cor D)\cand(D'\cor C'))\cor C\cor A}}}
\end{equation}
}%
where $p$ and $q$ are the canonical maps (composed of several switches,
twists, and associativity) that are determined by the *-autonomous 
structure.

One usually speaks of ``coherence'' \cite{maclane:71} if all diagrams
of this kind commute. Very often a ``coherence theorem'' is based on
so-called ``coherence graphs''
\cite{kelly:maclane:71,dosen:petric:coherence-book}. In certain cases
(see, e.g., \cite{str:SD05}) the notion of coherence graph is too
restricted. For this reason, in \cite{lam:str:05:freebool}, the notion
of ``graphicality'' is introduced.

\begin{definition}\label{def:graphical}
  Let $\cC$ be a single-mixed \Bi-category, and let $\cCplus$ be the
  category obtained from $\cC$ by adding for each pair of objects $A$
  and $B$ a map $\mix[A,B]^{-1}\colon A\cor B\to A\cand B$ which is
  inverse to $\mix[A,B]$ (i.e., the two bifunctors $-\cand-$ and
  $-\cor-$ are naturally isomorphic in $\cCplus$). We say that $\cC$
  is \dfn{graphical} if the canonical forgetful functor
  $F\colon\cC\to\cCplus$ is faithful.
\end{definition}

Note that $\cCplus$ is a *-autonomous category in which the two
monoidal structures coincide, i.e., it is a compact closed
category. In fact, the whole point of Definition~\ref{def:graphical}
is to forget in a given *-autonomous category the difference between
$\cand$ and $\cor$. The actual problem is usually to find a canonical
way of making this collapse. But here, we can explore the fact that
$\cC$ is single-mixed and that the structure of a \Bi-category does
not induce any other natural maps $A\cand B\to A\cor B$ or $A\cor B\to
A\cand B$. Although in general inverting arrows in a category can
destroy the structure, it is harmless here since it only makes mix an
iso, and hence $\cCplus$  compact closed. We do not go into further
details of inverting arrows in categories because the paper is already
very long and Definition~\ref{def:graphical} does not play an
important role in the paper. Its main purpose is to provide the means
of formulating the following open problem.

\newtheorem{openproblem}{Open Problem}

\begin{openproblem}
  Let $\cE$ be a set of equations and let $\cC$ be the free
  \Bi-category that is generated from a set $\cA$ of generators (e.g.,
  propositional variables) and that obeys all of $\cE$. Is $\cC$
  graphical?  This question is equivalent to asking for a general
  coherence result for Boolean categories. The present paper exhibits
  many equations that have to hold, but it gives no clue whether they
  are enough, or what could be missing. 
\end{openproblem}

Note that for example the freely generated *-autonomous category
without units
\cite{lam:str:05:freebool,houston:etal:unitless,dosen:petric:05} is
graphical. This can be shown by using traditional proof nets for
multiplicative linear logic. However, the work of
\cite{lam:str:freestar} can be used to show that the freely generated
*-autonomous category with units is not graphical.

Clearly, in a graphical \Biv-category the equations
\eqref{eq:mix-med-tens}, \eqref{eq:med-tens-swi}, and
\eqref{eq:medsq-swi-medsq} all hold. However, at the current state of
the art it is not known whether they hold in every
\Biv-category.\footnote{The conjecture is that it is not the case, but
so far no counterexample could be constructed.}  But what can easily
be shown is the following proposition.

\begin{proposition}
  In every \Biv-category
  \begin{enumerate}[\rm(i)]
  \item the equation \eqref{eq:mckinley} holds if
    and only if equation \eqref{eq:mix-med-tens} holds, and
  \item the equation \eqref{eq:switchy}
    holds if and only if equation \eqref{eq:med-tens-swi} holds.
  \end{enumerate}
\end{proposition}

\begin{proof}
  Since we do not need this later, we leave the proof as an exercise to the
  reader.   
\end{proof}

\begin{definition}\label{def:B5}
  A \dfn{\Bv-category} is a \Biv-category that obeys equations
  \eqref{eq:mix-med-tens}, \eqref{eq:med-tens-swi}, and
  \eqref{eq:medsq-swi-medsq} for all objects.
\end{definition}

The motivation for this definition is the following lemma which will be needed
in the next section.

\begin{lemma}\label{lem:B5}
  In a \Bv-category the following equation holds for all objects $A$,
  $A'$, $B$, $B'$, $C$, $C'$, $D$, and $D'$:
  {\small
    \def\cor{\mathord\vee}
    \def\cand{\mathord\wedge}
  $$
  \diagcol=-7em
  \diagrow=7ex
  \def\lastarga{\labupleft{7}{13}
    {A'\cor B'\cor \medial[B,D',D,C']\cor C\cor A}}
  \def\lastargb{A'\cor B'\cor ((B\cor D)\cand(D'\cor C'))\cor C\cor A}
  \vcdiagdecagonhexa
      {{(A'\cor A)\cand(B'\cor B)\cand(C'\cor C)\cand(D'\cor D)}
	{\labdownright{7}{15}{\tens[A',A,B',B]\cand\tens[C',C,D',D]}}}
      {{(A'\cor B'\cor(A\cand B))\cand(C'\cor D'\cor(C\cand D))}
	{\tens[A'\cor B',A\cand B,C'\cor D',C\cand D]}}
      {{((A'\cor B')\cand(C'\cor D'))\cor(D\cand C)\cor(B\cand A)}
	{((A'\cor B')\cand(C'\cor D'))\cor\medial[D,C,B,A]}}
      {{((A'\cor B')\cand(C'\cor D'))\cor((D\cor B)\cand(C\cor A))}
	{\medial[A'\cor B',C'\cor D',D\cor B,C\cor A]}}
      {{(A'\cor B'\cor B\cor D)\cand(D'\cor C'\cor C\cor A)}
	{\labupright{7}{13}{\tens[A'\cor B',B\cor D,D'\cor C',C\cor A]}}}
      {{\labdownleft{7}{15}{(A'\cor A)\cand\tens[B',B,C',C]\cand(D'\cor D)}}
	{(A'\cor A)\cand(B'\cor (B\cand C')\cor C)\cand(D'\cor D)}}
      {{\mix[A'\cor A,B'\cor (B\cand C')\cor C]\cand(D'\cor D)}
	{(A'\cor B'\cor (B\cand C')\cor C\cor A)\cand(D'\cor D)}}
      {{\switch[A'\cor B'\cor C\cor A,B\cand C',D'\cor D]}
	{A'\cor B'\cor (B\cand(D'\cor D)\cand C')\cor C\cor A}}
      {{A'\cor B'\cor \cotens[B,D',D,C']\cor C\cor A}
	{A'\cor B'\cor (B\cand D')\cor(D\cand C')\cor C\cor A}}
  $$
  }%
\end{lemma}

\begin{proof}
  Chase the following diagram:
  {\scriptsize \diagcol=-7.6em \diagrow=7ex
    \def\cor{\mathord\vee}
    \def\cand{\mathord\wedge}
  $$
  \vcdiagram{
    \arpto[dd]
	  {(A'\cor A)\cand(B'\cor (B\cand C')\cor C)\cand(D'\cor D)}
	  {\mix}
    \arbto[dr]{}{\switch}
    &&
    \arpto[ll]
	  {(A'\cor A)\cand(B'\cor B)\cand(C'\cor C)\cand(D'\cor D)}
	  {\tens}
    \arbto[rr]{}{\tens\cand\tens}
    &&
    \arbto[dd]
	  {(A'\cor B'\cor(A\cand B))\cand(C'\cor D'\cor(C\cand D))}
	  {\tens}
    \\
    &
    \arbto[dl]
	  {(((A'\cor A)\cand(B'\cor C))\cor(B\cand C'))\cand(D'\cor D)}
	  {\mix}
    \arbto[dd]{}{\medial}
    \arbto[rr]{}{\switch}
    &&
    \arpto[dd]
	  {(((A'\cor A)\cand(B'\cor C))\cor(B\cand C'\cand(D'\cor D))}
	  {\cotens}
    \\
    \arpto[dd]
	  {(A'\cor B'\cor (B\cand C')\cor C\cor A)\cand(D'\cor D)}
	  {\switch}
    &&&&
    \arbto[dd]
	  {((A'\cor B')\cand(C'\cor D'))\cor(D\cand C)\cor(B\cand A)}
	  {\medial}
    \\
    &
    \arbto[dd]
	  {(A'\cor A\cor B)\cand(B'\cor C\cor C')\cand(D'\cor D)}
	  {\cotens}
    \arpto[ul]{}{\tens}
    &&
    \arpto[dd]
	  {((A'\cor A)\cand(B'\cor C))\cor(B\cand D')\cor(C'\cand D)}
	  {\comedialsq}
    \\
    \arpto[dd]
	  {A'\cor B'\cor (B\cand(D'\cor D)\cand C')\cor C\cor A}
	  {\cotens}
    &&&&
    \arbto[dd]
	  {((A'\cor B')\cand(C'\cor D'))\cor((B\cor D)\cand(C\cand A))}
	  {\medial}
    \\
    &
    \arbto[rr]
          {((A'\cor A\cor B)\cand D')\cor(D\cand(B'\cor C\cor C'))}
	  {\medial}
    \arbto[dl]{}{\switch\cor\switch}
    &&
    \arbto[dl]
	  {(A'\cor A\cor B\cor D)\cand (D'\cor B'\cor C\cor C')}
	  {\tens}
    \\
    \arpto[rr]
	  {A'\cor B'\cor (B\cand D')\cor(D\cand C')\cor C\cor A}
	  {\medial}
    &&
    \noar{A'\cor B'\cor ((B\cor D)\cand(D'\cor C'))\cor C\cor A}
    &&
    \arbto[ll]
	  {(A'\cor B'\cor B\cor D)\cand (D'\cor C'\cor C\cor A)}
	  {\tens}    
  }
  $$ }
  The little triangle in the upper left commutes because of
  \eqref{eq:mix-assoc}. The little triangle below it is just
  \eqref{eq:mix-med-tens}, and the pentagon below commutes because of the
  coherence in *-autonomous categories\footnote{It even commutes in the
  setting of weakly distributive categories.} \cite{BCST,lam:str:freestar}.
  The big square in the center is \eqref{eq:med-tens-swi} and the small
  parallelogram at the bottom is just two applications of
  \eqref{eq:med-switch} plugged together, and the big horse-shoe shape on the
  left is \eqref{eq:medsq-swi-medsq}.   
\end{proof}

\section{Beyond medial}\label{sec:beyond}

The definition of monoidal categories settles how the maps $\assoc[A,B,C]$,
$\twist[A,B]$, $\runit[A]$, and $\lunit[A]$ behave with respect to each other,
and how the maps $\coassoc[A,B,C]$, $\cotwist[A,B]$, $\corunit[A]$, and
$\colunit[A]$ behave with respect to each other. The notion of *-autonomous
category then settles via the bijection \eqref{eq:star} how the two monoidal
structures interact.  Then, the structure of a \Bi-category adds
$\cor$-monoids and $\cand$-comonoids, and the structure of \Bii-categories
allows the $\cor$-monoidal structure to go well with the $\cor$-monoids and
the $\cand$-monoidal structure to go well with the $\cand$-comonoids.
Finally, the structure of \Biv-categories ensures that \emph{both} monoidal
structures go well with the $\cor$-monoids \emph{and} the $\cand$-comonoids.

However, what has been neglected so far is how the $\cor$-monoids and
the $\cand$-comonoids go along with each other. Recall that in any
\Bii-category the maps $\codiag$ and $\coproj$ preserve the $\cor$-monoid
structure and the maps $\diag$ and $\proj$ preserve the $\cand$-comonoid
structure (Theorem~\ref{thm:B4}).

\newtheorem{combat}{
    Compatibility of $\cor$-monoids and $\cand$-comonoids}

\begin{combat}
  \label{para:poss}
  We have the following possibilities:
  \begin{enumerate}[\rm(i)]
  \item\label{l:pq} The maps $\proj$ and $\coproj$ are quasientropies.
  \item\label{l:pc} The maps $\proj$ and $\coproj$ are clonable.
  \item\label{l:dq} The maps $\diag$ and $\codiag$ are quasientropies.
  \item\label{l:dc} The maps $\diag$ and $\codiag$ are clonable.
  \end{enumerate}
\end{combat}

Condition \eqref{l:pq} says in particular that the following diagram commutes
\begin{equation}
  \label{eq:pq}
  \vcdiagtriangleup
      {{\cfalse}{\coproj[A]}}
      {{A}{\proj[A]}}
      {{\proj[\cfalse]}{\ctrue}}
\end{equation}
Every \Bi-category obeying \eqref{eq:B2a} and \eqref{eq:pq} is
not only single-mixed but also for every object $A$ the composition
$\cfalse\stackrel{\;\coproj[A]}{\to}A\stackrel{\;\proj[A]}{\to}\ctrue$ yields the same result. In \cite{lam:str:05:freebool}
the equation~\eqref{eq:pq} was used as basic axiom, and the mix map was
constructed from that without the use proper units.

The next observation to make is that \eqref{l:pc} and \eqref{l:dq} of
\ref{para:poss} are equivalent, provided \eqref{eq:nmedial} and \eqref{eq:B3a}
are present:

\begin{proposition}\label{prop:B3-diag-proj}
  In a \Bii-category with nullary medial and \eqref{eq:B3a} the following
  are equivalent for every object $A$:
  \begin{enumerate}[\rm(i)]
  \item\label{l:pm} The map $\proj[A]$ preserves
    the $\cor$-multiplication.
  \item\label{l:du} The map $\codiag[A]$
    preserves the $\cand$-counit.\vadjust{\vskip2pt}
  \item The map $\coproj[\cneg A]$ preserves the
    $\cand$-comultiplication.
  \item The map $\diag[\cneg A]$ preserves the
    $\cor$-unit.
  \end{enumerate}
\end{proposition}

\begin{proof}
  The equivalence of \eqref{l:du} and \eqref{l:pm} follows from
  $$
  \diagsquare{{A\cor A\arbto[dr]{}{\proj[A\cor A]}}{\codiag[A]}}
  {{A}{\proj[A]}}
  {{\proj[A]\cor\proj[A]}{\ctrue\cor\ctrue}}{{\codiag[\ctrue]}{\ctrue}}
  $$
  The lower triangle is \eqref{eq:nmedial} together with
  \eqref{eq:B3a}. The upper triangle is \eqref{l:du}, and the square
  is \eqref{l:pm}. The other equivalences follow by duality.
\end{proof}

Condition \ref{para:poss} \eqref{l:dc} exhibits yet another example of
a ``creative tension'' between algebra and proof theory. From the
viewpoint of algebra, it makes perfect sense to demand that the
$\cor$-monoid structure and the $\cand$-comonoid structure be
compatible with each other, i.e.,
that~\ref{para:poss}~\eqref{l:pq}--\eqref{l:dc} do all hold (see
\cite{lamarche:gap}). However, from the proof-theoretic point of view
it is reasonable to make some fine distinctions: We have to keep in
mind that in the sequent calculus it is the
``contraction-contraction-case''
$$
\vcenter
{\ddernote{\cutr}{}{\sqn{\Gamma,\Delta}}
  {\root{\conr}{\sqn{\Gamma,A}}
    {\leaf{\sqnsmallderi{\Gamma,A,A}{\pi_1}}}}
  {\root{\conr}{\sqn{\cneg A,\Delta}}
    {\leaf{\sqnsmallderi{\cneg A,\cneg A,\Delta}{\pi_2}}}}} 
$$ which spoils the confluence of cut elimination and which causes the
exponential blow-up of the size of the proof. This
questions~\ref{para:poss}~\eqref{l:dc}, i.e., the commutativity of the
diagram
\begin{equation}
  \label{eq:delta-nabla}
  \vcdiagsquare
  {{A\cor A}{\codiag[A]}}
  {{A}{\diag[A]}}
  {{\diag[A]\cor\diag[A]}{(A\cand A)\cor(A\cand A)}}
  {{\codiag[A\cand A]}{A\cand A}}
\end{equation}
motivates the distinction made in the following definition.

\begin{definition}\label{def:flat}
  We say a \Bi-category is \emph{weakly flat} if for every object $A$, the
  maps $\proj[A]$ and $\coproj[A]$ are strong and the maps $\diag[A]$ and
  $\codiag[A]$ are quasientropies (i.e.,
  \ref{para:poss}~\eqref{l:pq}--\eqref{l:dq} hold), and it is \emph{flat} if
  for every object $A$, the maps $\proj[A]$, $\coproj[A]$, $\diag[A]$ and
  $\codiag[A]$ are all strong (i.e., all of
  \ref{para:poss}~\eqref{l:pq}--\eqref{l:dc} do hold).
\end{definition}

\begin{corollary}
  A \Biii-category is weakly flat, if and only if $\proj[A]$ is a
  $\cor$-monoid morphism for every object $A$.
\end{corollary}

To understand the next (and final) axiom of this paper, recall that in every
*-auto\-no\-mous category we have
\begin{equation}
  \label{eq:star-cut}
  \vcdiagram{
    \arbto[r]{\ctrue}{\nid[A]\cand\nid[A]}
    &
    \arbto[d]{(\cneg A\cor A)\cand(\cneg A\cor A)}{\tens}
    \\
    \arbfrom[u]{\nid[A]}{\cneg A\cor A}
    &
    \arbto[l]{\cneg A\cor (A\cand \cneg A)\cor A}
	  {\cneg A\cor\conid[A]\cor A}
    }
\end{equation}
and that this equation is the reason why the cut elimination for
multiplicative linear logic (proof nets as well as sequent calculus) works so
well. The motivation for the following definition is to obtain something
similar for classical logic (cf.~\cite{lam:str:05:freebool}).

\begin{definition}
  A \Bi-category is \dfn{contractible} if the following diagram commutes for
  all objects~$A$.
  \begin{equation}\label{eq:loopkill}
    \vcdiagram{
      \arbto[r]{\ctrue}{\nid[A]}
      &
      \arbto[d]{\cneg A\cor A}{\diag[\cneg A\cor A]}
      \\
      &
      \arbto[d]{(\cneg A\cor A)\cand(\cneg A\cor A)}{\tens}
      \\
      \arbfrom[uu]{\nid[A]}{\cneg A\cor A}
      &
      \arbto[l]{\cneg A\cor (A\cand \cneg A)\cor A}
      {\cneg A\cor\conid[A]\cor A}
      }
  \end{equation}
\end{definition}

The following theorem states one of the main results of this paper. It
explains the deep reasons why the cut elimination for the proof nets
of \cite{lam:str:05:naming} is not confluent in the general case. It
also shows that the combination of equations \eqref{eq:delta-nabla}
and \eqref{eq:loopkill} together with the \Bv-structure leads to a
certain collapse, which can be compared to the collapse made by an
\LK-category. Nonetheless, even with this collapse we can find
reasonable models for proofs of Boolean logic, as it is shown in the
next section.

\begin{theorem}\label{thm:almost-idem}
  In a \Bv-category that is flat and contractible, we have
  $$
  \idf[A]+\idf[A]=\idf[A]
  $$
  for all objects $A$.
\end{theorem}

\begin{proof}
  \def\nA{\cneg A}
  The proof idea here is the same as in the proof of Theorem~2.4.7 in
  \cite{lam:str:05:freebool}. The novelty is that here we do not need the
  sledge-hammer axiom of graphicality. Instead we make use of
  Lemma~\ref{lem:B5}.  We proceed by showing that
  $\nid[A]+\nid[A]=\nid[A]\colon\ctrue\to\cneg A\cor A$ for all objects
  $A$. From this the result follows by Proposition~\ref{prop:semigroup-iso}.
  Note that in particular we have that $\nid[A]+\nid[A]$ is the map
  $$
  \vcdiagram{
    \arbto[r]{\ctrue}{\nid[A\cand A]}
    &
    \arbto[r]{\nA\cor\nA\cor(A\cand A)}{\nA\cor\nA\cor\mix[A,A]}
    &
    \arbto[r]{\nA\cor\nA\cor A\cor A}{\codiag[\nA]\cor\codiag[A]}
    &
    \noar{\nA\cor A}
  }
  $$
  which is because of \eqref{eq:mix-assoc} and the *-autonomous structure the
  same as the left-most down path in the following diagram.
  {\small 
    \def\cor{\mathord\vee}
    \def\cand{\mathord\wedge}
  \diagrow=7.3ex
  \diagcol=2.3em
  $$
  \vcdiagram{
    \arbto[rrr]{\ctrue}{\nid[A]\cand\nid[A]\cand\nid[A]\cand\nid[A]}
    \arbto[dr]{}{\labdownright{5}{5}{\nid[A]\cand\nid[A]\cand\nid[A]}}
    \arpto[ddd]{}{\nid[A]\cand\nid[A]}
    &&&
    \arbto[d]
    {(\nA\cor A)\cand(\nA\cor A)\cand(\nA\cor A)\cand(\nA\cor A)}
    {\tens[\nA,A,\nA,A]}
    \\
    &
    \arpto[d]
    {(\nA\cor A)\cand(\nA\cor A)\cand(\nA\cor A)}
    {\tens[\nA,A,\nA,A]}
    \arbto[urr]{}{\nid[A]}
    &&
    \arbto[d]
    {(\nA\cor A)\cand(\nA\cor(A\cand\nA)\cor A)\cand(\nA\cor A)}
    {\mix[\nA\cor A,\nA\cor(A\cand\nA)\cor A]}
    \\
    &
    \arpto[d]
    {(\nA\cor A)\cand(\nA\cor(A\cand\nA)\cor A)}
    {\mix}
    \arbto[urr]{}{\nid[A]}
    \arbto[dl]{}{\conid[A]}
    &&
    \arbto[d]
    {(\nA\cor\nA\cor(A\cand\nA)\cor A\cor A)\cand(\nA\cor A)}
    {\switch[\nA\cor\nA\cor A\cor A,A\cand\nA,\nA\cor A]}
    \\
    \arpto[d]
    {(\nA\cor A)\cand(\nA\cor A)}
    {\mix[\nA\cor A,\nA\cor A]}
    &
    \arpto[rr]
    {\nA\cor\nA\cor(A\cand\nA)\cor A\cor A}
    {\nid[A]}
    \arbto[urr]{}{\nid[A]}
    \arbto[dl]{}{\conid[A]}
    &&
    \arbto[d]
    {\nA\cor\nA\cor(A\cand(\nA\cor A)\cand\nA)\cor A\cor A}
    {\cotens[A,\nA,A,\nA]}
    \\
    \arpto[d]
    {\nA\cor\nA\cor A\cor A}
    {\codiag[\nA]\cor\codiag[A]}
    &
    \arpto[d]
    {\nA\cor\nA\cor(A\cand\nA)\cor A\cor A}
    {\codiag[\nA]\cor\codiag[A]}
    \arbto[l]{}{\conid[A]}
    &&
    \arbto[d]
    {\nA\cor\nA\cor(A\cand\nA)\cor(A\cand\nA)\cor A\cor A}
    {\medial[A,\nA,A,\nA]}
    \arpto[ll]{}{\codiag[A\cand\nA]}
    \\
    \noar{\nA\cor A}
    &
    \arbto[l]
    {\nA\cor(A\cand\nA)\cor A}
    {\conid[A]}
    &&
    \arbto[ll]
    {\nA\cor\nA\cor((A\cor A)\cand(\nA\cor\nA))\cor A\cor A}
    {\codiag[\nA]\cor(\codiag[A]\cand\codiag[\nA])\cor\codiag[A]}
    \arpto[ull]{}{\codiag[A]\cand\codiag[\nA]}
    }
  $$}%
  The upper triangle commutes because of functoriality of $\cand$, the square
  in the lower left corner because of functoriality of $\cor$, and the
  parallelograms because of naturality of $\mix$ and $\tens$. The quadrangle
  in the upper left commutes because of \eqref{eq:star-cut}, and the
  little triangle in the right center is just \eqref{eq:swi-unit} 
  together with naturality of switch. The pentagon below it is just the dual
  of \eqref{eq:loopkill}, and the two little triangles at the lower right
  corner are \eqref{eq:comedial} and functoriality of $\cor$.
  Therefore, this diagram gives us a complicated way of writing just
  $\nid[A]+\nid[A]$. Similarly, the next diagram gives us a complicated way of
  writing $\nid[A]$:
  {\small 
    \def\cor{\mathord\vee}
    \def\cand{\mathord\wedge}
  \diagrow=7.3ex
  \diagcol=4em
  $$
  \vcdiagram{
    \arbto[rrr]{\ctrue}{\nid[A]\cand\nid[A]\cand\nid[A]\cand\nid[A]}
    \arpto[ddr]{}{\nid[A]}
    \arbto[ddrrr]{}{\qquad\nid[(A\cand A)\cor(A\cand A)]}
    &&&
    \arbto[d]
    {(\nA\cor A)\cand(\nA\cor A)\cand(\nA\cor A)\cand(\nA\cor A)}
    {\tens[\nA,A,\nA,A]\cand\tens[\nA,A,\nA,A]}
    \\
    &&&
    \arbto[d]
    {(\nA\cor\nA\cor(A\cand A))\cand(\nA\cor\nA\cor(A\cand A))}
    {\tens[\nA\cor\nA,A\cand A,\nA\cor\nA,A\cand A]}
    \\
    &
    \arpto[d]{\nA\cor A}{\diag[\nA]\cor\diag[A]}
    &&
    \arbto[d]
    {((\nA\cor\nA)\cand(\nA\cor\nA))\cor(A\cand A)\cor(A\cand A)}
    {\medial[A,A,A,A]}
    \arpto[dll]{}{(\codiag[\nA]\cand\codiag[\nA])\cor\codiag[A\cand A]\qquad}
    \\
    &
    \arpto[d]
    {(\nA\cand\nA)\cor(A\cand A)}
    {\medial[\nA,\nA,A,A]}
    &&
    \arbto[d]
    {((\nA\cor\nA)\cand(\nA\cor\nA))\cor((A\cor A)\cand(A\cor A))}
    {\medial[\nA\cand\nA,\nA\cor\nA,A\cor A,A\cor A]}
    \arbto[ll]{}
    {(\codiag[\nA]\cand\codiag[\nA])\cor(\codiag[A]\cand\codiag[A])}    
    \\
    &
    \arpto[d]
    {(\nA\cor A)\cand(\nA\cor A)}
    {\tens[\nA,A,\nA,A]}
    &&
    \arbto[d]
    {(\nA\cor\nA\cor A\cor A)\cand(\nA\cor\nA\cor A\cor A)}
    {\tens[\nA\cand\nA,A\cor A,\nA\cor\nA,A\cor A]}
    \arbto[ll]{}
    {(\codiag[\nA]\cor\codiag[A])\cand(\codiag[\nA]\cor\codiag[A])}    
    \\
    \arbfrom[uuuuu]{\nid[A]}{\nA\cor A}
    &
    \arbto[l]
    {\nA\cor(A\cand\nA)\cor A}
    {\conid[A]}
    &&
    \arbto[ll]
    {\nA\cor\nA\cor((A\cor A)\cand(\nA\cor\nA))\cor A\cor A}
    {\codiag[\nA]\cor(\codiag[A]\cand\codiag[\nA])\cor\codiag[A]}    
    }
  $$}%
  Here the big upper right ``triangle'' commutes because of the
  *-autonomous structure. The irregular quadrangle in the center is a
  transposed version of \eqref{eq:delta-nabla}, the little triangle below it
  is \eqref{eq:comedial}, the two squares at the bottom are naturality of
  $\medial$ and $\tens$, and the left-most part of the diagram commutes
  because of \eqref{eq:loopkill} and \eqref{eq:medial}.  Finally, we apply
  Lemma~\ref{lem:B5} to paste the two diagrams together, which yields
  $\nid[A]+\nid[A]=\nid[A]$ as desired.   
\end{proof}

In Figure~\ref{fig:almost-idem} the basic idea of this proof is shown. The
first four equations in that figure express the idea behind the first big
diagram in the proof of Theorem~\ref{thm:almost-idem}, and the last three
equations in Figure~\ref{fig:almost-idem} express the idea of the second
diagram. More explanations on this will follow in the next section.

\begin{corollary}
  Let $\cA$ be a set of propositional variables and let $\cC$ be the free
  flat and contractible \Bv-category generated by $\cA$. Then $\cC$ is
  idempotent.
\end{corollary}

\section{A concrete example: proof nets}\label{sec:pn}

In this section we will construct a concrete example of a category
which has almost all the properties discussed in this paper. Its
existence shows that this paper actually makes sense: The equations
presented here do not lead to the collapse into a Boolean algebra. In
fact, this category was the main source of motivation for 
introducing the equations presented in Sections
\ref{sec:star}, \ref{sec:monoids}, \ref{sec:medial}, and
\ref{sec:beyond}. 

We are going to present two versions of proof nets:
\begin{enumerate}[1.]
\item The \emph{simple proof nets} are a slight modification of the proof nets
  introduced in \cite{lam:str:05:naming,lam:str:05:freebool}. The difference
  is that the categories of proof nets defined in these papers had only weak
  units, while here we are assuming from the beginning that
  $\ctrue$ and $\cfalse$ are proper unit objects.
\item The \emph{extended proof nets} have a richer structure than the simple
  nets. From the algebraic point of view the main difference to the simple
  nets is that the category of extended nets does not obey equation
  \eqref{eq:delta-nabla} and is not idempotent (and is therefore not an
  \LK-category). From the proof theoretic point of view, the extended nets
  keep more information about the proofs. In particular the size of proofs can
  captured.
\end{enumerate}

\begin{definition}
  Let $\cA$ be a set of propositional variables. The
  set $\cF$ of \dfn{formulae} is generated via
  $$
  \cF \grammareq \cA \mid \cneg\cA \mid \lone \mid \lbot \mid
  \cF\ltens\cF \mid \cF\lpar\cF
  \quadfs
  $$
  A \dfn{sequent} is a finite list of formulae, separated by comma.
  A formula can be seen as a tree and a sequent as a forest whose leaves
  are labeled by elements of the set
  $\cA\cup\cneg\cA\cup\set{\lbot,\lone}$ and whose inner nodes are
  labeled by elements of $\set{\cand,\cor}$. For a sequent $\Gamma$,
  let $\Leaf{\Gamma}$ denote the set of its leaves. For a leaf
  $i\in\Leaf{\Gamma}$ let
  $\lab{i}\in\cA\cup\cneg\cA\cup\set{\lbot,\lone}$ denote its labeling.
  A \dfn{linking} for a sequent $\Gamma$ is a binary relation
  $P\subseteq\Leaf{\Gamma}\times\Leaf{\Gamma}$ such that 
  \begin{enumerate}[(i)]
  \item for every $i\in\Leaf{\Gamma}$ with $\lab{i}=\ctrue$ we have
    $(i,i)\in P$, and
  \item if $(i,j)\in P$, then one of the following cases must hold:
    \begin{itemize}
    \item $i=j$ and $\lab{i}=\ctrue$, or
    \item $i\neq j$ and $\lab{i}=\cneg a$ and
      $\lab{j}=a$ for some $a\in\cA$.
    \end{itemize}
  \end{enumerate}
  A \dfn{simple prenet}\footnote{What we call \dfn{prenet} is in the
    literature sometimes also called a \dfn{proof structure.}}  consists of a
  sequent $\Gamma$ and a linking $P$ for it.  It will be denoted by
  $\prfnet{P}{\Gamma}$.
\end{definition}

In this paper, we will write prenets by simply
writing down the sequent and by putting the linking as (directed)
graph above it,  as in these two examples\footnote{Here we make two
  modifications to the proof nets used in
  \cite{lam:str:05:naming,lam:str:05:freebool}: (i) We force every $\ctrue$ to
  be linked to itself and we do not allow links between $\ctrue$ and
  $\cfalse$. The reason is that we deal in this paper with proper units in the
  categorical sense, while \cite{lam:str:05:naming,lam:str:05:freebool} used
  ``weak units'' (see also the introduction). The observation that 
  linking every
  $\ctrue$ to itself and disallowing $\ctrue$-$\cfalse$-links is enough to get
  proper units is due to Fran\c{c}ois Lamarche. (ii) We use here directed
  links between complementary pairs of atoms (instead of undirected links as
  in \cite{lam:str:05:naming,lam:str:05:freebool}). This brings a slight
  simplification of cut elimination via path composition. The idea for this
  has been taken from Dominic Hughes
  \cite{hughes:simple-mult,hughes:freestar}.}:
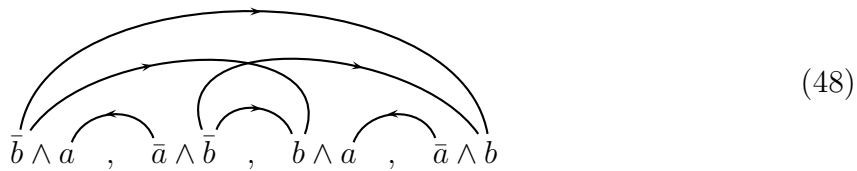
\begin{equation}
  \label{eq:exa1}
  \vcenter{\hbox{
    \begin{psmatrix}[rowsep=3\baselineskip]
      \quad\\
      $\rnode{nb1}{\cneg b}\cand\rnode{a1}{a}
      \seqsep
      \rnode{na1}{\cneg a}\cand\rnode{nb2}{\cneg b}
      \seqsep
      \rnode{b1}{b}\cand\rnode{a2}{a}
      \seqsep
      \rnode{na2}{\cneg a}\cand\rnode{b2}{b}
      $
      \udvecanglesheight{nb1}{b2}{80}{100}{.7}
      \vecanglesposheight{nb1}{b1}{55}{70}{.4}{.8}
      \vecanglesposheight{nb2}{b2}{110}{125}{.6}{.8}
      \uldvec{na1}{a1}
      \urdvec{nb2}{b1}
      \uldvec{na2}{a2}
    \end{psmatrix}
    }}
\end{equation}
and
\begin{equation}
  \label{eq:exa2}
  \vcenter
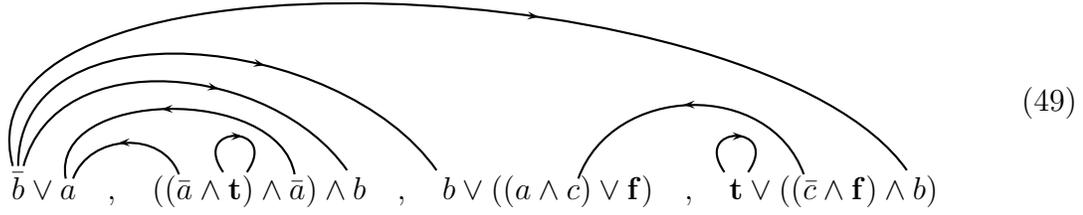
{\hbox{
    \begin{psmatrix}[rowsep=4\baselineskip]
      \quad\\
      $\rnode{nb1}{\cneg b}\cor\rnode{a1}{a}
      \seqsep
      ((\rnode{na1}{\cneg a}\cand\rnode{t1}{\ctrue})
      \cand\rnode{na2}{\cneg a})\cand\rnode{b1}{b}
      \seqsep
      \rnode{b2}{b}\cor((\rnode{a2}{a}\cand\rnode{c1}{c})
      \cor\rnode{f1}{\cfalse})
      \seqsep
      \rnode{t2}{\ctrue}\cor
      ((\rnode{nc1}{\cneg c}\cand\rnode{f2}{\cfalse})\cand\rnode{b3}{b})
      $
      \uloop{t1}
      \uloop{t2}
      \uldvec{na1}{a1}
      \vecanglesheight{na2}{a1}{100}{100}{.8}
      \uldvecheight{nc1}{c1}{.9}
      \vecanglesposheight{nb1}{b1}{75}{125}{.6}{.8}
      \vecanglesposheight{nb1}{b2}{90}{125}{.6}{.8}
      \vecanglesposheight{nb1}{b3}{105}{125}{.6}{.55}
    \end{psmatrix}
    }}
\end{equation}

Now we will define when a simple prenet is correct, i.e., comes from an
actual proof. A \dfn{conjunctive pruning} of a prenet
$\prfnet{P}{\Gamma}$ is the result of removing one of the two
subformulae of each $\cand$ in $\Gamma$ and restricting the linking
$P$ accordingly. Here are two (of 16 possible) conjunctive prunings of
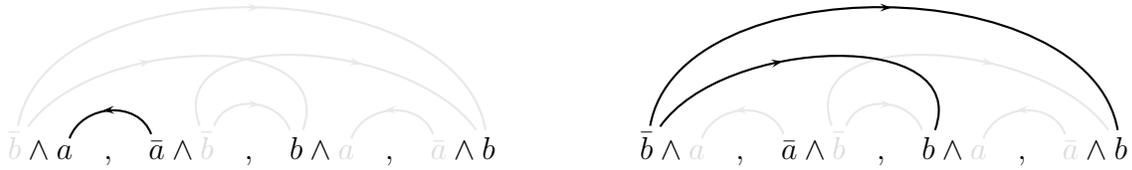
\eqref{eq:exa1}:
$$
  \vcenter{\hbox{
    \begin{psmatrix}[rowsep=3\baselineskip]
      \quad\\
      $\prune{\rnode{nb1}{\cneg b}}\cand\rnode{a1}{a}
      \seqsep
      \rnode{na1}{\cneg a}\cand\prune{\rnode{nb2}{\cneg b}}
      \seqsep
      \rnode{b1}{b}\cand\prune{\rnode{a2}{a}}
      \seqsep
      \prune{\rnode{na2}{\cneg a}}\cand\rnode{b2}{b}
      $
      \prune{%
        \udvecanglesheight{nb1}{b2}{80}{100}{.7}
        \vecanglesposheight{nb1}{b1}{55}{70}{.4}{.8}
        \vecanglesposheight{nb2}{b2}{110}{125}{.6}{.8}
        \urdvec{nb2}{b1}
        \uldvec{na2}{a2}
        }
      \uldvec{na1}{a1}
    \end{psmatrix}
    \qquad
    \qquad
    \begin{psmatrix}[rowsep=3\baselineskip]
      \quad\\
      $\rnode{nb1}{\cneg b}\cand\prune{\rnode{a1}{a}}
      \seqsep
      \rnode{na1}{\cneg a}\cand\prune{\rnode{nb2}{\cneg b}}
      \seqsep
      \rnode{b1}{b}\cand\prune{\rnode{a2}{a}}
      \seqsep
      \prune{\rnode{na2}{\cneg a}}\cand\rnode{b2}{b}
      $
      \prune{%
        \vecanglesposheight{nb2}{b2}{110}{125}{.6}{.8}
        \uldvec{na1}{a1}
        \urdvec{nb2}{b1}
        \uldvec{na2}{a2}
        }
      \udvecanglesheight{nb1}{b2}{80}{100}{.7}
      \vecanglesposheight{nb1}{b1}{55}{70}{.4}{.8}
    \end{psmatrix}
    }}
$$ A simple prenet is \dfn{correct} if each of its conjunctive prunings
contains at least one link (i.e, the linking is not empty). A \dfn{simple
proof net} is a correct simple prenet. The two examples in \eqref{eq:exa1}
and~\eqref{eq:exa2} are proof nets. Here is a prenet, which is not a proof net
because there is a pruning in which all links disappear:
$$
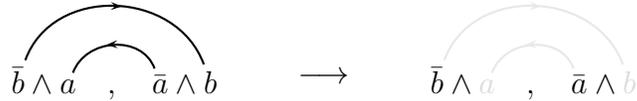
\hbox{
  \begin{psmatrix}[rowsep=1.5\baselineskip]
    \quad\\
    $\rnode{nb1}{\cneg b}\cand\rnode{a1}{a}
    \seqsep
    \rnode{na1}{\cneg a}\cand\rnode{b1}{b}
    $
    \urdvec{nb1}{b1}
    \uldvec{na1}{a1}
  \end{psmatrix}
  }
\qquad\longrightarrow\qquad
\hbox{
  \begin{psmatrix}[rowsep=1.5\baselineskip]
    \quad\\
    $\rnode{nb1}{\cneg b}\cand\prune{\rnode{a1}{a}}
    \seqsep
    \rnode{na1}{\cneg a}\cand\prune{\rnode{b1}{b}}
    $
    \prune{
      \urdvec{nb1}{b1}
      \uldvec{na1}{a1}
      }
  \end{psmatrix}
}
$$

Now we show how simple proof nets can be composed. As in
\cite{lam:str:05:naming,lam:str:05:freebool,str:SD05} this is done via
cut elimination. But we use here a notational trick to make it even
more intuitive: we allow to write proof nets in two-sided form:
Instead of putting the linking \emph{above} a sequent
$A_1,\ldots,A_n,B_1,\ldots,B_m$, we put it \emph{in between} $\cneg
A_n,\ldots,\cneg A_1$ and $B_1,\ldots,B_m$, where the negation $\cneg
A$ of a formula $A$ is inductively defined as follows\footnote{We
  invert the order when taking the negation in order to reduce the
  number of crossings in the pictures.}:
$$
\cneg a=\cneg a\,,\quad\cneg{\cneg a}=a\,,\quad 
\cneg\ctrue=\cfalse\,,\quad\cneg\cfalse=\ctrue\,,\quad
\overline{(A\cand B)}=\cneg B\cor\cneg A\,,\quad
\overline{(A\cor B)}=\cneg B\cand \cneg A\,.
$$
Here are three different ways of writing example \eqref{eq:exa1} in
two-sided form:
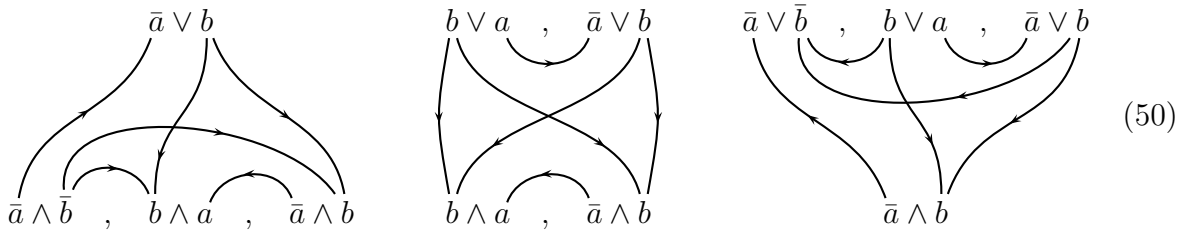
\begin{equation}
  \label{eq:exa1-two}
  \vcenter{\hbox{
      \begin{psmatrix}[rowsep=4\baselineskip]
        $\rnode{a1}{\cneg a}\cor\rnode{nb1}{b}$
        \\
        $\rnode{na1}{\cneg a}\cand\rnode{nb2}{\cneg b}
        \seqsep
        \rnode{b1}{b}\cand\rnode{a2}{a}
        \seqsep
        \rnode{na2}{\cneg a}\cand\rnode{b2}{b}
        $
        \dvecpos{nb1}{b1}{.8}
        \vecanglesheight{nb1}{b2}{-70}{100}{.8}
        \vecanglesposheight{nb2}{b2}{95}{125}{.6}{.7}
        \vecanglesheight{na1}{a1}{80}{-110}{.85}
        \urdvec{nb2}{b1}
        \uldvec{na2}{a2}
      \end{psmatrix}
      \qquad\;
      \begin{psmatrix}[rowsep=4\baselineskip]
        $\rnode{nb2}{b}\cor\rnode{na1}{a}
        \seqsep
        \rnode{a1}{\cneg a}\cor\rnode{nb1}{b}$
        \\
        $\rnode{b1}{b}\cand\rnode{a2}{a}
        \seqsep
        \rnode{na2}{\cneg a}\cand\rnode{b2}{b}
        $
        \vecangles{nb1}{b2}{-80}{80}
        \vecanglesposheight{nb1}{b1}{-105}{75}{.75}{.7}
        \vecanglesposheight{nb2}{b2}{-75}{105}{.75}{.7}
        \vecangles{nb2}{b1}{-100}{100}
        \druvec{na1}{a1}
        \uldvec{na2}{a2}
      \end{psmatrix}
      \qquad\;
      \begin{psmatrix}[rowsep=4\baselineskip]
        $\rnode{a2}{\cneg a}\cor\rnode{b1}{\cneg b}
        \seqsep
        \rnode{nb2}{b}\cor\rnode{na1}{a}
        \seqsep
        \rnode{a1}{\cneg a}\cor\rnode{nb1}{b}$
        \\
        $\rnode{na2}{\cneg a}\cand\rnode{b2}{b}
        $
        \vecanglesheight{nb1}{b2}{-100}{70}{.9}
        \vecanglesposheight{nb1}{b1}{-125}{-95}{.4}{.7}
        \dvecpos{nb2}{b2}{.7}
        \dluvec{nb2}{b1}
        \druvec{na1}{a1}
        \vecanglesheight{na2}{a2}{110}{-80}{.85}
      \end{psmatrix}
      }}
\end{equation}
And here is a different way of writing example \eqref{eq:exa2}:
\begin{equation}
  \label{eq:exa2-two}
  \vcenter{\hbox{
    \begin{psmatrix}[rowsep=4\baselineskip]
      $\rnode{a1}{\cneg a}\cand\rnode{nb1}{b}
      \seqsep
      (\rnode{b3}{\cneg b}\cor(\rnode{f2}{\ctrue}\cor\rnode{nc1}{c})
      \cand\rnode{t2}{\cfalse}
      $
      \\
      $((\rnode{na1}{\cneg a}\cand\rnode{t1}{\ctrue})
      \cand\rnode{na2}{\cneg a})\cand\rnode{b1}{b}
      \seqsep
      \rnode{b2}{b}\cor((\rnode{a2}{a}\cand\rnode{c1}{c})
      \cor\rnode{f1}{\cfalse})
      $
      \uloop{t1}
      \dloop{t2}
      \vecanglesheight{na1}{a1}{90}{-100}{.9}
      \vecanglesheight{na2}{a1}{90}{-80}{.9}
      \dvec{nc1}{c1}
      \vecanglesheight{nb1}{b1}{-100}{90}{.9}
      \vecanglesheight{nb1}{b2}{-85}{90}{.9}
      \druvec{nb1}{b3}
    \end{psmatrix}
    }}
\end{equation}

Note that for defining the direction of the links and for checking
correctness we always have to consider the negation of the formula on
the top. This is equivalent to pretending we had not taken the
negation of that formula when going from the one-sided to the two-sided
version. One can make this formally precise by using polarities
\cite{lamarche:01}. What is important is the fact that the objects in
\eqref{eq:exa1} and in \eqref{eq:exa1-two} denote \emph{the same} net.
Similarly, \eqref{eq:exa2} and in \eqref{eq:exa2-two} are just
different ways of drawing the same proof net.

Cut elimination can now be defined by plugging nets together, as in
the following example which is a composition of (the middle) net in
\eqref{eq:exa1-two} and the one in \eqref{eq:exa2-two}:
{\small
\begin{equation}
  \label{eq:exa1-exa2-two}
  \qlapm{\vcenter{\hbox{
      \begin{psmatrix}[rowsep=3\baselineskip]
        $\rnode{xnb2}{b}\cor\rnode{xna1}{a}
        \seqsep
        \rnode{xa1}{\cneg a}\cor\rnode{xnb1}{b}$
        \\
        $\rnode{xb1}{b}\cand\rnode{xa2}{a}
        \seqsep
        \rnode{cna2}{\cneg a}\cand\rnode{cb2}{b}
        \seqsep
        (\rnode{b3}{\cneg b}\cor(\rnode{f2}{\ctrue}\cor\rnode{nc1}{c})
        \cand\rnode{t2}{\cfalse}
        $
        \\
        $((\rnode{na1}{\cneg a}\cand\rnode{t1}{\ctrue})
        \cand\rnode{na2}{\cneg a})\cand\rnode{b1}{b}
        \seqsep
        \rnode{b2}{b}\cor((\rnode{a2}{a}\cand\rnode{c1}{c})
        \cor\rnode{f1}{\cfalse})
        $
        \vecanglesheight{xnb1}{cb2}{-80}{80}{.7}
        \vecanglesposheight{xnb1}{xb1}{-105}{75}{.75}{.5}
        \vecanglesposheight{xnb2}{cb2}{-85}{95}{.75}{.7}
        \vecanglesheight{xnb2}{xb1}{-100}{100}{.7}
        \druvec{xna1}{xa1}
        \uldvec{cna2}{xa2}
        \uloop{t1}
        \dloop{t2}
        \vecanglesheight{na1}{cna2}{90}{-100}{.9}
        \vecanglesheight{na2}{cna2}{90}{-80}{.9}
        \vecanglesheight{nc1}{c1}{-100}{90}{.9}
        \vecanglesheight{cb2}{b1}{-100}{90}{.9}
        \vecanglesheight{cb2}{b2}{-85}{90}{.9}
        \druvec{cb2}{b3}
      \end{psmatrix}
      }}
  \;\;\longrightarrow
  \vcenter{\hbox{
      \begin{psmatrix}[rowsep=6\baselineskip]
        $\rnode{xnb2}{b}\cor\rnode{xna1}{a}
        \seqsep
        \rnode{xa1}{\cneg a}\cor\rnode{xnb1}{b}
        \seqsep
        (\rnode{b3}{\cneg b}\cor(\rnode{f2}{\ctrue}\cor\rnode{nc1}{c})
        \cand\rnode{t2}{\cfalse}
        $
        \\
        $
        \rnode{xb1}{b}\cand\rnode{xa2}{a}
        \seqsep
        ((\rnode{na1}{\cneg a}\cand\rnode{t1}{\ctrue})
        \cand\rnode{na2}{\cneg a})\cand\rnode{b1}{b}
        \seqsep
        \rnode{b2}{b}\cor((\rnode{a2}{a}\cand\rnode{c1}{c})
        \cor\rnode{f1}{\cfalse})
        $
        \vecanglesposheight{xnb1}{xb1}{-120}{75}{.65}{.9}
        \vecanglesheight{xnb2}{xb1}{-100}{90}{1}
        \druvec{xna1}{xa1}
        \uloop{t1}
        \dloop{t2}
        \uldvec{na1}{xa2}
        \vecanglesheight{na2}{xa2}{100}{90}{.8}
        \dvec{nc1}{c1}
        \vecanglesposheight{xnb1}{b1}{-105}{90}{.45}{.9}
        \vecanglesheight{xnb1}{b2}{-85}{90}{.8}
        \druvec{xnb1}{b3}
        \vecanglesposheight{xnb2}{b1}{-90}{115}{.6}{.8}
        \vecanglesheight{xnb2}{b2}{-75}{115}{.75}
        \vecanglesheight{xnb2}{b3}{-65}{-90}{.6}
    \end{psmatrix}
    }}}
\end{equation}}%
There is a link in the resulting net if and only if there is a
corresponding path in the non-reduced composition. Writing it in the
two-sided version makes it more intuitive than in the one-sided
version, where \eqref{eq:exa1-exa2-two} would be written as:
{\small
\def\seqsep{\,\,,\,\,}
$$
\begin{array}{c}
  {\hbox{
      \begin{psmatrix}[rowsep=14ex]
        \quad\\
        $\rnode{xnb1}{\cneg b}\cand\rnode{xa1}{a}
        \seqsep
        \rnode{xna1}{\cneg a}\cand\rnode{xnb2}{\cneg b}
        \seqsep
        \rnode{xb1}{b}\cand\rnode{xa2}{a}
        \seqsep
        (\rnode{xna2}{\cneg a}\cand\rnode{xb2}{b})\ccut
        (\rnode{nb1}{\cneg b}\cor\rnode{a1}{a})
        \seqsep
        ((\rnode{na1}{\cneg a}\cand\rnode{t1}{\ctrue})
        \cand\rnode{na2}{\cneg a})\cand\rnode{b1}{b}
        \seqsep
        \rnode{b2}{b}\cor((\rnode{a2}{a}\cand\rnode{c1}{c})
        \cor\rnode{f1}{\cfalse})
        \seqsep
        \rnode{t2}{\ctrue}\cor
        ((\rnode{nc1}{\cneg c}\cand\rnode{f2}{\cfalse})\cand\rnode{b3}{b})
        $
        \uloop{t1}
        \uloop{t2}
        \uldvec{na1}{a1}
        \vecanglesheight{na2}{a1}{100}{100}{.8}
        \uldvecheight{nc1}{c1}{.9}
        \vecanglesposheight{nb1}{b1}{75}{125}{.6}{.8}
        \vecanglesposheight{nb1}{b2}{90}{125}{.6}{.8}
        \vecanglesposheight{nb1}{b3}{105}{125}{.6}{.55}
        \udvecanglesheight{xnb1}{xb2}{80}{100}{.7}
        \vecanglesposheight{xnb1}{xb1}{55}{70}{.4}{.8}
        \vecanglesposheight{xnb2}{xb2}{110}{125}{.6}{.8}
        \uldvec{xna1}{xa1}
        \urdvec{xnb2}{xb1}
        \uldvec{xna2}{xa2}
      \end{psmatrix}
      }}
  \\[2ex]
  \big\downarrow
  \\[5ex]
  {\hbox{
      \begin{psmatrix}[rowsep=12ex]
        \quad\\
        $\rnode{xnb1}{\cneg b}\cand\rnode{xa1}{a}
        \seqsep
        \rnode{xna1}{\cneg a}\cand\rnode{xnb2}{\cneg b}
        \seqsep
        \rnode{xb1}{b}\cand\rnode{xa2}{a}
        \seqsep
        ((\rnode{na1}{\cneg a}\cand\rnode{t1}{\ctrue})
        \cand\rnode{na2}{\cneg a})\cand\rnode{b1}{b}
        \seqsep
        \rnode{b2}{b}\cor((\rnode{a2}{a}\cand\rnode{c1}{c})
        \cor\rnode{f1}{\cfalse})
        \seqsep
        \rnode{t2}{\ctrue}\cor
        ((\rnode{nc1}{\cneg c}\cand\rnode{f2}{\cfalse})\cand\rnode{b3}{b})
        $
        \uloop{t1}
        \uloop{t2}
        \uldvec{na1}{xa2}
        \vecanglesheight{na2}{xa2}{100}{100}{.8}
        \uldvecheight{nc1}{c1}{.9}
        \uldvec{xna1}{xa1}
        \vecanglesposheight{xnb1}{xb1}{55}{90}{.4}{.7}
        \vecanglesposheight{xnb1}{b1}{70}{110}{.3}{.55}
        \vecanglesposheight{xnb1}{b2}{85}{110}{.4}{.6}
        \vecanglesposheight{xnb1}{b3}{100}{115}{.5}{.5}
        \urdvec{xnb2}{xb1}
        \vecanglesposheight{xnb2}{b1}{85}{130}{.6}{.7}
        \vecanglesposheight{xnb2}{b2}{100}{130}{.8}{.7}
        \vecanglesposheight{xnb2}{b3}{115}{130}{.6}{.5}
      \end{psmatrix}
      }}
\end{array}
$$}%
This is the way it has been done in \cite{lam:str:05:naming} where it has
been shown that this operation is associative and preserves correctness.  In
\cite{lam:str:05:naming} it has also been shown how sequent calculus
derivations are translated into proof nets. After what has been said here, it
might be more intuitive to think of them as \emph{flow graphs}
\cite{buss:91,carbone:97} of derivations in
$\SKS$~\cite{brunnler:tiu:01}. From the historical perspective it should be
mentioned that the basic idea of the simple proof nets discussed here appeared
in the literature already in \cite{andrews:76} as \emph{matings} and in
\cite{bibel:81} as \emph{matrix proofs}\footnote{Note, however, that there is
a subtle but crucial difference between our simple nets and the work of
\cite{andrews:76,bibel:81}: in this early work links between atoms in a
conjunction relation were not allowed (because they are irrelevant for
correctness), but these links are crucial for obtaining an associative cut
elimination operation (see \cite{lam:str:05:naming}).}, and that this
idea goes even back to the \emph{coherence graphs} of \cite{kelly:maclane:71}.

If we now restrict ourselves to proof nets with only two conclusions,
then we have a category\footnote{Without the restriction to two
  formulae we would obtain a polycategory \cite{lambek:69,szabo:75}.}:
the objects are the formulae and the arrows $A\to B$ are the proof
nets $\prfnet{P}{\cneg A,B}$. Arrow composition is defined as above,
and the identities are the trivial proof nets $\prfnet{P}{\cneg A,A}$.
Let us call this category $\SNet(\cA)$ where $\cA$ is the set of
propositional variables from which we started.

\begin{theorem}\label{thm:net-B4}
  The category $\SNet(\cA)$ is a \Bv-category that is flat and
  contractible.
\end{theorem}

\begin{proof}
  The maps $\assoc$, $\twist$, $\runit$, $\lunit$, $\switch$, $\medial$,
  $\proj$, and $\diag$ are given by the obvious nets. We show here as example
  the nets for $\medial[A,B,C,D]$, $\runit[A]$, $\proj[A]$, and $\diag[A]$
  (the others being similar, cf. \cite{str:SD05}):
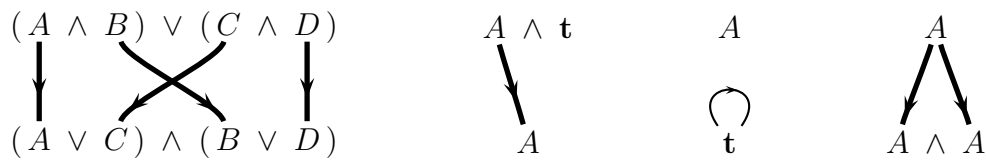
  $$
  \def\normalvecheight{.3}  
  \def\normalarrpos{.7}  
  \vcenter{\hbox{
  \begin{psmatrix}[rowsep=2\baselineskip]
    $\klam{\rnode{nA}{A}\medcand\rnode{nB}{B}}\medcor
    \klam{\rnode{nC}{C}\medcand\rnode{nD}{D}}$
    \\
    $\klam{\rnode{A}{A}\medcor\rnode{C}{C}}\medcand
    \klam{\rnode{B}{B}\medcor\rnode{D}{D}}$
    \boldvec
    \dvecpos{nA}{A}{.7}
    \drvecpos{nB}{B}{.9}
    \dlvecpos{nC}{C}{.9}
    \dvecpos{nD}{D}{.7}
  \end{psmatrix}
  }}
  \qquad\qquad
  \vcenter{\hbox{
  \begin{psmatrix}[rowsep=2\baselineskip]
    $\rnode{nA}{A}\medcand\rnode{tt}{\ctrue}$
    \\
    $\rnode{A}{A}$
    \boldvec
    \drvec{nA}{A}
  \end{psmatrix}
  }}
  \qquad\qquad
  \vcenter{\hbox{
  \begin{psmatrix}[rowsep=2\baselineskip]
    $\rnode{nA}{A}$
    \\
    $\rnode{tt}{\ctrue}$
    \uloop{tt}
  \end{psmatrix}
  }}
  \qquad\qquad
  \vcenter{\hbox{
  \begin{psmatrix}[rowsep=2\baselineskip]
    $\rnode{nA}{A}$
    \\
    $\rnode{A1}{A}\medcand\rnode{A2}{A}$
    \boldvec
    \dlvecpos{nA}{A1}{.9}
    \drvecpos{nA}{A2}{.9}
  \end{psmatrix}
  }}
  $$ In these drawings the bold links between formulae represent bundles of
  several links, one for each leaf of the formula tree. Note that the arrows
  have to have the right direction, and that there are no links connecting
  $\cfalse$ and $\ctrue$. There are four cases:
  \begin{equation}
    \def\normalarrpos{.7}  
    \label{eq:boldlinkdown}
    \vcenter{\hbox{
	\begin{psmatrix}[rowsep=2\baselineskip]
	  $\rnode{nA}{A}$
	  \\
	  $\rnode{A}{A}$
	  \boldvec
	  \dvec{nA}{A}
	\end{psmatrix}
    }}
    \qquad\qquad\leadsto\qquad\qquad 
    \vcenter{\hbox{
	\begin{psmatrix}[rowsep=2\baselineskip]
	  $\rnode{nA}{a}$
	  \\
	  $\rnode{A}{a}$
	  \dvec{nA}{A}
	\end{psmatrix}
    }}
    \qquad\qquad
    \vcenter{\hbox{
	\begin{psmatrix}[rowsep=2\baselineskip]
	  $\rnode{A}{\cneg a}$
	  \\
	  $\rnode{nA}{\cneg a}$
	  \uvec{nA}{A}
	\end{psmatrix}
    }}
    \qquad\qquad
    \vcenter{\hbox{
	\begin{psmatrix}[rowsep=2\baselineskip]
	  $\rnode{nA}{\ctrue}$
	  \\
	  $\rnode{A}{\ctrue}$
	  \uloop{A}
	\end{psmatrix}
    }}
    \qquad\qquad
    \vcenter{\hbox{
	\begin{psmatrix}[rowsep=2\baselineskip]
	  $\rnode{A}{\cfalse}$
	  \\
	  $\rnode{nA}{\cfalse}$
	  \dloop{A}
	\end{psmatrix}
    }}
  \end{equation}
  It is
  an easy exercise to check that all the equations demanded by the
  definitions do indeed hold. We show here only the case of the 
  contractibility axiom \eqref{eq:loopkill}:
  $$
  \def\normalvecheight{.3}  
  \def\normalarrpos{.7}  
  \vcenter{\hbox{
  \begin{psmatrix}[rowsep=2\baselineskip]
    $\ctrue$
    \\
    $\rnode{na1}{\cneg A}\quad\cor\quad\rnode{a1}{A}$
    \boldvec
    \urdvec{na1}{a1}
  \end{psmatrix}
  }}
  \qquad\qquad=\qquad\qquad
  \vcenter{\hbox{
  \begin{psmatrix}[rowsep=1.5\baselineskip]
    $\ctrue$
    \\
    $\rnode{na1}{\cneg A}\medcor\rnode{a1}{A}$
    \\
    $(\,\rnode{na21}{\cneg A}\medcor\rnode{a21}{A}\,)\medcand
    (\,\rnode{na22}{\cneg A}\medcor\rnode{a22}{A}\,)$
    \\
    $\rnode{na31}{\cneg A}\medcor(\,\rnode{a31}{A}\medcand
    \rnode{na32}{\cneg A}\,)\medcor\rnode{a32}{A}$
    \\
    $\rnode{na4}{\cneg A}\widecor\rnode{a4}{A}$
    \boldvec
    \urdvec{na1}{a1}
    \vecangles{na21}{na1}{60}{-120}
    \vecanglespos{na22}{na1}{120}{-60}{.4}
    \vecanglespos{a1}{a21}{-120}{60}{.9}
    \vecangles{a1}{a22}{-60}{120}
    \uvec{na31}{na21}
    \urvec{na32}{na22}
    \drvec{a21}{a31}
    \dvec{a22}{a32}
    \druvec{a31}{na32}
    \vecangles{na4}{na31}{120}{-60}
    \vecangles{a32}{a4}{-120}{60}
  \end{psmatrix}
  }}
  $$
  As above, the bold links represent bundles of normal links:
  \begin{equation}
    \def\normalarrpos{.7}  
    \label{eq:boldlinkloop}
    \vcenter{\hbox{
	\begin{psmatrix}[rowsep=1.5\baselineskip]
	  \\
	  $\rnode{nA}{\cneg A}\verywidecor\rnode{A}{A}$
	  \boldvec
	  \urdvec{nA}{A}
	\end{psmatrix}
    }}
    \quad\leadsto\quad 
    \vcenter{\hbox{
	\begin{psmatrix}[rowsep=1.5\baselineskip]
	  \\
	  $\rnode{nA}{\cneg a}\verywidecor\rnode{A}{a}$
	  \urdvec{nA}{A}
	\end{psmatrix}
    }}
    \qquad 
    \vcenter{\hbox{
	\begin{psmatrix}[rowsep=1.5\baselineskip]
	  \\
	  $\rnode{A}{a}\verywidecor\rnode{nA}{\cneg a}$
	  \uldvec{nA}{A}
	\end{psmatrix}
    }}
    \qquad 
    \vcenter{\hbox{
	\begin{psmatrix}[rowsep=1.5\baselineskip]
	  \\
	  $\rnode{nA}{\cfalse}\verywidecor\rnode{A}{\ctrue}$
	  \uloop{A}
	\end{psmatrix}
    }}
    \qqquad 
    \vcenter{\hbox{
	\begin{psmatrix}[rowsep=1.5\baselineskip]
	  \\
	  $\rnode{A}{\ctrue}\verywidecor\rnode{nA}{\cfalse}$
	  \uloop{A}
	\end{psmatrix}
    }}
  \end{equation}
  The transposition \eqref{eq:star} of proof nets is obtained by simply
  drawing the net in the transposed way, as it has been done with
  \eqref{eq:exa1} and \eqref{eq:exa1-two}. That this has the desired
  properties should be clear from inspecting \eqref{eq:boldlinkdown} and
  \eqref{eq:boldlinkloop}.  
\end{proof}

\begin{theorem}\label{thm:net-graphical}
  The category $\SNet(\cA)$ is graphical.
\end{theorem}

\begin{proof}
  Trivial (cf.~\cite{lam:str:05:freebool}).
\end{proof}

Note that in $\SNet(\cA)$ the sum $f+g\colon A\to B$ of two proofs
$f,g\colon A\to B$ is obtained by taking the (set-theoretic) union
of the two corresponding linkings. Hence, $\SNet(\cA)$ is idempotent,
and we have that $f\le g$ iff the linking for $f$ is a subset of the
linking for $g$.

\begin{theorem}\label{thm:net-LK}
  The category $\SNet(\cA)$ is an \LK-category.
\end{theorem}

\begin{proof}
  Let $f,g\colon A\to B$ be two maps in $\SNet(\cA)$. Let $f$ be given
  by the proof net $\prfnet{P}{\cneg A,B}$ and $g$ be given by
  $\prfnet{Q}{\cneg A,B}$. Then, by what has been said above, we
  define $f\fple g$ iff $Q\subseteq P$. After Theorem~\ref{thm:net-B4}
  it only remains to show that equation \eqref{eq:LK2} holds for all
  $f$. But this follows immediately from the definition of composition
  of proof nets.  
\end{proof}

It should be clear, that the category $\SNet(\cA)$ is quite a
degenerate model for proofs of Boolean propositional logic. The size
of a proof net is at most quadratic in the size of the sequent. This
means in particular, that the
information how often a certain link is used in a proof is not present
in the proof net. For this reason we will now allow more than one link
between a pair of complementary atoms.\footnote{In terms of
  \cite{lam:str:05:naming} this means stepping from the Boolean
  semiring of weights to the natural numbers semiring of weights.} But
as shown in \cite{lam:str:05:naming}, doing this naively means losing
confluence of cut elimination via path composition. I.e., we do not
get a category with associative arrow composition. A possible solution
has been suggested in \cite{str:SD05}:

\begin{definition}\label{def:extended}
An \dfn{extended prenet} consists of a sequent $\Gamma$, 
a finite set $K$ of \dfn{anchors}, an
\dfn{anchor labeling} $\anclab\colon K\to\cA$, and a linking (which
is now no longer a binary relation, but a binary function to the
naturals) $P\colon(\Leaf{\Gamma}\cup K)\times(\Leaf{\Gamma}\cup
K)\to\Nat$, such that
\begin{enumerate}[(i)]
\item for every $i\in\Leaf{\Gamma}$ with $\lab{i}=\ctrue$ we have
  $P(i,i)=1$,
\item for every $k\in K$ we have 
  $$
  \sum_{i\in\Leaf{\Gamma}\cup K} P(i,k)\ge 2
  \qquand
  \sum_{j\in\Leaf{\Gamma}\cup K} P(k,j)\ge 2
  $$ 
\item if $P(i,j)\ge 1$, then one of the following cases must hold:
  \begin{itemize}
  \item $i=j$ and $\lab{i}=\ctrue$ and $P(i,i)=1$, or
  \item $i\neq j$ and $i\in\Leaf{\Gamma}$ and $j\in\Leaf{\Gamma}\cup K$ and
    $\lab{i}=\cneg a$ and $\lab{j}=a$ for some $a\in\cA$, or
  \item $i\neq j$ and $i\in K$ and $j\in\Leaf{\Gamma}\cup K$ and 
    $\lab{i}=\lab{j}=a$ for some $a\in\cA$.
  \end{itemize}
\end{enumerate}
\end{definition}

As before, every $\ctrue$ has to be linked to itself. That we allow
only one and not many such links is due to Lemma~\ref{lem:tf-idem}
(which is a consequence of having proper units, cf.~\cite{str:SD05}).
But contrary to what we had before, we do now allow not only many links
between two atoms but also ``non-direct'' links visiting anchors on
their way. But each anchor can only serve links between atoms of the
same name. Furthermore, an anchor must have at least two incoming and
at least two outgoing links.  Here are two examples of extended
prenets:
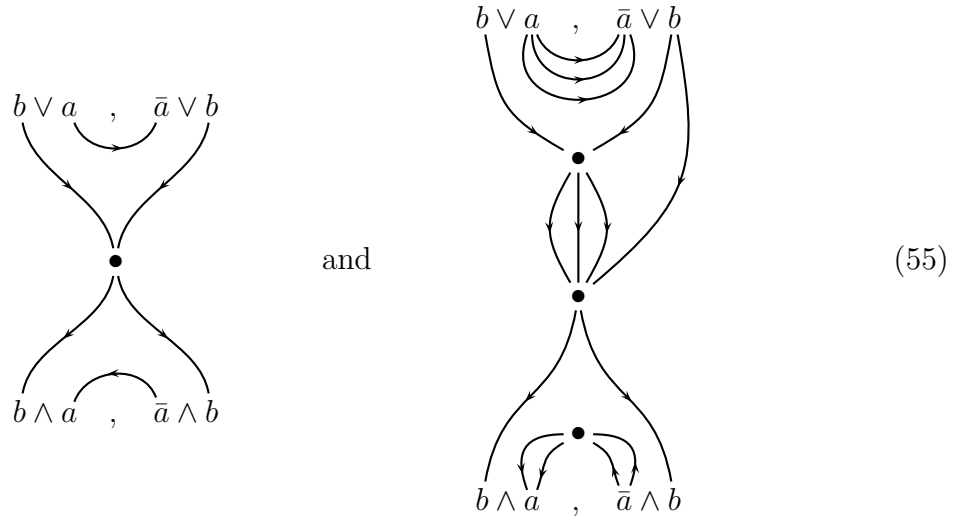
\begin{equation}
  \label{eq:extnet-exa}
  \vcenter{\hbox{
      \begin{psmatrix}[rowsep=3\baselineskip]
        $\rnode{nb2}{b}\cor\rnode{na1}{a}
        \seqsep
        \rnode{a1}{\cneg a}\cor\rnode{nb1}{b}$
        \\
	$\rnode{bb}{\anchor}$
	\\
        $\rnode{b1}{b}\cand\rnode{a2}{a}
        \seqsep
        \rnode{na2}{\cneg a}\cand\rnode{b2}{b}
        $
        \vecanglesheight{nb1}{bb}{-100}{80}{.7}
        \vecanglesheight{nb2}{bb}{-80}{100}{.7}
        \vecanglesheight{bb}{b1}{-100}{80}{.7}
        \vecanglesheight{bb}{b2}{-80}{100}{.7}
        \druvec{na1}{a1}
        \uldvec{na2}{a2}
      \end{psmatrix}
  }}
  \qqquand
  \vcenter{\hbox{
      \begin{psmatrix}[rowsep=.8\baselineskip]
        $\rnode{nb2}{b}\cor\rnode{na1}{a}
        \seqsep
        \rnode{a1}{\cneg a}\cor\rnode{nb1}{b}$
        \\ \\
	$\rnode{bb}{\anchor}$
	\\ \\
	$\rnode{bbb}{\anchor}$
	\\ \\
	$\rnode{aa}{\anchor}$
	\\
        $\rnode{b1}{b}\cand\rnode{a2}{a}
        \seqsep
        \rnode{na2}{\cneg a}\cand\rnode{b2}{b}
        $
        \vecangles{nb1}{bbb}{-80}{40}
        \vecanglespos{nb1}{bb}{-100}{30}{.8}
        \vecanglespos{nb2}{bb}{-80}{150}{.8}
        \vecangles{bb}{bbb}{-60}{60}
	\dvec{bb}{bbb}
        \vecangles{bb}{bbb}{-120}{120}
        \vecangles{bbb}{b1}{-100}{80}
        \vecangles{bbb}{b2}{-80}{100}
        \druvec{na1}{a1}
        \vecanglesheight{na1}{a1}{-90}{-90}{1.3}
        \vecanglesheight{na1}{a1}{-110}{-70}{1.8}
        \vecanglesposheight{na2}{aa}{70}{0}{.3}{1.2}
        \vecanglespos{na2}{aa}{110}{-30}{.4}
        \vecanglespos{aa}{a2}{-150}{70}{.8}
        \vecanglesposheight{aa}{a2}{180}{110}{.8}{1.2}
      \end{psmatrix}
  }}
\end{equation}
The labellings of the anchors are not shown because they are clear
from the linkings. Observe that while in a simple proof net the number
of links is at most quadratic in the size of the sequent, in an
extended net things can get arbitrarily large, as the second example
in \eqref{eq:extnet-exa} shows.

The correctness of extended nets is defined similarly as for simple
nets.  But now the condition is not that every pruning must contain a
link, but it must contain a complete path from one leaf to another (in
a conjunctive pruning the anchors remain and behave as big
disjunctions). A correct extended prenet is an \dfn{extended proof
net}. The examples in~\eqref{eq:extnet-exa} are extended proof nets.

\newtheorem{cutelim}{Cut elimination for extended nets}

\begin{cutelim}\label{para:cutelim}
  The composition of two extended proof nets is again defined via cut
  elimination, which can again be understood as path composition. But
  this time we have to be careful to treat the anchors correctly if we
  want a well-defined and associative composition.  To be formally
  precise, we define it in two steps:
  \begin{enumerate}[1.)]
  \item Replace every leaf of the cut formula by an anchor, while the links
    remain unchanged.
  \item Remove all anchors that have no right to exist (i.e., that have less
    than two incoming or outgoing links). This is done by repeatedly performing
    the following reduction steps until no further reduction is possible:
    \begin{itemize}
    \item Remove every anchor without any outcoming links, and remove all links
      coming into it.
    \item Remove every anchor without any incoming links, and remove all links
      coming out of it.
    \item If there is an anchor $k$ with only one link coming out, i.e., there
      is exactly one $i\in\Leaf{\Gamma}\cup K$ with $P(k,i)=1$, and $P(k,j)=0$
      for all $j\in\Leaf{\Gamma}\cup K$ with $j\ne i$, then 
      \begin{itemize}
      \item for every $l\in\Leaf{\Gamma}\cup K$ increase the value of $P(l,i)$
	by $P(l,k)$, i.e, all links going into $k$ are redirected to $i$, and
      \item remove $k$. 
      \end{itemize}
    \item If there is an anchor $k$ with only one link going in, i.e., there
      is exactly one $i\in\Leaf{\Gamma}\cup K$ with $P(i,k)=1$, and $P(j,k)=0$
      for all $j\in\Leaf{\Gamma}\cup K$ with $j\ne i$, then 
      \begin{itemize}
      \item for every $l\in\Leaf{\Gamma}\cup K$ increase the value of $P(i,l)$
	by $P(k,l)$, i.e, all links coming out of $k$ are replaced by links
	coming out of $i$, and
      \item remove $k$. 
      \end{itemize}
    \item Set $P(k,k)=0$ for all $k\in K$.
    \item Set $P(i,i)=1$ for all $i$ with $\lab{i}=\ctrue$.
    \end{itemize}
  \end{enumerate}
\end{cutelim}

This sounds more complicated than it actually is. As first example, we show
here the case of the contractibility axiom (applied to a single atom):
{\footnotesize \def\thisto{\quad\lower3ex\hbox{$\leadsto$}\quad}
  \def\A{a}
  \def\nA{\cneg a}
  \def\agap{\hskip1.5em}
  \def\widecor{\;\cor\;}
  \def\widecand{\;\cand\;}
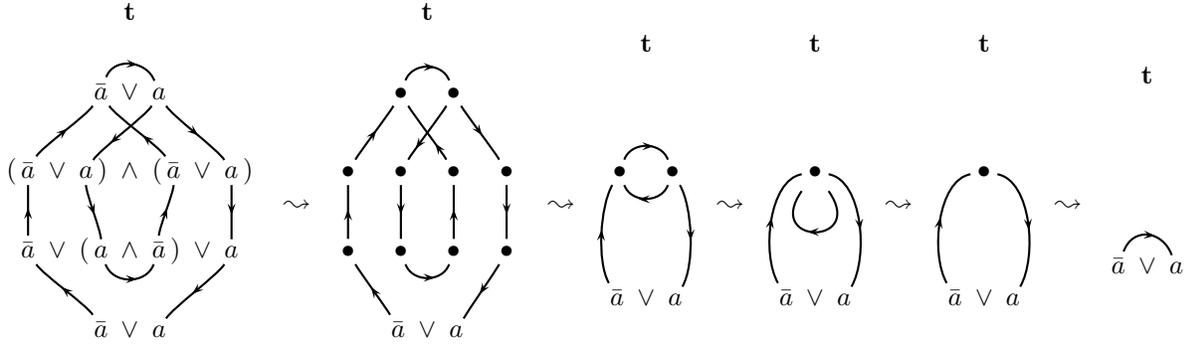
  $$
  \def\normalvecheight{.3}
  \vcenter{\hbox{
  \begin{psmatrix}[rowsep=1.5\baselineskip]
    $\ctrue$
    \\
    $\rnode{na1}{\nA}\medcor\rnode{a1}{\A}$
    \\
    $\klam{\rnode{na21}{\nA}\medcor\rnode{a21}{\A}}\medcand
    \klam{\rnode{na22}{\nA}\medcor\rnode{a22}{\A}}$
    \\
    $\rnode{na31}{\nA}\medcor
    \klam{\rnode{a31}{\A}\medcand\rnode{na32}{\nA}}
    \medcor\rnode{a32}{\A}$
    \\
    $\rnode{na4}{\nA}\widecor\rnode{a4}{\A}$
    \urdvec{na1}{a1}
    \vecangles{na21}{na1}{60}{-120}
    \vecanglespos{na22}{na1}{120}{-60}{.4}
    \vecanglespos{a1}{a21}{-120}{60}{.7}
    \vecangles{a1}{a22}{-60}{120}
    \uvec{na31}{na21}
    \uvec{na32}{na22}
    \dvec{a21}{a31}
    \dvec{a22}{a32}
    \druvec{a31}{na32}
    \vecangles{na4}{na31}{120}{-60}
    \vecangles{a32}{a4}{-120}{60}
  \end{psmatrix}
  }}
  \!\thisto\!
  \vcenter{\hbox{
  \begin{psmatrix}[rowsep=1.5\baselineskip]
    $\ctrue$
    \\
    $\rnode{na1}{\anchor}\agap\rnode{a1}{\anchor}$
    \\
    $\rnode{na21}{\anchor}\agap\rnode{a21}{\anchor}\agap
    \rnode{na22}{\anchor}\agap\rnode{a22}{\anchor}$
    \\
    $\rnode{na31}{\anchor}\agap\rnode{a31}{\anchor}\agap
    \rnode{na32}{\anchor}\agap\rnode{a32}{\anchor}$
    \\
    $\rnode{na4}{\nA}\widecor\rnode{a4}{\A}$
    \urdvec{na1}{a1}
    \vecangles{na21}{na1}{60}{-120}
    \vecanglespos{na22}{na1}{120}{-60}{.3}
    \vecanglespos{a1}{a21}{-120}{60}{.8}
    \vecangles{a1}{a22}{-60}{120}
    \uvec{na31}{na21}
    \uvec{na32}{na22}
    \dvec{a21}{a31}
    \dvec{a22}{a32}
    \druvec{a31}{na32}
    \vecangles{na4}{na31}{120}{-60}
    \vecangles{a32}{a4}{-120}{60}
  \end{psmatrix}
  }}
  \thisto
  \vcenter{\hbox{
  \begin{psmatrix}[rowsep=3\baselineskip]
    $\ctrue$
    \\
    $\rnode{na1}{\anchor}\agap\rnode{a1}{\anchor}$
    \\
    $\rnode{na4}{\nA}\widecor\rnode{a4}{\A}$
    \urdvec{na1}{a1}
    \dluvec{a1}{na1}
    \vecanglesheight{na4}{na1}{120}{-120}{.5}
    \vecanglesheight{a1}{a4}{-60}{60}{.5}
  \end{psmatrix}
  }}
  \thisto
  \vcenter{\hbox{
  \begin{psmatrix}[rowsep=3\baselineskip]
    $\ctrue$
    \\
    $\rnode{a1}{\strut\anchor}$
    \\
    $\rnode{na4}{\nA}\widecor\rnode{a4}{\A}$
    \dloop{a1}
    \vecanglesheight{na4}{a1}{120}{-170}{.6}
    \vecanglesheight{a1}{a4}{-10}{60}{.6}
  \end{psmatrix}
  }}
  \thisto
  \vcenter{\hbox{
  \begin{psmatrix}[rowsep=3\baselineskip]
    $\ctrue$
    \\
    $\rnode{a1}{\strut\anchor}$
    \\
    $\rnode{na4}{\nA}\widecor\rnode{a4}{\A}$
    \vecanglesheight{na4}{a1}{120}{-170}{.6}
    \vecanglesheight{a1}{a4}{-10}{60}{.6}
  \end{psmatrix}
  }}
  \thisto\hskip-.5ex
  \vcenter{\hbox{
  \begin{psmatrix}[rowsep=5\baselineskip]
    $\ctrue$
    \\
    $\rnode{na1}{\nA}\widecor\rnode{a1}{\A}$
    \urdvecheight{na1}{a1}{1.1}
  \end{psmatrix}
  }}
  $$
}%
Here are two other examples (compare with \eqref{eq:exa1-exa2-two}):
{\small
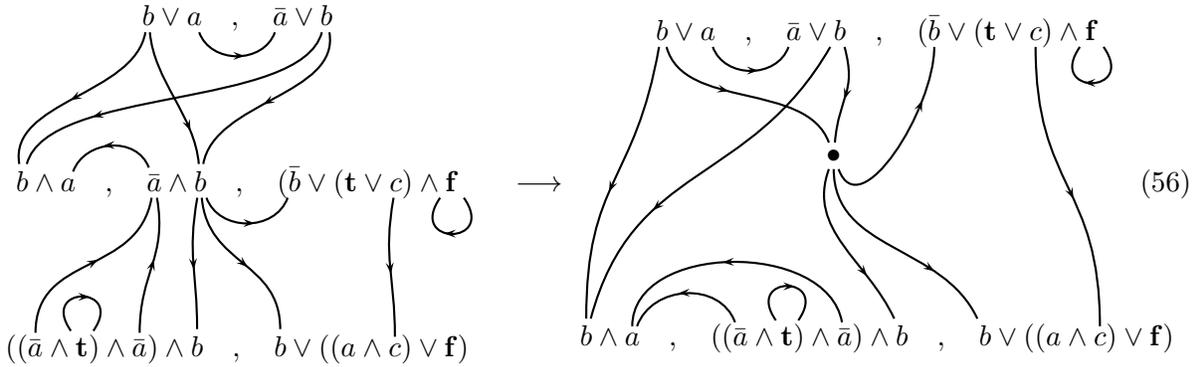
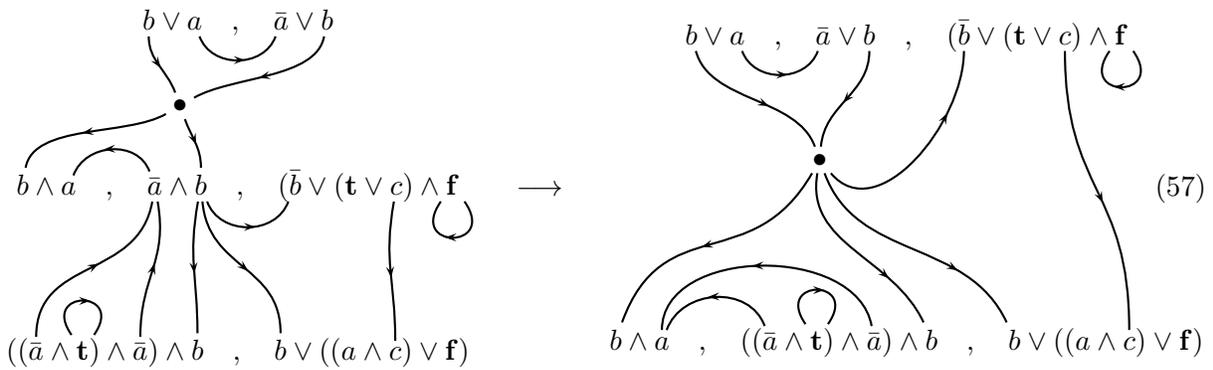
\begin{equation}
  \label{eq:extnet-cutreda}
  \qlapm{\vcenter{\hbox{
      \begin{psmatrix}[rowsep=1.3\baselineskip]
        $\rnode{xnb2}{b}\cor\rnode{xna1}{a}
        \seqsep
        \rnode{xa1}{\cneg a}\cor\rnode{xnb1}{b}$
        \\ \\
        $\rnode{xb1}{b}\cand\rnode{xa2}{a}
        \seqsep
        \rnode{cna2}{\cneg a}\cand\rnode{cb2}{b}
        \seqsep
        (\rnode{b3}{\cneg b}\cor(\rnode{f2}{\ctrue}\cor\rnode{nc1}{c})
        \cand\rnode{t2}{\cfalse}
        $
        \\ \\
        $((\rnode{na1}{\cneg a}\cand\rnode{t1}{\ctrue})
        \cand\rnode{na2}{\cneg a})\cand\rnode{b1}{b}
        \seqsep
        \rnode{b2}{b}\cor((\rnode{a2}{a}\cand\rnode{c1}{c})
        \cor\rnode{f1}{\cfalse})
        $
        \vecanglesheight{xnb1}{cb2}{-80}{80}{.7}
        \vecanglesposheight{xnb1}{xb1}{-105}{75}{.75}{.5}
        \vecanglesposheight{xnb2}{cb2}{-85}{95}{.75}{.7}
        \vecanglesheight{xnb2}{xb1}{-100}{100}{.7}
        \druvec{xna1}{xa1}
        \uldvec{cna2}{xa2}
        \uloop{t1}
        \dloop{t2}
        \vecanglesheight{na1}{cna2}{90}{-100}{.9}
        \vecanglesheight{na2}{cna2}{90}{-80}{.9}
        \vecanglesheight{nc1}{c1}{-100}{90}{.9}
        \vecanglesheight{cb2}{b1}{-100}{90}{.9}
        \vecanglesheight{cb2}{b2}{-85}{90}{.9}
        \druvec{cb2}{b3}
      \end{psmatrix}
      }}
  \quad\longrightarrow
  \vcenter{\hbox{
      \begin{psmatrix}[rowsep=.7\baselineskip]
        $\rnode{xnb2}{b}\cor\rnode{xna1}{a}
        \seqsep
        \rnode{xa1}{\cneg a}\cor\rnode{xnb1}{b}
        \seqsep
        (\rnode{b3}{\cneg b}\cor(\rnode{f2}{\ctrue}\cor\rnode{nc1}{c})
        \cand\rnode{t2}{\cfalse}
        $
        \\ \\
	$\rnode{bb}{\anchor}\qquad\quad$
	\\ \\ \\
        $
        \rnode{xb1}{b}\cand\rnode{xa2}{a}
        \seqsep
        ((\rnode{na1}{\cneg a}\cand\rnode{t1}{\ctrue})
        \cand\rnode{na2}{\cneg a})\cand\rnode{b1}{b}
        \seqsep
        \rnode{b2}{b}\cor((\rnode{a2}{a}\cand\rnode{c1}{c})
        \cor\rnode{f1}{\cfalse})
        $
        \vecanglesposheight{xnb1}{xb1}{-120}{75}{.65}{.9}
        \vecanglesheight{xnb2}{xb1}{-100}{90}{1}
        \druvec{xna1}{xa1}
        \uloop{t1}
        \dloop{t2}
        \uldvec{na1}{xa2}
        \vecanglesheight{na2}{xa2}{100}{90}{.8}
        \dvec{nc1}{c1}
        \vecanglesposheight{xnb2}{bb}{-75}{110}{.4}{.7}
        \vecanglesheight{xnb1}{bb}{-75}{85}{1}
        \vecanglesposheight{bb}{b1}{-110}{115}{.7}{.8}
        \vecanglesposheight{bb}{b2}{-90}{115}{.7}{.8}
        \vecanglesposheight{bb}{b3}{-70}{-90}{.7}{.8}
    \end{psmatrix}
    }}}
\end{equation}}%
and
{\small
\begin{equation}
  \label{eq:extnet-cutredb}
  \qqqqqlapm{\vcenter{\hbox{
      \begin{psmatrix}[rowsep=1.3\baselineskip]
        $\rnode{xnb2}{b}\cor\rnode{xna1}{a}
        \seqsep
        \rnode{xa1}{\cneg a}\cor\rnode{xnb1}{b}$
        \\
	$\rnode{bb}{\anchor}\qquad\qquad$
	\\
        $\rnode{xb1}{b}\cand\rnode{xa2}{a}
        \seqsep
        \rnode{cna2}{\cneg a}\cand\rnode{cb2}{b}
        \seqsep
        (\rnode{b3}{\cneg b}\cor(\rnode{f2}{\ctrue}\cor\rnode{nc1}{c})
        \cand\rnode{t2}{\cfalse}
        $
        \\ \\
        $((\rnode{na1}{\cneg a}\cand\rnode{t1}{\ctrue})
        \cand\rnode{na2}{\cneg a})\cand\rnode{b1}{b}
        \seqsep
        \rnode{b2}{b}\cor((\rnode{a2}{a}\cand\rnode{c1}{c})
        \cor\rnode{f1}{\cfalse})
        $
        \vecanglesheight{xnb2}{bb}{-90}{110}{.8}
        \vecanglesheight{xnb1}{bb}{-100}{30}{.6}
        \vecanglesheight{bb}{xb1}{-150}{80}{.6}
        \vecanglesheight{bb}{cb2}{-70}{90}{.8}
        \druvec{xna1}{xa1}
        \uldvec{cna2}{xa2}
        \uloop{t1}
        \dloop{t2}
        \vecanglesheight{na1}{cna2}{90}{-100}{.9}
        \vecanglesheight{na2}{cna2}{90}{-80}{.9}
        \vecanglesheight{nc1}{c1}{-100}{90}{.9}
        \vecanglesheight{cb2}{b1}{-100}{90}{.9}
        \vecanglesheight{cb2}{b2}{-85}{90}{.9}
        \druvec{cb2}{b3}
      \end{psmatrix}
      }}
  \quad\longrightarrow\quad
  \vcenter{\hbox{
      \begin{psmatrix}[rowsep=.7\baselineskip]
        $\rnode{xnb2}{b}\cor\rnode{xna1}{a}
        \seqsep
        \rnode{xa1}{\cneg a}\cor\rnode{xnb1}{b}
        \seqsep
        (\rnode{b3}{\cneg b}\cor(\rnode{f2}{\ctrue}\cor\rnode{nc1}{c})
        \cand\rnode{t2}{\cfalse}
        $
        \\ \\
	$\rnode{bb}{\anchor}\qquad\qquad\qquad$
	\\ \\ \\
        $
        \rnode{xb1}{b}\cand\rnode{xa2}{a}
        \seqsep
        ((\rnode{na1}{\cneg a}\cand\rnode{t1}{\ctrue})
        \cand\rnode{na2}{\cneg a})\cand\rnode{b1}{b}
        \seqsep
        \rnode{b2}{b}\cor((\rnode{a2}{a}\cand\rnode{c1}{c})
        \cor\rnode{f1}{\cfalse})
        $
        \druvec{xna1}{xa1}
        \uloop{t1}
        \dloop{t2}
        \uldvec{na1}{xa2}
        \vecanglesheight{na2}{xa2}{100}{90}{.8}
        \dvec{nc1}{c1}
        \vecanglesposheight{xnb2}{bb}{-75}{110}{.6}{.7}
        \vecanglesheight{xnb1}{bb}{-90}{85}{.8}
        \vecanglesheight{bb}{xb1}{-120}{70}{1}
        \vecanglesposheight{bb}{b1}{-100}{115}{.7}{.8}
        \vecanglesposheight{bb}{b2}{-70}{115}{.7}{.8}
        \vecanglesposheight{bb}{b3}{-50}{-90}{.7}{.8}
    \end{psmatrix}
    }}}
\end{equation}}

The whole point of the construction of the extended proof nets is, that we
again get a category, which we denote by $\ENet(\cA)$.\footnote{The
restrictions made to the anchors in Definition~\ref{def:extended} are chosen
such that we indeed get a well-defined and associative composition.}  This
category is again a \Bv-category and it is graphical. But it is \emph{not}
flat because the following two nets are obviously not equal:
\begin{equation}
  \label{eq:delta-nabla-pn}
  \vcenter{\hbox{
      \begin{psmatrix}[rowsep=4\baselineskip]
	$\rnode{na1}{a}\widecor\rnode{na2}{a}$
	\\
	$\rnode{a1}{a}\widecand\rnode{a2}{a}$
	\vecangles{na1}{a1}{-105}{105}
	\vecangles{na2}{a2}{-75}{75}
	\vecanglesposheight{na1}{a2}{-85}{95}{.75}{.7}
	\vecanglesposheight{na2}{a1}{-95}{85}{.75}{.7}
      \end{psmatrix}
  }}
  \qqqquand
  \def\normalvecheight{.7}
  \vcenter{\hbox{
      \begin{psmatrix}[rowsep=1.5\baselineskip]
	$\rnode{na1}{a}\widecor\rnode{na2}{a}$
	\\
	$\rnode{aa}{\anchor}$
	\\
	$\rnode{a1}{a}\widecand\rnode{a2}{a}$
	\vecangles{na1}{aa}{-95}{105}
	\vecangles{na2}{aa}{-85}{75}
	\vecanglespos{aa}{a1}{-105}{95}{.65}
	\vecanglespos{aa}{a2}{-75}{85}{.65}
      \end{psmatrix}
  }}
\end{equation}
which means that diagram \eqref{eq:delta-nabla} does not commute.
It is also easy to see that $\ENet(\cA)$ is not idempotent, and is
therefore not an \LK-category. However, we have:

\begin{theorem}\label{thm:enet-B4}
  The category $\ENet(\cA)$ is a \Bv-category that is weakly flat and
  contractible.
\end{theorem}

\begin{proof}
  Again, the maps $\assoc$, $\twist$ $\runit$, $\lunit$, $\switch$,
  $\medial$, $\proj$, and $\diag$ are given by the obvious nets. It is
  an easy exercise to check that the equations demanded by the
  definitions do still hold.
\end{proof}

\begin{theorem}\label{thm:enet-graphical}
  The category $\ENet(\cA)$ is graphical.
\end{theorem}

\begin{proof}
  Trivial.
\end{proof}

\begin{figure}[t]
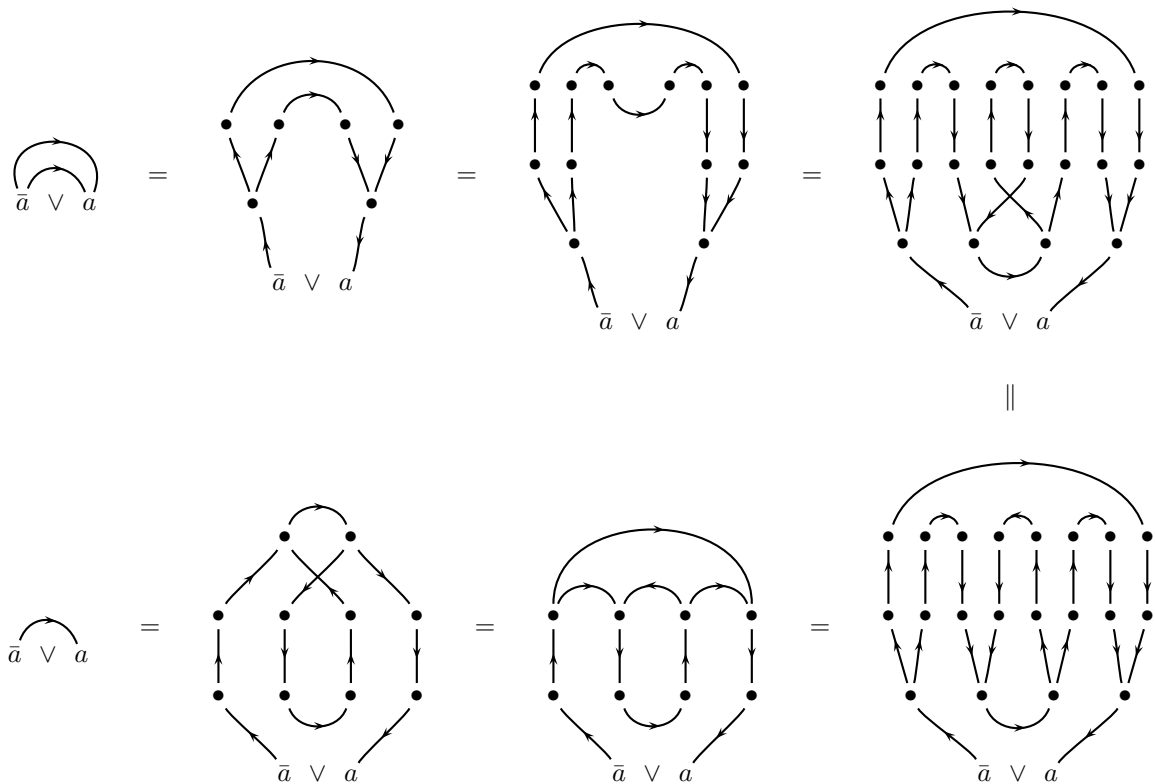

  \begin{center}
    {\footnotesize 
      \def\thisto{\lower.75ex\hbox to 4em{\hss$=$\hss}}
      \def\A{a}
      \def\nA{\cneg a}
      \def\normalvecheight{.3}
      $$ 
	\vcenter{\hbox{
	    \begin{psmatrix}[rowsep=1.5\baselineskip]
	      \\
	      $\rnode{na1}{\nA}\widecor\rnode{a1}{\A}$
	      \urdvecheight{na1}{a1}{1.1}
	      \vecanglesheight{na1}{a1}{110}{70}{1.8}
	    \end{psmatrix}
	}}
	\thisto
	\vcenter{\hbox{
	    \begin{psmatrix}[rowsep=1.5\baselineskip]
	      \\
	      $
	      \rnode{a21}{\anchor}\agapmed\rnode{a22}{\anchor}
	      \agap 
	      \rnode{a27}{\anchor}\agapmed\rnode{a28}{\anchor}
	      $
	      \\
	      $\rnode{a31}{\anchor}\agap\agap 
	      \rnode{a34}{\anchor}$
	      \\
	      $\rnode{na4}{\nA}\widecor\rnode{a4}{\A}$
	      \vecanglesheight{a21}{a28}{70}{110}{.9}
	      \urdvec{a22}{a27}
	      \ulvecpos{a31}{a21}{.8} \urvecpos{a31}{a22}{.8}
	      \drvecpos{a27}{a34}{.5} \dlvecpos{a28}{a34}{.5} 
	      \vecangles{na4}{a31}{120}{-60}
	      \vecangles{a34}{a4}{-120}{60}
	    \end{psmatrix}
	}}
	\thisto
	\vcenter{\hbox{
	    \begin{psmatrix}[rowsep=1.5\baselineskip]
	      \\
	      $
	      \rnode{a11}{\anchor}\agapdemi\rnode{a12}{\anchor}
	      \agapdemi
	      \rnode{a13}{\anchor}\agapdemi\agapdemi\rnode{a16}{\anchor}
	      \agapdemi
	      \rnode{a17}{\anchor}\agapdemi\rnode{a18}{\anchor}
	      $
	      \\
	      $
	      \rnode{a21}{\anchor}\agapdemi\rnode{a22}{\anchor}
	      \agapdemi
	      \phantom{{\anchor}\agapdemi\agapdemi{\anchor}}
	      \agapdemi
	      \rnode{a27}{\anchor}\agapdemi\rnode{a28}{\anchor}
	      $
	      \\
	      $\rnode{a31}{\anchor}\agapplus\agapplus 
	      \rnode{a34}{\anchor}$
	      \\
	      $\rnode{na4}{\nA}\widecor\rnode{a4}{\A}$
	      \vecanglesheight{a11}{a18}{70}{110}{.7}
	      \urdvec{a12}{a13}
	      \urdvec{a16}{a17}
	      \uvecpos{a21}{a11}{.8} \uvecpos{a22}{a12}{.8} 
	      \druvec{a13}{a16}
	      \dvecpos{a17}{a27}{.8} \dvecpos{a18}{a28}{.8} 
	      \vecanglespos{a31}{a21}{120}{-75}{.8} \uvecpos{a31}{a22}{.8}
	      \dvecpos{a27}{a34}{.5} \vecanglespos{a28}{a34}{-105}{60}{.5} 
	      \vecangles{na4}{a31}{120}{-60}
	      \vecangles{a34}{a4}{-120}{60}
	    \end{psmatrix}
	}}
	\thisto
	\vcenter{\hbox{
	    \begin{psmatrix}[rowsep=1.5\baselineskip]
	      \\
	      $
	      \rnode{a11}{\anchor}\agapdemi\rnode{a12}{\anchor}\agapdemi
	      \rnode{a13}{\anchor}\agapdemi\rnode{a15}{\anchor}\agapdemi
	      \rnode{a14}{\anchor}\agapdemi\rnode{a16}{\anchor}\agapdemi
	      \rnode{a17}{\anchor}\agapdemi\rnode{a18}{\anchor}
	      $
	      \\
	      $
	      \rnode{a21}{\anchor}\agapdemi\rnode{a22}{\anchor}\agapdemi
	      \rnode{a23}{\anchor}\agapdemi\rnode{a25}{\anchor}\agapdemi
	      \rnode{a24}{\anchor}\agapdemi\rnode{a26}{\anchor}\agapdemi
	      \rnode{a27}{\anchor}\agapdemi\rnode{a28}{\anchor}
	      $
	      \\
	      $\rnode{a31}{\anchor}\agapplus\rnode{a32}{\anchor}\agapplus
	      \rnode{a33}{\anchor}\agapplus\rnode{a34}{\anchor}$
	      \\
	      $\rnode{na4}{\nA}\widecor\rnode{a4}{\A}$
	      \vecanglesheight{a11}{a18}{70}{110}{.7}
	      \urdvec{a12}{a13}
	      \urdvec{a15}{a14}
	      \urdvec{a16}{a17}
	      \uvecpos{a21}{a11}{.8} \uvecpos{a22}{a12}{.8} 
	      \dvecpos{a13}{a23}{.8} \dvecpos{a14}{a24}{.8} 
	      \uvecpos{a25}{a15}{.8} \uvecpos{a26}{a16}{.8} 
	      \dvecpos{a17}{a27}{.8} \dvecpos{a18}{a28}{.8} 
	      \ulvecpos{a31}{a21}{.8} \urvecpos{a31}{a22}{.8}
	      \drvecpos{a23}{a32}{.5} \dlvecpos{a24}{a32}{.8} 
	      \ulvecpos{a33}{a25}{.4} \urvecpos{a33}{a26}{.8}
	      \drvecpos{a27}{a34}{.5} \dlvecpos{a28}{a34}{.5} 
	      \druvec{a32}{a33}
	      \vecangles{na4}{a31}{120}{-60}
	      \vecangles{a34}{a4}{-120}{60}
	    \end{psmatrix}
	}}
	$$
	$$ \hskip 32.5em\vbox{\vskip1em\hbox{$\Vert$}\vskip1em} $$
	$$
	\vcenter{\hbox{
	    \begin{psmatrix}[rowsep=1.5\baselineskip]
	      \\
	      $\rnode{na1}{\nA}\widecor\rnode{a1}{\A}$
	      \urdvecheight{na1}{a1}{1.1}
	    \end{psmatrix}
	}}
	\thisto
	\vcenter{\hbox{
	    \begin{psmatrix}[rowsep=1.5\baselineskip]
	      \\
	      $\rnode{na1}{\anchor}\agap\rnode{a1}{\anchor}$
	      \\
	      $\rnode{na21}{\anchor}\agap\rnode{a21}{\anchor}\agap
	      \rnode{na22}{\anchor}\agap\rnode{a22}{\anchor}$
	      \\
	      $\rnode{na31}{\anchor}\agap\rnode{a31}{\anchor}\agap
	      \rnode{na32}{\anchor}\agap\rnode{a32}{\anchor}$
	      \\
	      $\rnode{na4}{\nA}\widecor\rnode{a4}{\A}$
	      \urdvec{na1}{a1}
	      \vecangles{na21}{na1}{60}{-120}
	      \vecanglespos{na22}{na1}{120}{-60}{.3}
	      \vecanglespos{a1}{a21}{-120}{60}{.8}
	      \vecangles{a1}{a22}{-60}{120}
	      \uvec{na31}{na21}
	      \uvec{na32}{na22}
	      \dvec{a21}{a31}
	      \dvec{a22}{a32}
	      \druvec{a31}{na32}
	      \vecangles{na4}{na31}{120}{-60}
	      \vecangles{a32}{a4}{-120}{60}
	    \end{psmatrix}
	}}
	\thisto
	\vcenter{\hbox{
	    \begin{psmatrix}[rowsep=1.5\baselineskip]
	      \\ \\
	      $\rnode{na21}{\anchor}\agap\rnode{a21}{\anchor}\agap
	      \rnode{na22}{\anchor}\agap\rnode{a22}{\anchor}$
	      \\
	      $\rnode{na31}{\anchor}\agap\rnode{a31}{\anchor}\agap
	      \rnode{na32}{\anchor}\agap\rnode{a32}{\anchor}$
	      \\
	      $\rnode{na4}{\nA}\widecor\rnode{a4}{\A}$
	      \vecanglesheight{na21}{a22}{90}{90}{1}
	      \urdvec{na21}{a21}
	      \uldvec{na22}{a21}
	      \urdvec{na22}{a22}
	      \uvec{na31}{na21}
	      \uvec{na32}{na22}
	      \dvec{a21}{a31}
	      \dvec{a22}{a32}
	      \druvec{a31}{na32}
	      \vecangles{na4}{na31}{120}{-60}
	      \vecangles{a32}{a4}{-120}{60}
	    \end{psmatrix}
	}}
	\thisto
	\vcenter{\hbox{
	    \begin{psmatrix}[rowsep=1.5\baselineskip]
	      \\
	      $
	      \rnode{a11}{\anchor}\agapdemi\rnode{a12}{\anchor}\agapdemi
	      \rnode{a13}{\anchor}\agapdemi\rnode{a14}{\anchor}\agapdemi
	      \rnode{a15}{\anchor}\agapdemi\rnode{a16}{\anchor}\agapdemi
	      \rnode{a17}{\anchor}\agapdemi\rnode{a18}{\anchor}
	      $
	      \\
	      $
	      \rnode{a21}{\anchor}\agapdemi\rnode{a22}{\anchor}\agapdemi
	      \rnode{a23}{\anchor}\agapdemi\rnode{a24}{\anchor}\agapdemi
	      \rnode{a25}{\anchor}\agapdemi\rnode{a26}{\anchor}\agapdemi
	      \rnode{a27}{\anchor}\agapdemi\rnode{a28}{\anchor}
	      $
	      \\
	      $\rnode{a31}{\anchor}\agapplus\rnode{a32}{\anchor}\agapplus
	      \rnode{a33}{\anchor}\agapplus\rnode{a34}{\anchor}$
	      \\
	      $\rnode{na4}{\nA}\widecor\rnode{a4}{\A}$
	      \vecanglesheight{a11}{a18}{70}{110}{.7}
	      \urdvec{a12}{a13}
	      \uldvec{a15}{a14}
	      \urdvec{a16}{a17}
	      \uvecpos{a21}{a11}{.8} \uvecpos{a22}{a12}{.8} 
	      \dvecpos{a13}{a23}{.8} \dvecpos{a14}{a24}{.8} 
	      \uvecpos{a25}{a15}{.8} \uvecpos{a26}{a16}{.8} 
	      \dvecpos{a17}{a27}{.8} \dvecpos{a18}{a28}{.8} 
	      \ulvecpos{a31}{a21}{.8} \urvecpos{a31}{a22}{.8}
	      \drvecpos{a23}{a32}{.5} \dlvecpos{a24}{a32}{.5} 
	      \ulvecpos{a33}{a25}{.8} \urvecpos{a33}{a26}{.8}
	      \drvecpos{a27}{a34}{.5} \dlvecpos{a28}{a34}{.5} 
	      \druvec{a32}{a33}
	      \vecangles{na4}{a31}{120}{-60}
	      \vecangles{a34}{a4}{-120}{60}
	    \end{psmatrix}
	}}
    $$ 
    }
    \caption{The idea of the proof of Theorem~\ref{thm:almost-idem}}
    \label{fig:almost-idem}
  \end{center}
\end{figure}

In Figure~\ref{fig:almost-idem} we use the notation of the extended
proof nets to illustrate the idea behind the proof of
Theorem~\ref{thm:almost-idem}. The middle equation in the second line
is \eqref{eq:delta-nabla}, i.e, the identity of the two nets in
\eqref{eq:delta-nabla-pn}. The left-most equation in the second line
and the right-most equation in the first line are both the
contractibility equation \eqref{eq:loopkill}. Everything else in
Figure~\ref{fig:almost-idem} is rather trivial from the viewpoint of
proof nets. However, since we do not have a ``coherence theorem'',
Figure~\ref{fig:almost-idem} cannot tell us whether the equations are
really consequences of the axioms. For this, the proper proof in
Section~\ref{sec:beyond} is necessary.

\section{More thoughts on order enrichment}

Although $\ENet(\cA)$ is not an \LK-category, we can enrich it with a
partial order which is induced by cut elimination according to the
ideas of \cite{fuhrmann:pym:oecm,fuhrmann:pym:dj}. This means that
$f\fple g$ if $g$ is obtained from $f$ via cut elimination in
some formal system (not necessarily \LK\ or another sequent system).

In category-theoretic terms, this is achieved by keeping properties
\eqref{l:LK-monoton} and \eqref{l:LK-iso} in Definition~\ref{def:LK},
but by dropping \eqref{eq:LK1} and \eqref{eq:LK2}. It should be clear
that there is a wide range of possibilities of providing such a
partial order. As an example we will sketch here the idea which has
been proposed in~\cite{str:SD05}.

Let $f\colon A\to B$ be a map in $\ENet(\cA)$, i.e., an extended proof
net, and let $k\in K_f$ be an anchor in $f$, and let $P_f$ be the
linking of $f$.  Then we can remove $k$ according to the cut
elimination for simple proof nets (as defined in
\cite{lam:str:05:naming}).  Let $K'$ be $K_f\setminus\set{k}$ and for
all $i,j\in\Leaf{\Gamma}\cup K'$ let $P'(i,j)=P_f(i,j)+P_f(i,k)\cdot
P_f(k,j)$. Now define $g\colon A\to B$ to be the result of applying
the second step of \ref{para:cutelim} to $P',K'$.

For example the net on the left in \eqref{eq:delta-nabla-pn} is the result
of eliminating the anchor of the net on the right in
\eqref{eq:delta-nabla-pn}.

Note that this anchor elimination process is not confluent, i.e., in a
net with many anchors, the result of eliminating all of them depends
on the order in which they are eliminated. This has been shown in
\cite{lam:str:05:naming}, but morally it is a consequence of
Theorem~\ref{thm:almost-idem}.

There is also a close relationship to cut elimination in the calculus
of structures. There is work in progress to nail down the precise
relation between the anchor elimination for proof nets defined above
and the splitting technique \cite{gug:SIS} for elimination the cuts in
system $\SKS$ \cite{brunnler:phd}.

Let us finish this paper by proposing yet another way of enriching
$\ENet(\cA)$ with a partial order: Since maps in that category are just
directed graphs, we can define $f\fple g$ if $g$ is a minor of~$f$ in the
graph-theoretic sense. We have to leave it as problem for future work to
investigate the proof-theoretic implications of this.

\section*{Acknowledgments}

Some essential parts of the research described here have been carried out
while the author was working at Saarland University, Programming Systems Lab.

I am grateful to Fran\c{c}ois Lamarche for many fruitful discussions
and helpful comments on early drafts of this paper. Kai Br\"unnler,
Alessio Guglielmi, and Richard McKinley contributed indirectly to this
work by asking the right questions at the right time. Furthermore I
would thank the anonymous referee for several significant comments
that helped to improve the paper.

\bibliographystyle{alpha} \bibliography{my_lit}

\end{document}